\documentclass[preprint]{aa} 
\usepackage{graphicx}
\usepackage{txfonts}
\newcommand{\1}{{~\sc i}}
\newcommand{\2}{{~\sc ii}}
\newcommand{\3}{{~\sc iii}}

\newcommand{\wm}{{\,W\,m$^{-2}$}}
\newcommand{\wmsr}{{\,W\,m$^{-2}$\,sr$^{-1}$}}
\newcommand{\kms}{{\,km\,s$^{-1}$}}
\newcommand{\cc}{{\,cm$^{-3}$}}
\newcommand{\mic}{{\,$\mu$m}}

\begin{document}
   \title{Physical conditions in the gas phases of the giant H\2\ region LMC-N\,11 unveiled by \textit{Herschel}\thanks{{\it Herschel} is an ESA space observatory with science instruments provided by European-led Principal Investigator consortia and with important participation from NASA.}}

   \subtitle{I. Diffuse [C\2] and [O\3] emission in LMC-N\,11B}
   
   \titlerunning{Physical conditions in the gas phases of the H\2\ region LMC-N\,11B unveiled by Herschel}
   \authorrunning{Lebouteiller et al.}

   \author{V.\ Lebouteiller\inst{1}, D.\ Cormier\inst{1}, S.\,C.\ Madden\inst{1}, F.\ Galliano\inst{1}, R.\ Indebetouw\inst{2}, N.\ Abel\inst{3}, M.\ Sauvage\inst{1}, S.\ Hony\inst{1}, A.\ Contursi\inst{4}, A.\ Poglitsch\inst{4}, A.\ R{\'e}my\inst{1}, E.\ Sturm\inst{4}, R.\ Wu\inst{1}
          }

   \institute{$1:$ Laboratoire AIM, CEA/DSM-CNRS-Universit\'e Paris Diderot DAPNIA/Service d'Astrophysique B\^at. 709, CEA-Saclay F-91191 Gif-sur-Yvette C\'edex, France \email{vianney.lebouteiller@cea.fr} \\
   $2:$ Department of Astronomy, University of Virginia, P.O. Box 3818, Charlottesville, VA 22903, USA \\
   $3:$ Department of Physics, University of Cincinnati, Cincinnati, OH 45221, USA \\
   $4:$ Max-Planck-Institute for Extraterrestrial Physics (MPE), Giessenbachstra§e 1, 85748 Garching, Germany 
             }

   \date{Received 20/01/2012; accepted 26/09/212}

 
  \abstract
   {The Magellanic Clouds provide a nearby laboratory for metal-poor dwarf galaxies. The low dust abundance enhances the penetration of UV photons into the interstellar medium (ISM), resulting in a relatively larger filling factor of the ionized gas. Furthermore, there is very likely a molecular gas reservoir probed by the [C\2] 157\mic\ line not traced by CO(1-0), the so-called ``dark" gas. The \textit{Herschel} Space Telescope allows us to observe far-infrared (FIR) cooling lines and to examine the physical conditions in the gas phases of a low-metallicity environment to unprecedented, small spatial scales.  }
   {Our objective is to interpret the origin of the diffuse emission of FIR cooling lines in the H\2\ region N\,11B in the Large Magellanic Cloud. We first investigate the filling factor of the ionized gas. We then constrain the origin of the [C\2] line by comparing to tracers of the low-excitation ionized gas and of photodissociation regions (PDRs). }
   {We present \textit{Herschel}/PACS maps of N\,11B in several tracers, [C\2] 157\mic, [O\1] 63\mic\ and 145\mic, [N\2] 122\mic, [N\3] 57\mic, and [O\3] 88\mic. Optical images in H$\alpha$ and [O\3] 5007\AA\ were used as complementary data to investigate the effect of dust extinction. Observations were interpreted with photoionization models to infer the gas conditions and estimate the ionized gas contribution to the [C\2] emission. PDRs were probed through polycyclic aromatic hydrocarbons (PAHs) observed with the \textit{Spitzer} Space Telescope. }
   {[O\3] 88\mic\ is dominated by extended emission from the high-excitation diffuse ionized gas. This is the brightest FIR line throughout N\,11B, $\sim4$ times brighter than [C\2]. We find that about half of the emission from the ionized gas is extinguished by dust. We modeled [O\3] around each O-type star and find that the density of the ISM is $\lesssim16$\cc\ on large scales. The extent of the [O\3] emission suggests that the medium is rather fragmented, allowing far-UV photons to permeates the ISM to scales of $\gtrsim30$\,pc. Furthermore, by comparing [C\2] with [N\2] 122\mic, we find that $95$\%\ of [C\2] arises in PDRs, except toward the stellar cluster for which as much as $15$\%\ could arise in the ionized gas. 
We find a remarkable correlation between [C\2]+[O\1] and PAH emission, with [C\2] dominating the cooling in diffuse PDRs and [O\1] dominating in the densest PDRs. The combination of [C\2] and [O\1] provides a proxy for the total gas cooling.  Our results suggest that PAH emission describes better the gas heating in PDRs as compared to the total infrared emission.
 }
   {}


   \maketitle
%

\section{Introduction}

Dwarf galaxies are important testbeds for understanding star formation in metal-poor environments that are intermediate between primordial galaxies and evolved systems such as the Milky Way. The low dust abundance in a metal-poor interstellar medium (ISM) allows far-UV (FUV) photons to permeate on larger physical scales in interstellar clouds. This effect reduces the size of the CO cores and pushes the C$^+$/C$^0$/CO interface deeper into the cloud, hence the very low CO luminosities (e.g., Leroy et al.\ 2006) and the difficulty of accurately determining the molecular hydrogen reservoir. Since molecular hydrogen (H$_2$) is more efficiently self-shielded from photodissociation than is CO, this potentially results in a thick layer where H$_2$ exists along with C$^+$ and C$^0$ (Grenier et al.\ 2005; Roellig et al.\ 2006; Wolfire et al.\ 2010). This so-called ``dark gas" might represent a significant fraction of the molecular mass in metal-poor galaxies (Madden et al.\ 1997). 

The ratio of the far-IR (FIR) line [C\2] 157\mic\ to CO(1-0) is found to be much higher in dwarf galaxies than in normal galaxies (Madden et al.\ 1997; Pineda et al.\ 2008), hinting at a possibly significant dark gas reservoir. However, many uncertainties remain regarding the exact origin of the [C\2] line as a function of star-formation activity and metallicity. Given the ionization potential of C$^0$ (11.3\,eV) and the low excitation temperature of the 157\mic\ line ($91$\,K), [C\2] can originate in the cold and warm neutral atomic and molecular medium and in the warm ionized gas (see also Madden et al.\ 1993; Heiles 1994). These phases coexist in a single beam for most extragalactic observations, complicating the diagnostics drawn from [C\2]. It is therefore essential to probe the smallest physical scale possible in the closest galaxies in order to constrain the conditions of [C\2] emission. This has proven to be a challenge until now, especially for dwarf galaxies. 

The relatively larger mean free path of FUV photons in low-metallicity environments also results in a large filling factor of ionized gas. The best probe for estimating this filling factor is an optically thin tracer with a low critical density (i.e., the density for which the spontaneous radiation rate equals the collision rate) and an ionization potential permitting its presence outside the dense H\2\ regions. [O\3] 88\mic, [N\2] 122\mic, and [N\2] 205\mic\ are the best probes available (see critical densities and ionization potentials in Table\,\ref{tab:lines}). 

The PACS instrument (Poglitsch et al.\ 2010) on board the \textit{Herschel} Space Telescope (Pillbratt et al.\ 2010) observes with a spatial resolution of $\sim10-13\arcsec$, which allows (1) examining the distribution of [C\2] in nearby galaxies and (2) comparing the local excitation conditions of this important coolant to those of complementary tracers of the high-excitation ionized gas ([N\3] 57\mic, [O\3] 88\mic), of the low-excitation ionized gas (e.g., [N\2] 122\mic, 205\mic), and of the neutral and molecular gas ([O\1] 63\mic, 145\mic). 

\begin{figure*}
\includegraphics[angle=0,scale=0.99,clip=false]{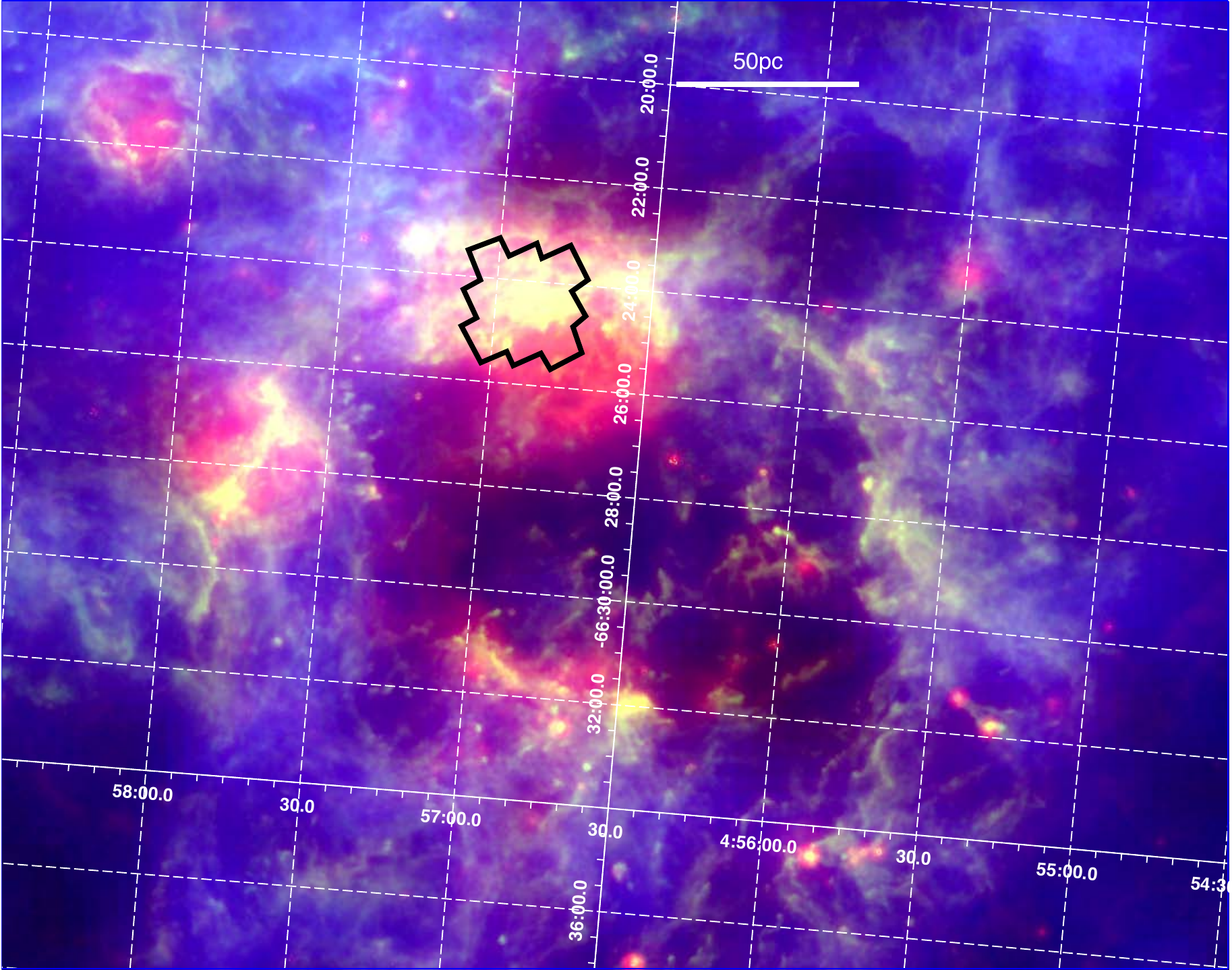}
\caption{The giant H\2\ region N\,11, with \textit{Spitzer}/MIPS 24\mic\ shown in red, \textit{Spitzer}/IRAC  $8.0$\mic\ in green, and H\1\ column density (from Kim et al.\ 2003) in blue. The \textit{Spitzer}/MIPS $24$\mic\ emission is dominated by very small grains that are stochastically heated and by big, warm, dust grains in thermal equilibrium with the interstellar radiation field. \textit{Spitzer}/IRAC $8.0$\mic\ emission is dominated by PAHs. The polygon shows the \textit{Herschel}/PACS spectroscopic observations of N\,11B presented in this paper.}
\label{fig:n11_shell}
\end{figure*}

The Magellanic Clouds provide both a local dwarf galaxy template and a moderately metal-poor environment, with a metallicity of $\approx1/2$\,Z$_\odot$ for the Large Magellanic Cloud (LMC) and $\approx1/5$\,Z$_\odot$ for the Small Magellanic Cloud (SMC). The presence of massive star-forming regions, such as 30\,Doradus and N\,11 in the LMC, makes it possible to constrain the origin of [C\2] in the various phases involved in star formation. The \textit{Herschel} SHINING Guaranteed Time Key Program (P.I.\ E.\ Sturm) observed several H\2\ regions in N\,11 with PACS. The first objective is to unveil the particular morphology of the ISM in a metal-poor environment through the study of the ionized gas distribution and filling factor. The second objective is to understand and constrain the origin of the [C\2] emission in order to account for the molecular gas reservoir that is not seen in CO. In the present paper, we investigate the large-scale emission-line structures in the N\,11B H\2\ region that is the northern and brightest H\2\ region in the N\,11 complex (Fig.\,\ref{fig:n11_shell}). In a second paper (Lebouteiller et al.\ in preparation), we examine the other H\2\ regions in the N\,11 complex, with a focus on the physical conditions in the photodissociation regions (PDRs).

In Sect.\,\ref{sec:environment}, we describe the environment and morphology of N\,11B. We present the observations in Sect.\,\ref{sec:obs} and the data analysis in Sect.\,\ref{sec:da}. The spatial distribution of the FIR tracers is examined in Sect.\,\ref{sec:distrib}, and the physical conditions of the ionized gas are investigated in Sect.\,\ref{sec:physcond}. The ionized gas spatial distribution is discussed in Sect.\,\ref{sec:spatialOIII}. Finally, the origin of the [C\2] emission is discussed in Sect.\,\ref{sec:cii_origin}. We present the \textit{Herschel}/PACS data reduction and analysis tool \texttt{PACSman} in the Appendix.

\section{Environment}\label{sec:environment}

\subsection{Morphology}\label{sec:morphology}

N\,11 (also DEM\,34; Henize 1956; Davies, Elliot \&\ Meaburn 1976) is the second largest giant H\2\ region in the Large Magellanic Cloud after 30\,Doradus (Kennicutt \&\ Hodge 1986). The morphology of N\,11 consists of nine distinct nebulae (N11A-N11I; Rosado et al.\ 1996) distributed around the OB association LH\,9 (NGC\,1760), which is dominated by the compact cluster HD\,32228 (also Radcliffe\,64, Sk-66-28, or Breysacher\,9). The LH\,9 association carved out a cavity and triggered a secondary, peripheral starburst (Lucke \& Hodge 1970; Parker et al.\ 1992; Walborn \&\ Parker 1992). The starburst shell created by LH\,9 reached $\sim120$\,pc in diameter (80x60\,pc, Lucke \& Hodge 1970). Figure\,\ref{fig:n11_shell} shows the H\1\ hole corresponding to the cavity around LH\,9. The nebulae are distributed around the H\1\ shell; N\,11B is the brightest nebula to the north.

\begin{figure*}
\includegraphics[angle=0,scale=0.9,clip=true]{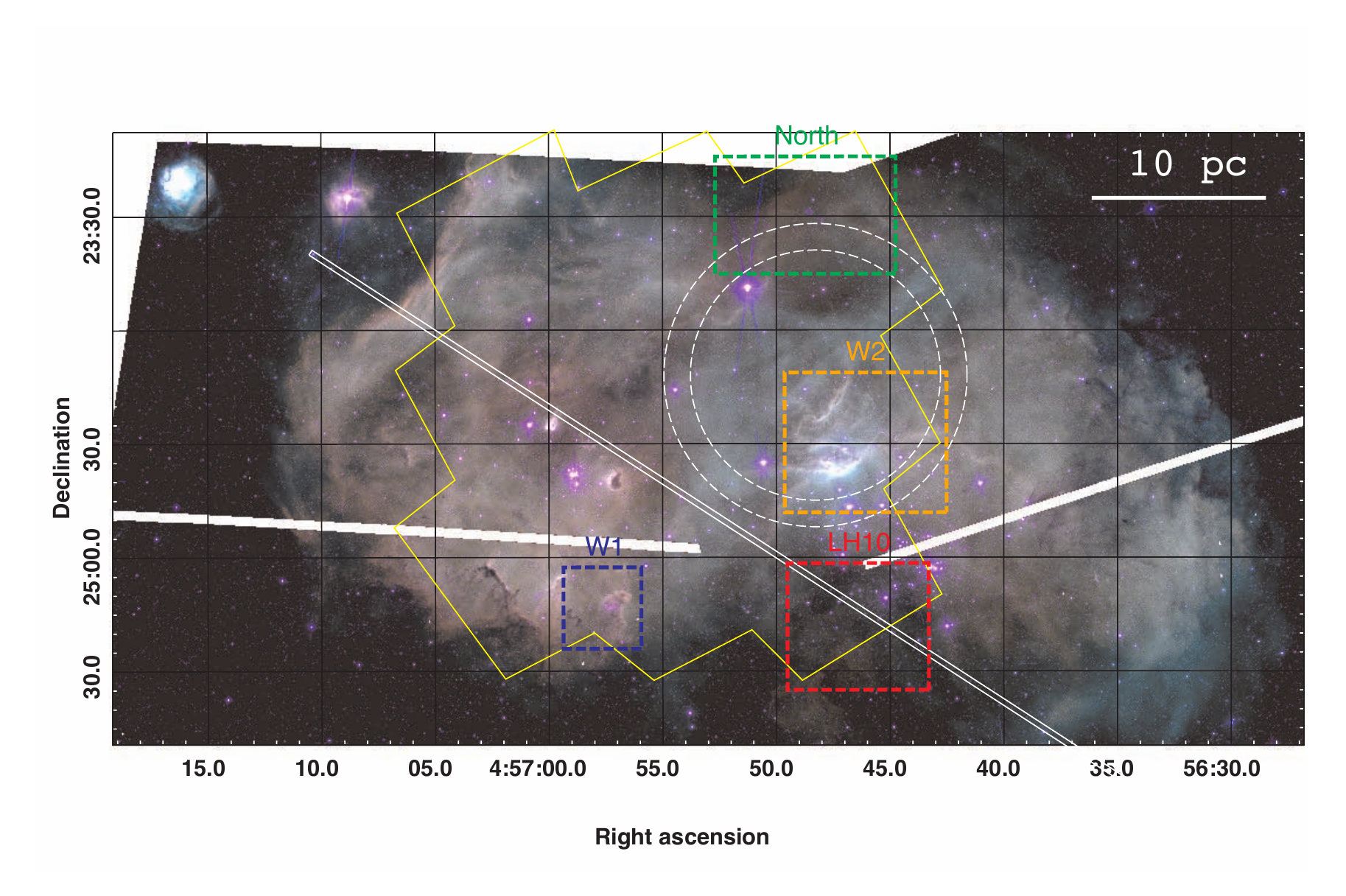}
\centering
\includegraphics[angle=0,scale=0.46,clip=true]{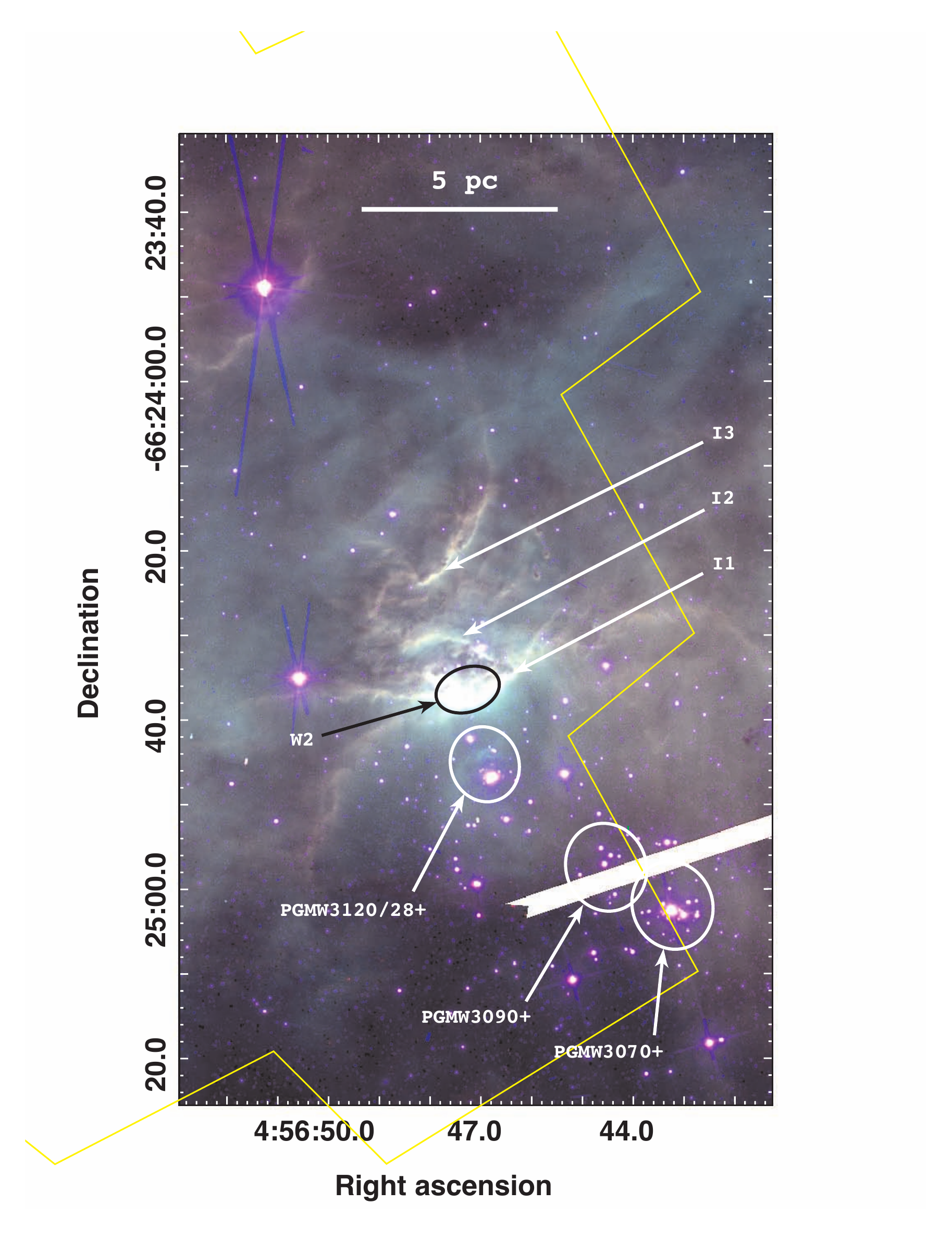}
\includegraphics[angle=0,scale=0.44,clip=true]{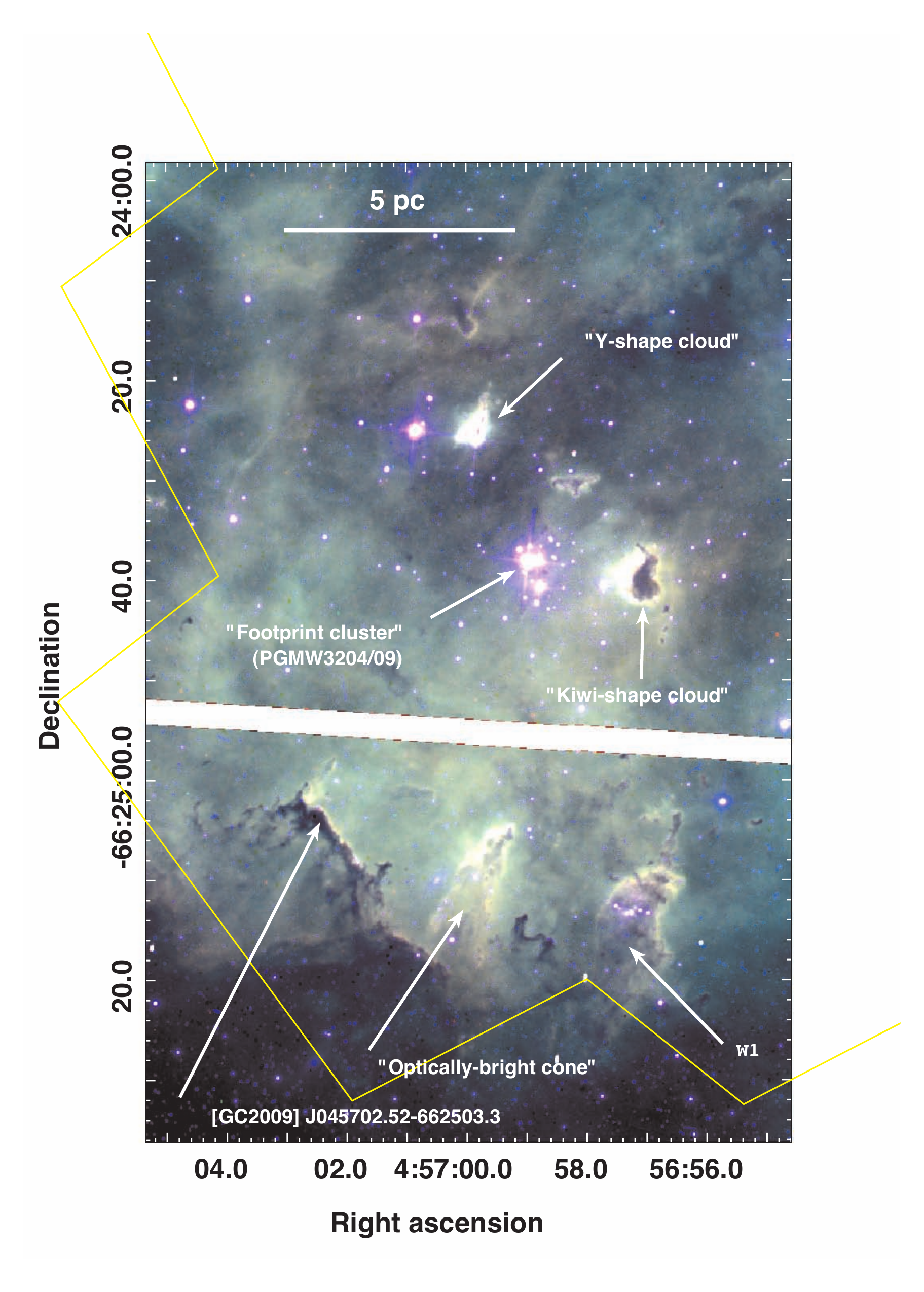}
\caption{\textit{Top} $-$ HST/ACS color-composite image of the N\,11B region (blue: F814; green: F658N/H$\alpha$; red: F660N/[N\2]). Images were downloaded from the \textit{Hubble} Legacy Archive (\textit{http://hla.stsci.edu/}) and later stitched with Montage (\textit{http://montage.ipac.caltech.edu/}). 
The white stripes correspond to a gap between individual exposures.
The yellow polygon indicates the field-of-view of the \textit{Herschel}/PACS observations ($\approx2.3\arcmin\times2.3\arcmin$). The $4$ rectangles in color indicate specific regions discussed in the text. The dashed circles represent the ISO-LWS observation (from $65"$ to $80"$ FWHM beam) and the slit represents the optical observation by Tsamis et al.\ (2001). \textit{Bottom} $-$ Close-up of the eastern (\textit{left panel}) and western (\textit{right panel}) sides. The stellar complexes are labeled as the brightest star with a `+' sign appended.}
\label{fig:acs_all}
\end{figure*}

There is evidence of ongoing star formation in N\,11B. About 20 HAeBe stars, which are intermediate-mass (3 to 7\,M$_\odot$) pre-main sequence stars (1 to 3\,Myr), have been detected (Barba et al.\ 2003; Hatano et al.\ 2006). A methanol maser of 0.3\,Jy was observed by Ellingsen et al.\ (1994). 
Finally, Hatano et al.\ (2006) identified two ultracompact H\2\ regions, one at ($04$h$56$m$48$s, $-66^\circ24'34''$), corresponding to the radio source known as W2 (or B0456-6629) in Indebetouw et al.\ (2004), and the other at ($04$h$56$m$57$s $-66^\circ25'13''$), which is known as W1 (or B0456-6629).

From the \textit{Hubble} Space Telescope (HST) Advanced Camera for Surveys (ACS) archival images (Fig.\,\ref{fig:acs_all}), two regions can be distinguished within N\,11B with distinct morphological attributes. The eastern half has sharp edges toward the east and the south, indicating the presence of relatively dense interfaces. 
A lower density zone is seen in the middle of N\,11B, with the OB association PGMW\,3204/09 (hereafter the ``footprint" cluster; Sect.\ \ref{sec:ionizing_sources}) surrounded by several small clouds with bright ionization fronts, including the kiwi-shaped cloud and the Y-shaped cloud described in Naz{\'e} et al.\ (2001). The western half of N\,11B is dominated by PDR structures north of the local OB association LH\,10. Several ionization fronts (I1, I2, and I3 in Barb{\'a} et al.\ 2003) are seen above LH\,10. 

In the following we concentrate on four distinct regions that were selected based on their extreme nature among N\,11B, translating into significant differences in line ratio diagnostics:\\
$-$ The fragmented region within the orange box in Fig.\,\ref{fig:acs_all}, which we refer to as W2 from now on, which contains relatively dense interfaces under the influence of nearby massive stars from the LH\,10 cluster. \\
$-$ The stellar cluster below W2, which we refer to from now on as LH\,10 (red box). LH\,10 is embedded in a relatively diffuse gas. \\
$-$ The northern region (green box), where the absence of optical edge or arcs suggests a smooth transition into the general, lower density, LMC environment. This region is little, if at all, affected by known massive stars. \\
$-$ The compact region W1 (blue box), which contains a small stellar cluster surrounded by dense shells.

\subsection{Main ionizing sources}\label{sec:ionizing_sources}

The stellar cluster associated with N\,11B is the LH\,10 OB association (also NGC\,1763, IC\,2115, Lucke \&\ Hodge 1970). 
LH\,10 is the youngest cluster in N\,11 ($\lesssim3$\,Myr; Walborn et al.\ 1999). Most of the stars in LH\,10 are gathered into several compact complexes located south of the W2 star-forming region.
The PGMW\,3128 and the PGMW\,3120 complexes are located near W2, while the PGMW\,3090 and PGMW\,3070 are farther away to the south (Parker et al.\ 1992; Fig.\,\ref{fig:acs_all}). From the high-resolution HST/ACS image, we count at least 260 stars south of W2. Including faint sources increases the number to more than 300 stars. This is a lower limit since some complexes are unresolved. At least two O3 stars were identified in the southern LH\,10 cluster, PGMW\,3058 and PGMW\,3061, slightly off the area covered by our PACS observations (Parker et al.\ 1992). The presence of ``intact" (not yet evolved) O3 stars suggests that no supernova event has occurred yet. 

The ``footprint" cluster is located to the east of N\,11B (see lower left panel in Fig.\,\ref{fig:acs_all}) and close to the kiwi-shaped cloud, with several massive stars and concentrations identified by Parker et al.\ (1992). The main sources are PGMW\,3204-3209-3211, with at least 30 stars (Fig.\,\ref{fig:acs_all}). An O3\,III star was discovered in PGMW\,3209 (star ``A" in Walborn \&\ Parker 1992). Another concentration is located nearby, to the east of the Y-shaped cloud with at least a dozen stars.

\subsection{Electron density}\label{sec:density}

Mac\ Low et al.\ (1998) find an electron density of $n_e\approx38$\cc\ from the H$\alpha$ emission measure, assuming N\,11B is a uniform ionized sphere. By estimating the physical size of structures within the nebula, Naz{\'e} et al.\ (2001) find that the emission measure is consistent with $n_e\sim15$\cc\ in the lowest surface brightness regions. It reaches $50-150$\cc\ in moderately bright arcs and up to $200-500$\cc\ in the bright ridges of dust clouds (e.g., the kiwi-shape cloud). The bright optical arcs studied by Naz{\'e} et al.\ (2001) correspond to the highest densities and also show low [O\3] 5007\AA/H$\alpha$ ratios, which is attributed to lower ionization and excitation. 

Tsamis et al.\ (2001) observed the [S\2] $\lambda6731/\lambda6716$ and [O\2] $\lambda3729/\lambda3726$ doublets in long-slit spectroscopy (see slit position in Fig.\,\ref{fig:acs_all}) and find $n_e=80$\cc\ and $110$\cc,  respectively. They measure a significantly higher density from the [Cl\3] $\lambda5537/\lambda5517$ doublet, with $1700$\cc, which they attribute to the higher critical densities involved in the [Cl\3] transitions. Strong density variations are therefore expected in N\,11B. We tentatively expect a density of $\sim25$\cc\ for the low-density nebula and densities between $100-1000$\cc\ for the bright optical arcs.

\subsection{Chemical abundances}\label{sec:abundances}

Chemical abundances in N\,11B were derived by Tsamis et al.\ (2003), who find $12+\log({\rm O/H}) = 8.41$ and $12+\log({\rm N/H}) = 6.93$.  Using the oxygen solar abundance $12+\log({\rm O/H}) = 8.69$ from Asplund (2009), the metallicity is $0.52$\,Z$_\odot$. The N/O abundance ratio is significantly lower than the solar value since N/H is $\approx7$ times below the solar abundance. This nitrogen deficiency is observed in several H\2\ regions of the LMC and is likely due to depletion of N on dust grains (e.g., Garnett 1999). For the carbon abundance, we assume $12+\log({\rm C/H})=7.9$ which is the average abundance in LMC H\2\ regions (Garnett 1999).

\section{Observations}\label{sec:obs}

N\,11B was observed on January 4, 2010 with the PACS spectrometer on board \textit{Herschel}. The main FIR fine-structure lines were mapped in the wavelength-switching mode (later decommissioned and replaced by the unchopped scan mode): [N\3] 57\mic, [O\1] 63\mic, [N\2] 122\mic, [O\1] 145\mic, [C\2] 157\mic, [N\2] 205\mic\ (OBSID 1342188940), and [O\3] 88\mic\ (OBSID 1342188941). Two follow-up observations were performed in unchopped scan mode, one pointing toward W2 in [N\2] 122\mic\ on April 4, 2011 (OBSID 1342219439), and one pointing toward W1 in [N\2] 122\mic\ and [O\1] 145\mic\ on July 19, 2011 (OBSID 1342225175). The [O\3] 52\mic\ was not observed with PACS because of the low transmission at this wavelength, but the line was detected with ISO (Sect.\,\ref{sec:observations_ancillary}). The observed lines are listed in Table\,\ref{tab:lines} along with a summary of the main transition parameters. From now on we use the line notation with the wavelength as subscript for ambiguous cases (e.g., [O\1]$_{63}$ for [O\1] 63\mic).

\begin{table*}
\caption{Main parameters of the transitions used in this study.
\label{tab:lines}}
\centering
\begin{tabular}{llllllll}
\hline\hline
Line & $\lambda$  & IP$_1$\tablefootmark{a} & IP$_2$\tablefootmark{b} & Configuration & $n_{\rm cr,e}$\tablefootmark{c} & $n_{\rm cr,H}$\tablefootmark{d} & $T_{\rm exc}$ \\
     &      & (eV)  & (eV)  &  & (\cc)  & (\cc) & (K)  \\
\hline
$[$O\1$]_{63}$  &  63\mic\  & & $13.62$ & ${^3}P_1-{^3}P_2$   &   & $\approx(4-10)\times10^5$   & $228$ \\
$[$O\1$]_{145}$  &  145\mic\ &  & $13.62$  & ${^3}P_0-{^3}P_1$ &  & $\approx(1-20)\times10^5$  &   $327$ \\
$[$C\2$]$  &  157\mic\ & $11.26$ & $24.38$  & ${^2}P_{3/2}-{^2}P_{1/2}$ &  $\approx80$ & $\approx3000-5000$  &   $91$ \\
$[$O\3$]_{52}$  &  52\mic\ &  $35.12$ & $54.93$ & ${^3}P_2-{^3}P_1$   & $\approx3\,500$   &  & $441$ \\
$[$O\3$]_{88}$  &  88\mic\ &  $35.12$ & $54.93$ & ${^3}P_1-{^3}P_0$   & $\approx510$   &  & $163$ \\
$[$N\3$]$  &  57\mic\ & $29.60$ & $47.45$  & ${^2}P_{3/2}-{^2}P_{1/2}$   & $\approx3000$ &  & $251$ \\
$[$N\2$]_{122}$  &  122\mic\ & $14.53$ & $29.60$   & ${^3}P_2-{^3}P_1$   &  $\approx400$ &  & $188$ \\
$[$N\2$]_{205}$  &  205\mic\ & $14.53$ & $29.60$ & ${^3}P_1-{^3}P_0$   & $\approx180$ &  & $70$ \\
\hline
$[$O\3$]_{\rm opt}$  &  5007\AA\   & $35.12$ & $54.93$ & ${^1}D_2-{^3}P_2$  & $\approx7\times10^5$ &  & $2.9\times10^4$ \\
 \hline
\end{tabular}\\
\tablefoottext{a}{Ionization potential to create the species.}
\tablefoottext{b}{Ionization potential to ionize the species.}
\tablefoottext{c}{Critical density for collisions with electrons; i.e., the density for which the spontaneous radiation rate equals the collision rate.}
\tablefoottext{d}{Critical density for collisions with H\1\ and H$_2$. }
\end{table*}

As explained in Poglitsch et al.\ (2010), the PACS spectrometer is an integral-field spectrometer that samples the spatial direction with $25$ pixels and the spectral direction with $16$ pixels. Each spectral pixel ``sees" a distinct wavelength range that is scanned by varying the grating angle. The resulting projection of the PACS array on the sky is a footprint of $5\times5$ spatial pixels (``spaxels"), corresponding to a $47\arcsec\times47\arcsec$ field-of-view. 
At a distance of the LMC ($48.5$\,kpc), the footprint size $47\arcsec$ corresponds to 11\,pc. The map observations of N\,11B were designed as $3\times3$ raster map, with an overlap of $\approx1.5$ spaxels. Figure\,\ref{fig:acs_all} shows the PACS map coverage (about $2.3\arcmin\times2.3\arcmin$) on the optical image. According to the PACS Observers Manual\footnote{\textit{http://herschel.esac.esa.int/Docs/PACS/html/pacs\_om.html}}, the point spread function full width at half maximum (FWHM) is  $\approx9.5\arcsec$ ($\approx2.2$\,pc) between $55$\mic\ and $100$\mic, and it increases to about $\approx13\arcsec$ ($\approx3$\,pc) at $180$\mic.

\section{Data analysis}\label{sec:da}

\subsection{\textit{Herschel}/PACS}

The data reduction was performed in HIPE track number \texttt{4.0.700} (Ott 2010) using a combination of the default wavelength-switching and unchopped scan scripts. 

\subsubsection{Wavelength-switching}

A spatially offset position is usually observed at the beginning and at the end of the observation to subtract the telescope background emission. In our mapping observations, the offset was located on the source itself (central raster position in the map) so that it was not subtracted but instead used to increase the signal-to-noise ratio (S/N) by $\sim\sqrt{2}$ at that location. Based on the unchopped scan observation, for which an offset was available (Sect.\,\ref{sec:unchopped}), we estimate that the lack of an offset in the wavelength-switching observation resulted in a lower S/N in the final spectra. No line contamination from the Milky Way was observed in the offset in the unchopped scan observation, implying that our line measurements are not affected by foreground line emission, as expected from the low galactic latitude of LMC-N\,11 ($\sim-66.4^\circ$).

The wavelength-switching mode was initially designed to remove the variations in the telescope background emission on small timescales by switching wavelengths rapidly, resulting in a differential spectral line profile. The wavelength modulation of the grating follows the \textbf{AABBBBAA} pattern, where \textbf{A} observes at the nominal wavelength and  \textbf{B} observes the modulated wavelength (with a shift of about half of the FWHM). 

The differential line profile resulting from the wavelength-switching proved itself difficult to fit, especially for low S/Ns. We decided to consider the line profile in the combined spectrum \textbf{A}+\textbf{B} (i.e., without subtracting the switched wavelength spectrum). This is effectively a similar approach to the unchopped scan mode.
The final spectrum was divided by a factor $1.3$ for the blue camera and $1.1$ for the red camera in order to account for differences between ground and in-flight performances (PACS ICC calibration document PICC-KL-TN-041). 

\subsubsection{Unchopped scan}\label{sec:unchopped}

Follow-up pointed observations of [N\2]$_{122}$ and [O\1]$_{145}$ were performed with the unchopped scan mode. The offset exposures were checked for line emission and were averaged before subtraction from the ``on-source" observations. The unchopped scan observations took longer (per line and for one raster position) than the wavelength-switching, and transient cosmic rays were observed on the timeline of several photoconductor pixels. We corrected the signal by using a multiresolution algorithm that is part of the data reduction and analysis \texttt{PACSman} (see Appendix). 

\subsubsection{Line fitting and map projection}\label{sec:fitandmap}

\begin{figure*}
\includegraphics[angle=0,scale=0.38,clip=true]{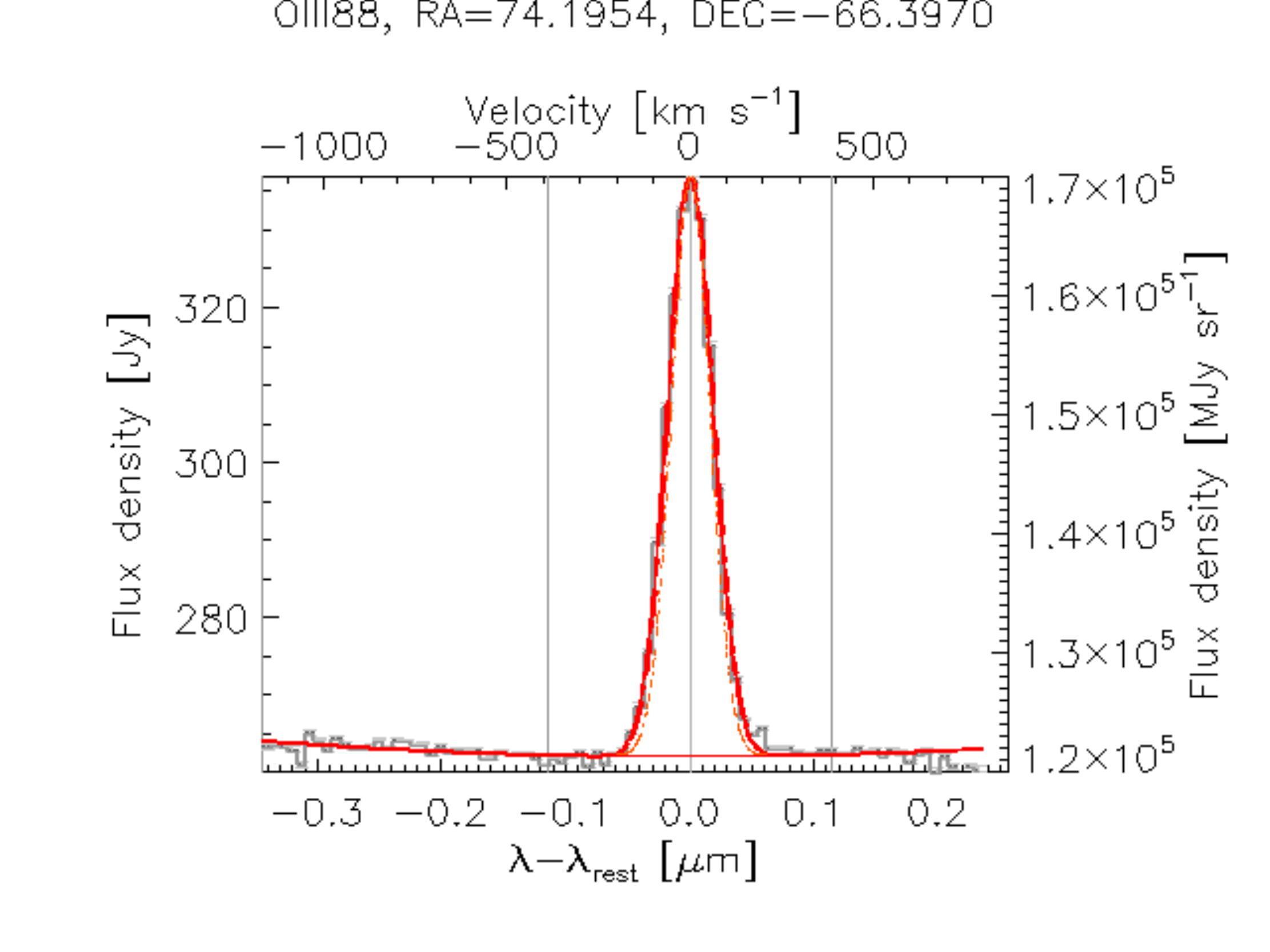}
\includegraphics[angle=0,scale=0.38,clip=true]{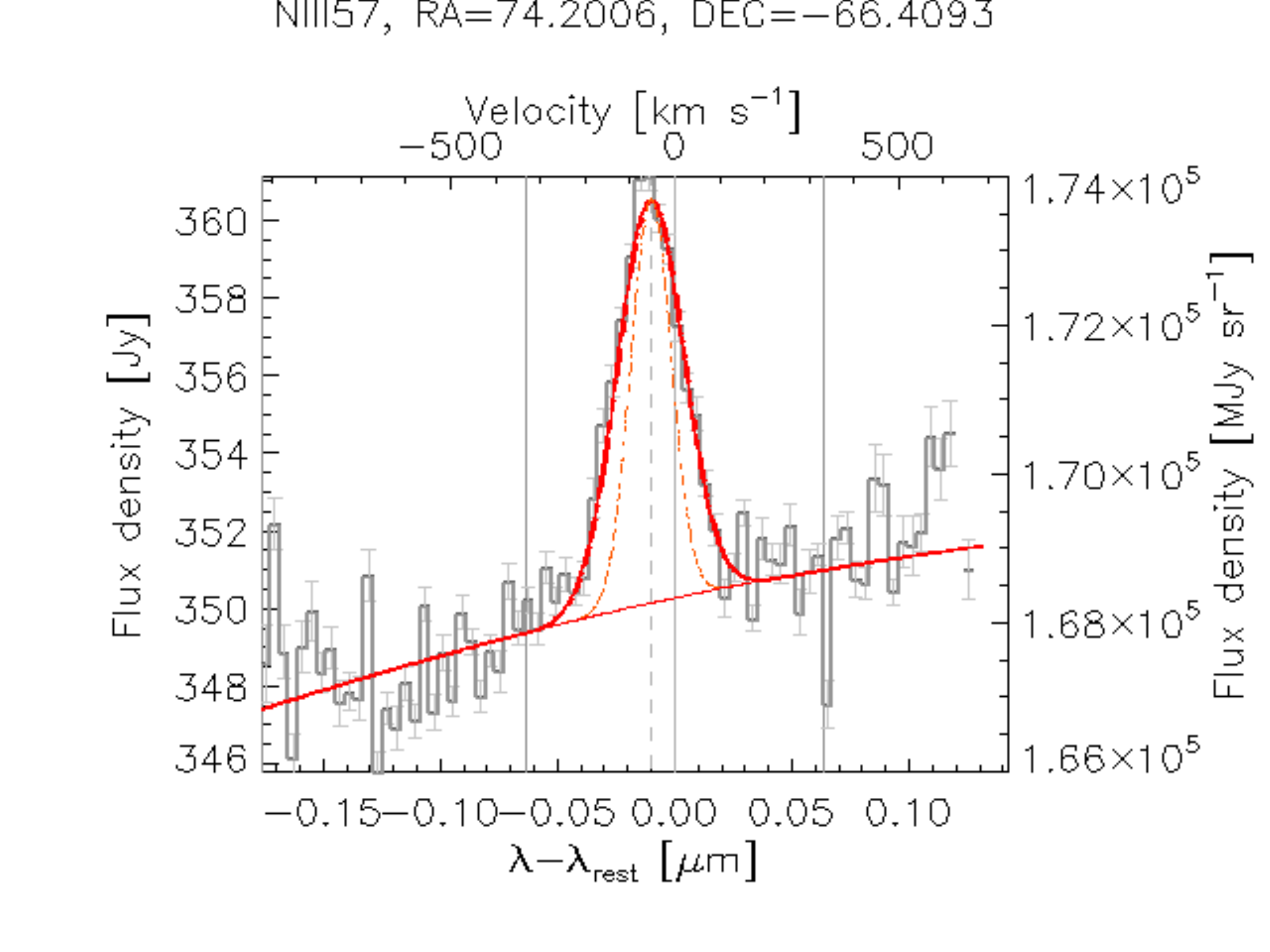}\\
\includegraphics[angle=0,scale=0.38,clip=true]{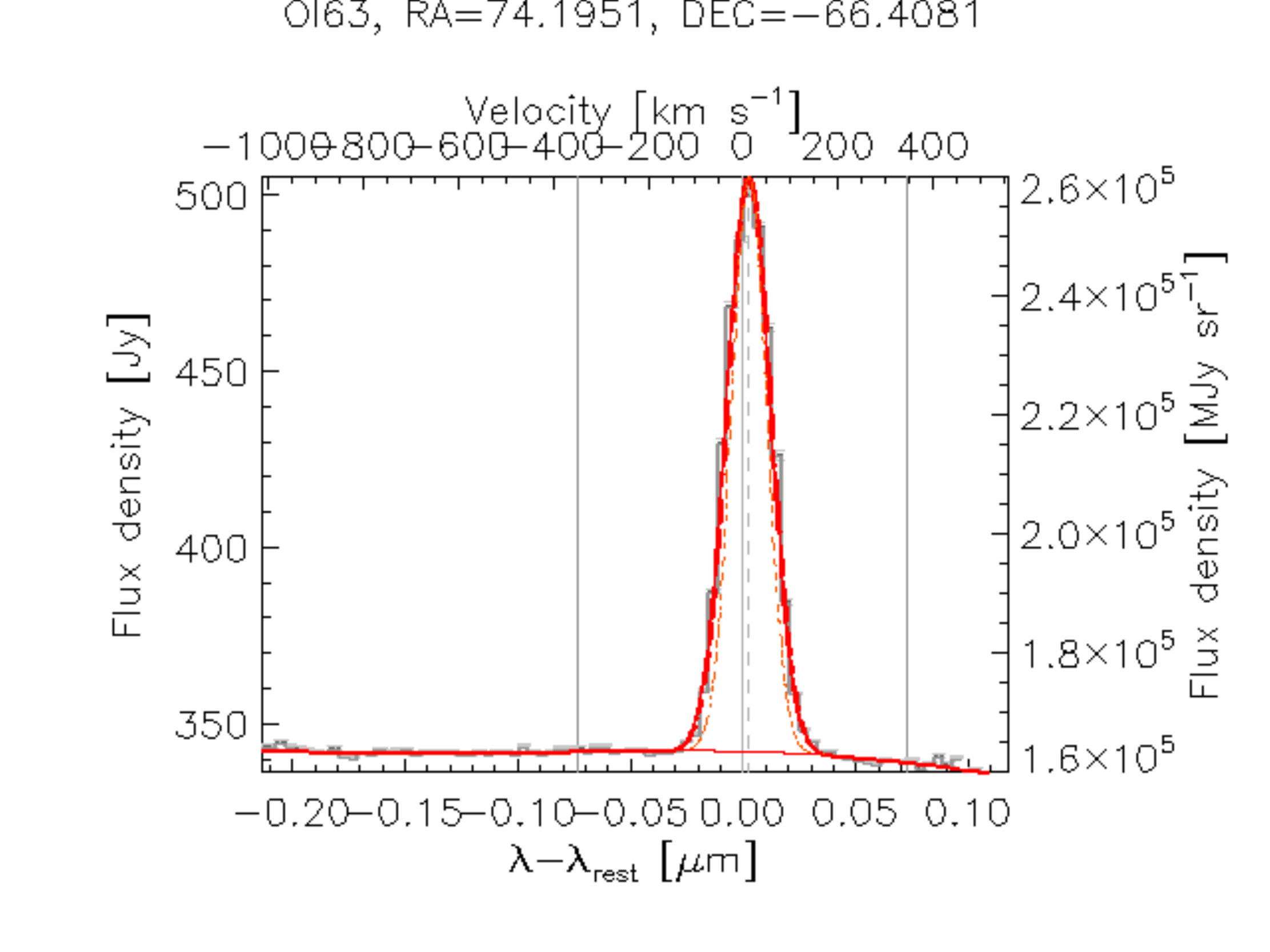}
\includegraphics[angle=0,scale=0.38,clip=true]{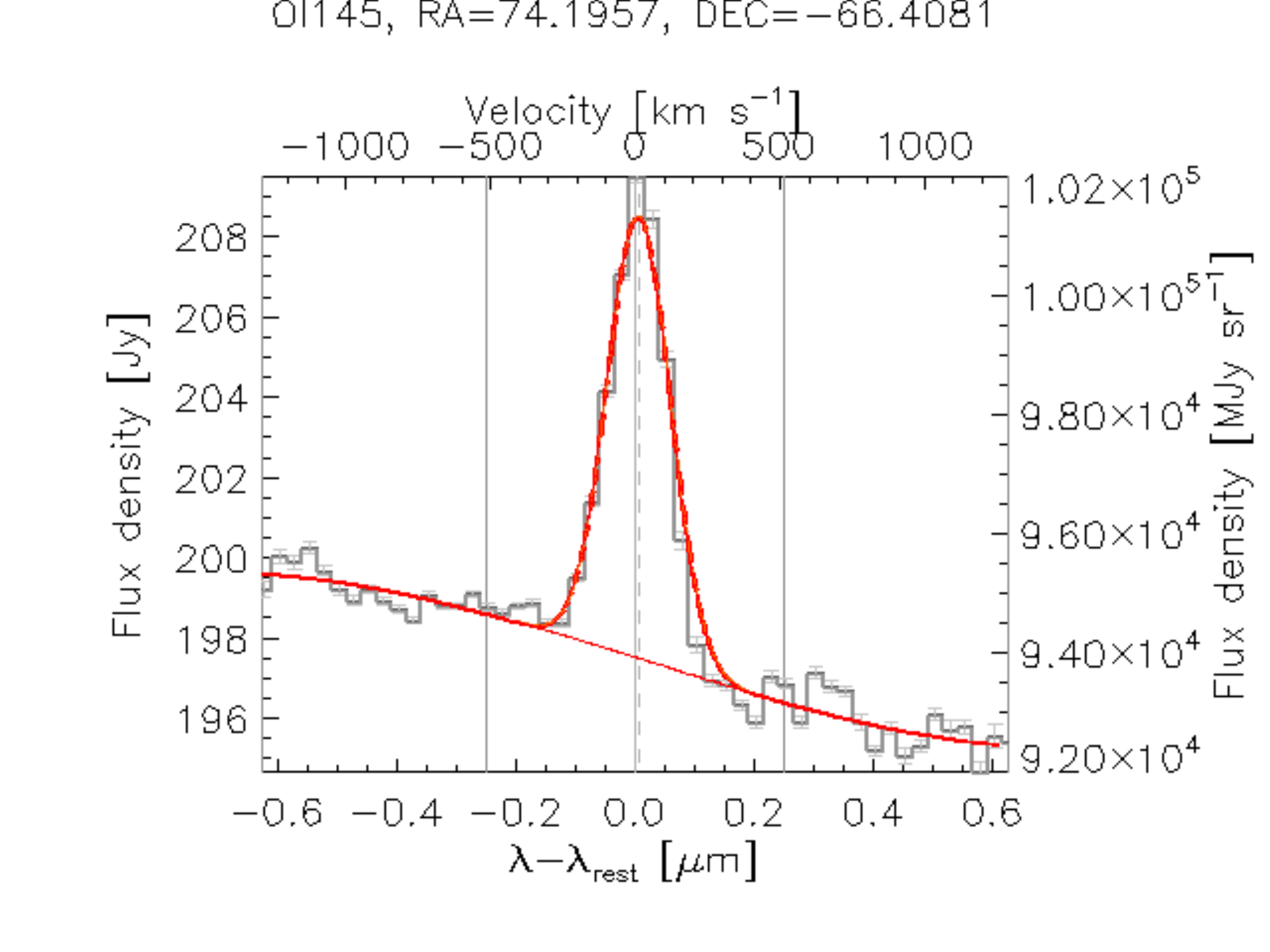}\\
\includegraphics[angle=0,scale=0.38,clip=true]{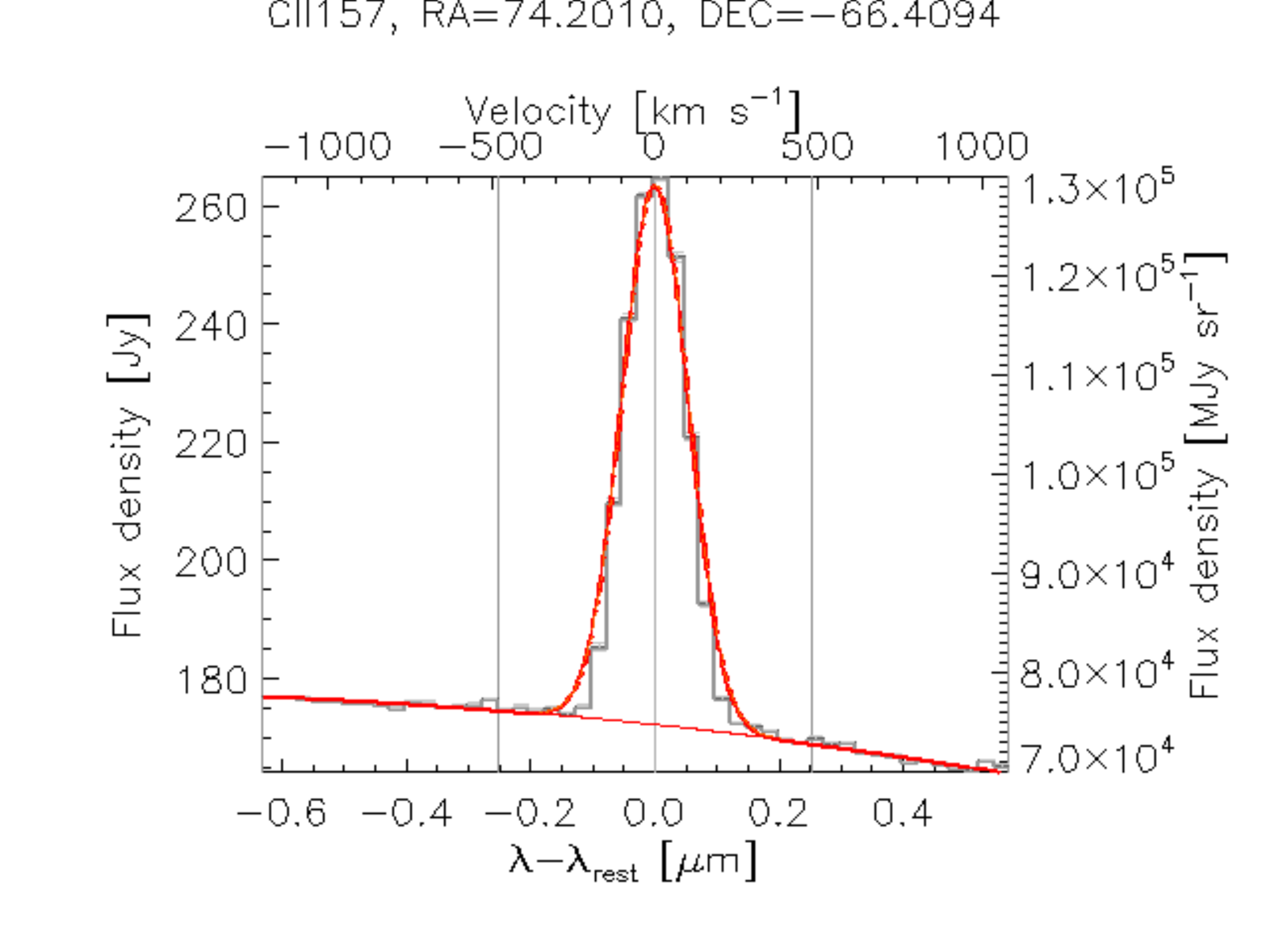}
\includegraphics[angle=0,scale=0.38,clip=true]{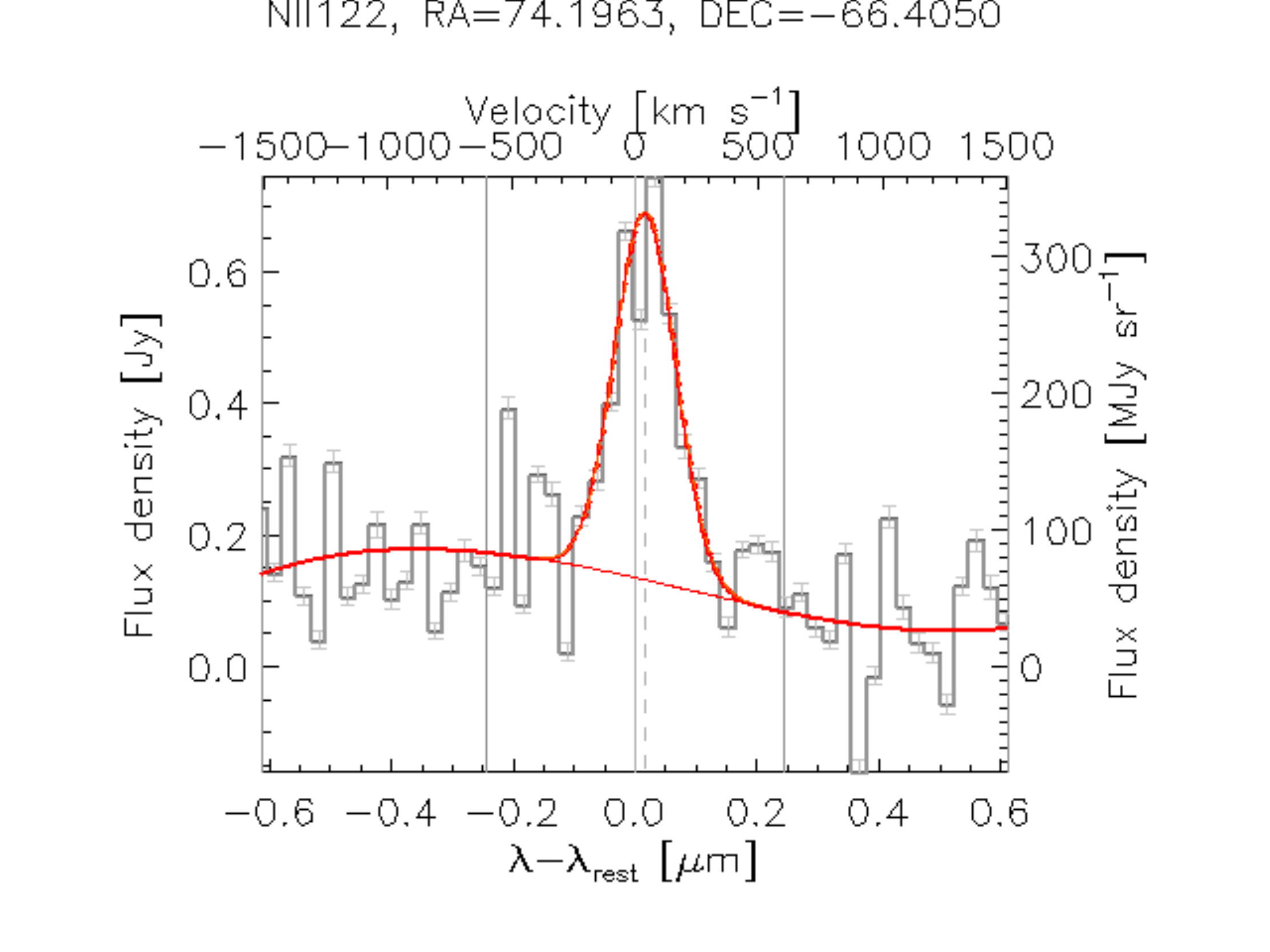}
\caption{Line fits are shown for one spaxel observing the W2 region. The histogram shows the rebinned data, the thin curve shows the continuum, and the thick curve shows the fit. The fit was performed on the data cloud, which contains $\approx140000$ points. We show the rebinned spectrum for display purposes.
\label{fig:fits}}
\end{figure*}

Line fits were performed by \texttt{PACSman} (see Appendix) for each spaxel at each raster position. A Gaussian profile was adjusted simultaneously with a polynomial continuum of degrees one to three. All lines have heliocentric radial velocities of $305\pm20$\,\kms\ with no significant variations across the map. This velocity is consistent with the value $292\pm23$\,\kms\ found by Rosado et al.\ (1995) for the entire N\,11B nebula. 
The instrument spectral resolution ranges from $\approx55$\kms\ to $\approx320$\kms\ depending on the band/order and wavelength. No significant additional broadening is observed for the long-wavelength lines [O\1]$_{145}$ and [C\2], while a broadening of $70-90$\,\kms\ is observed for [N\3], [O\1]$_{63}$, and [O\3]$_{88}$ (for which the spectral resolution is finer), which might be real. Example of line fits are shown in Fig.\,\ref{fig:fits}. 

For the mapping observations, the line fluxes of all the spaxels are projected on a subpixel grid (see Appendix). The subpixel size ($1/3$ of the spaxel size, i.e., $\approx3\arcsec$ or $\approx0.7$\,pc at the distance of the LMC) is chosen in order to recover the best possible spatial sampling in the footprint overlap regions. 
For comparison, the map of the Orion bar in Bernard-Salas et al.\ (2011) is 0.25\,pc in size, i.e., a fraction of one pixel in our maps.

\begin{figure*}
\centering
\includegraphics[angle=0,scale=0.48,clip=true]{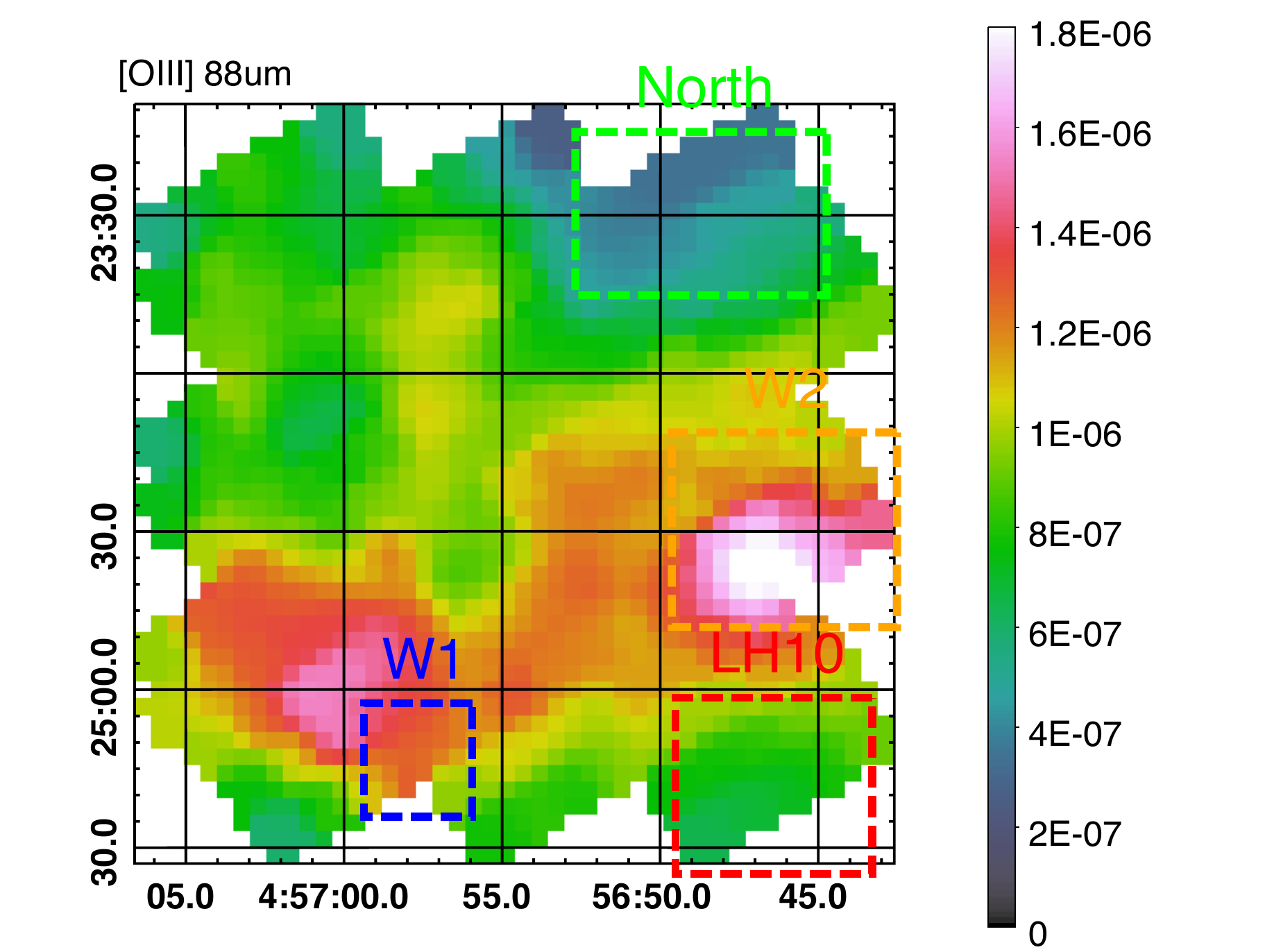}
\includegraphics[angle=0,scale=0.48,clip=true]{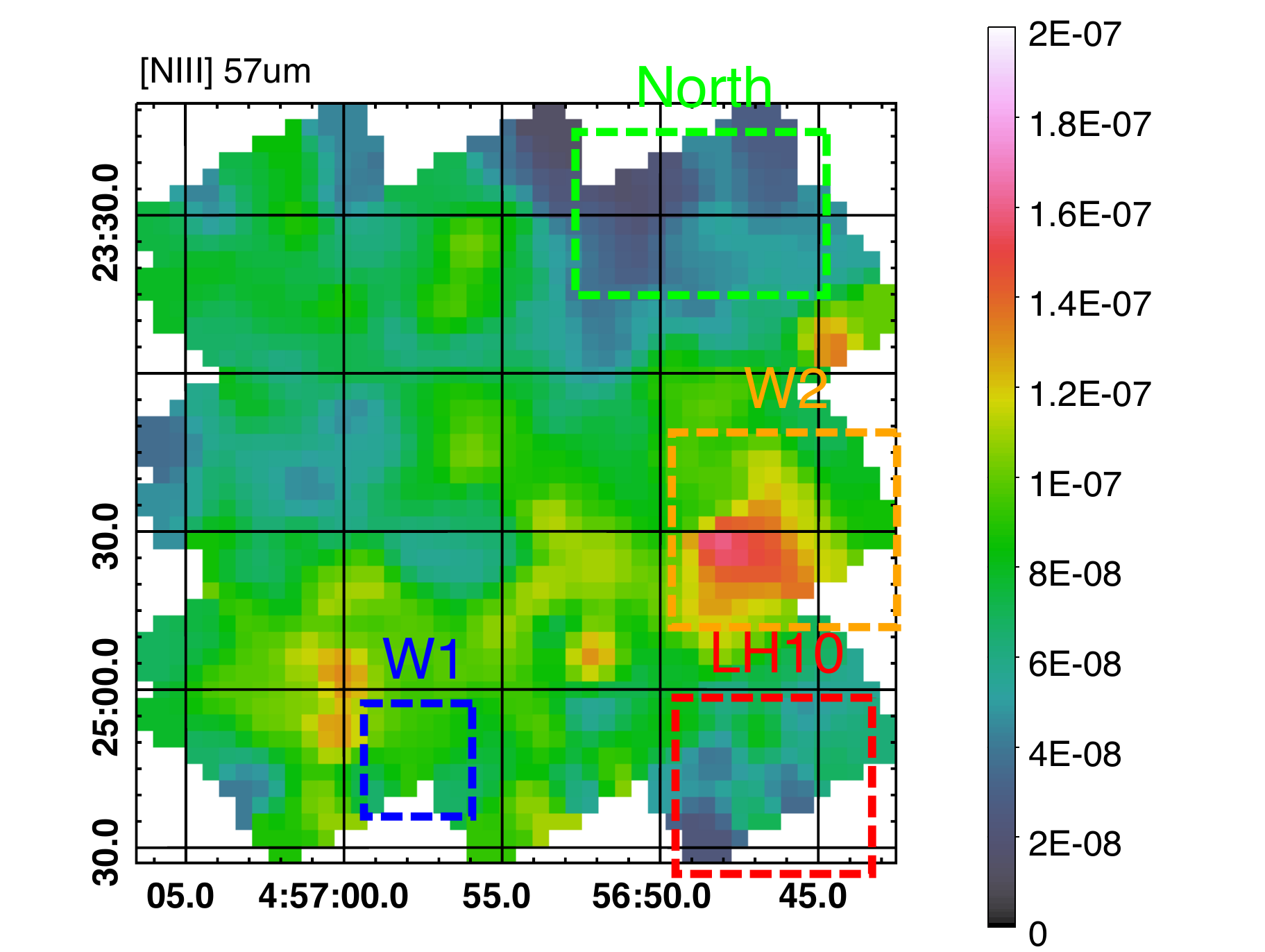}\\
\includegraphics[angle=0,scale=0.48,clip=true]{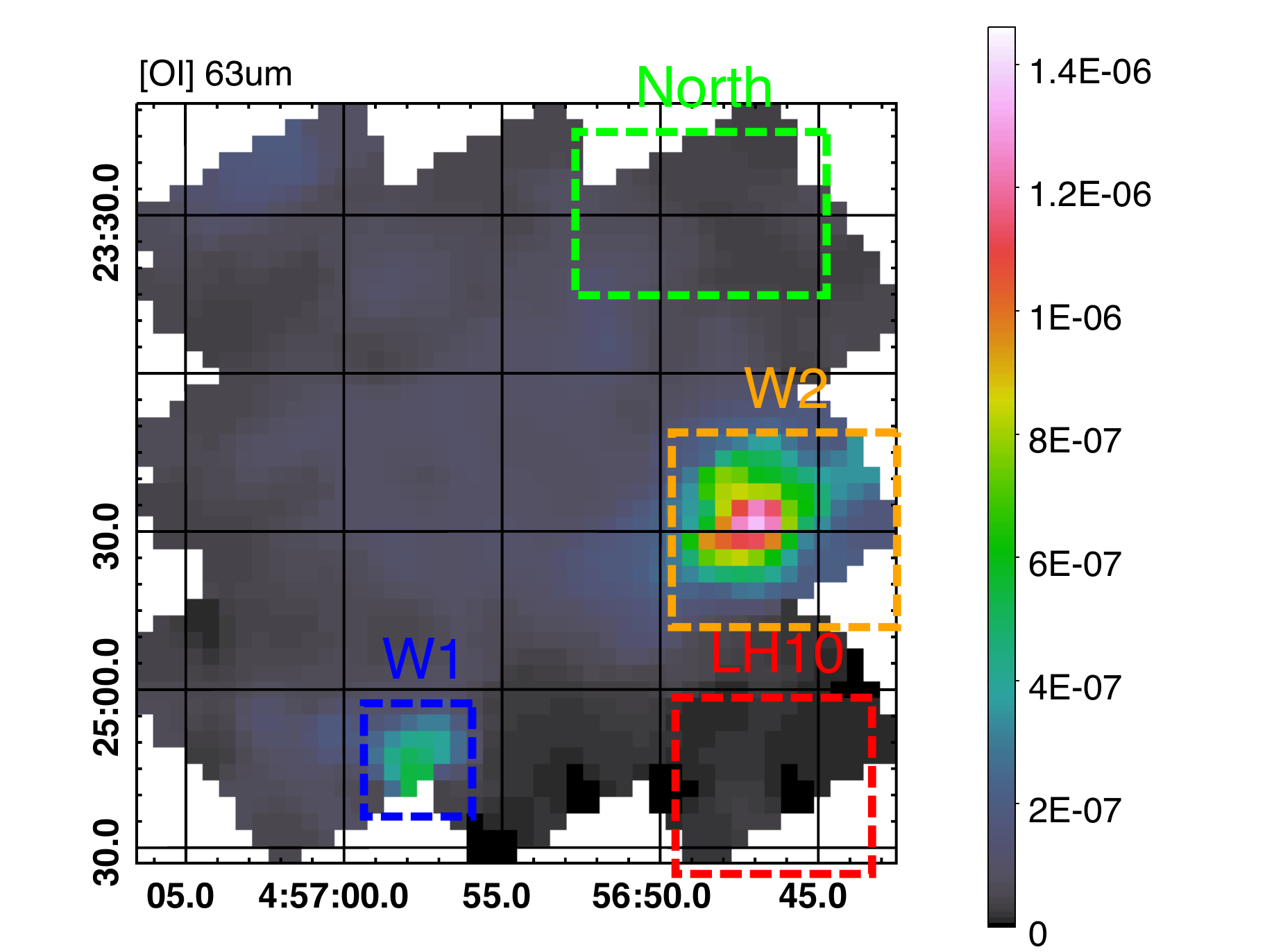}
\includegraphics[angle=0,scale=0.48,clip=true]{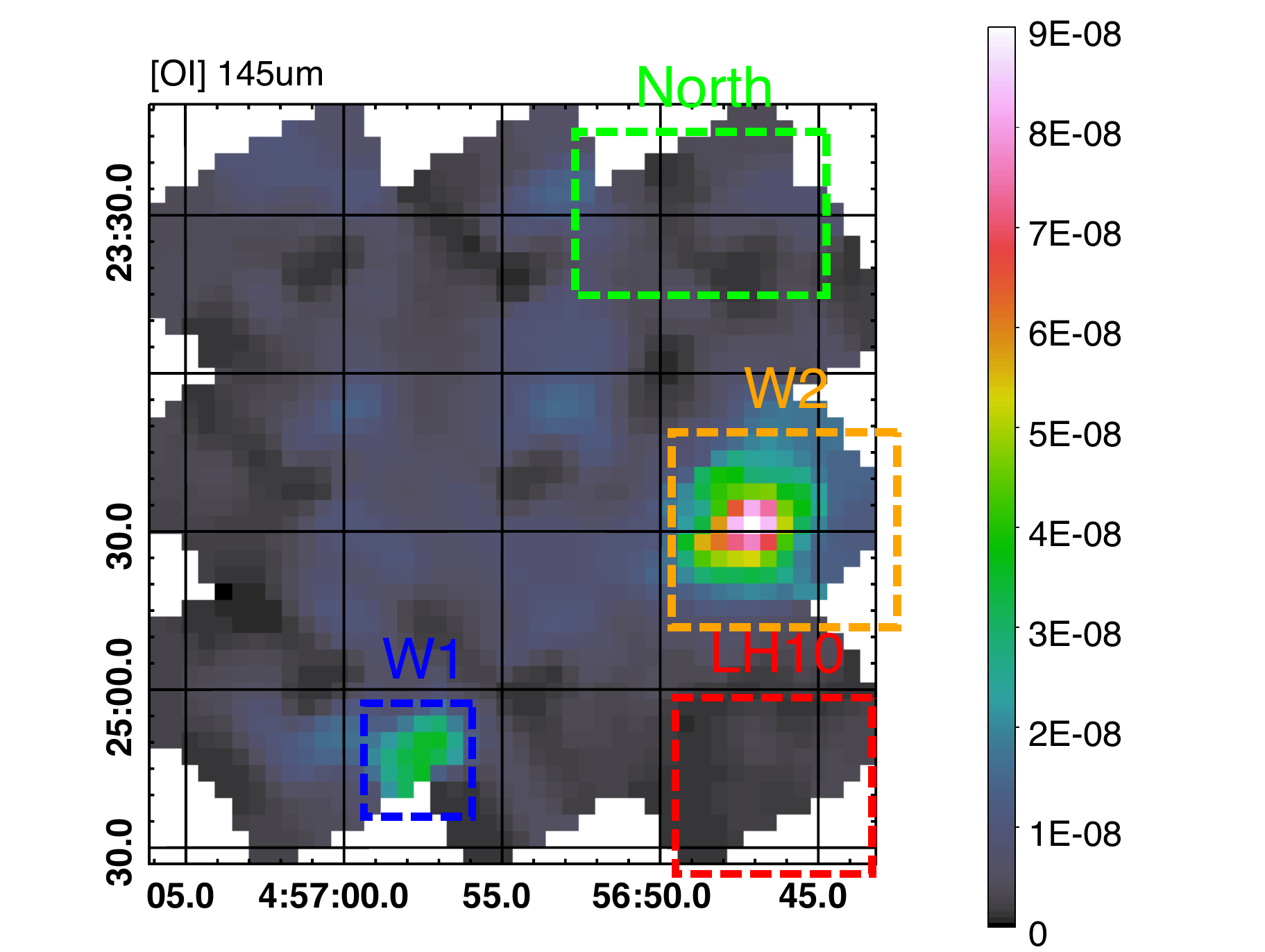}\\
\includegraphics[angle=0,scale=0.48,clip=true]{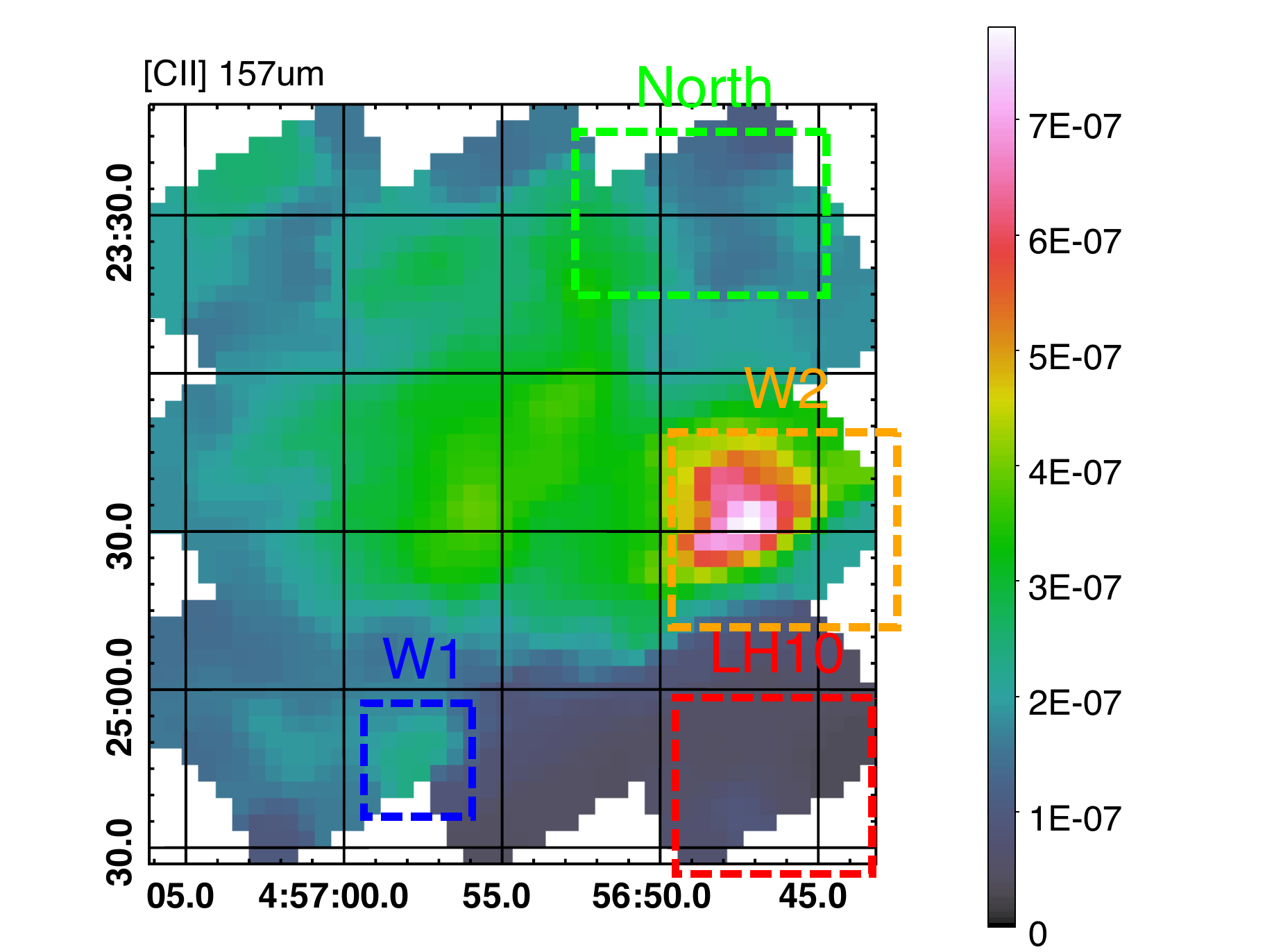}
\includegraphics[angle=0,scale=0.48,clip=true]{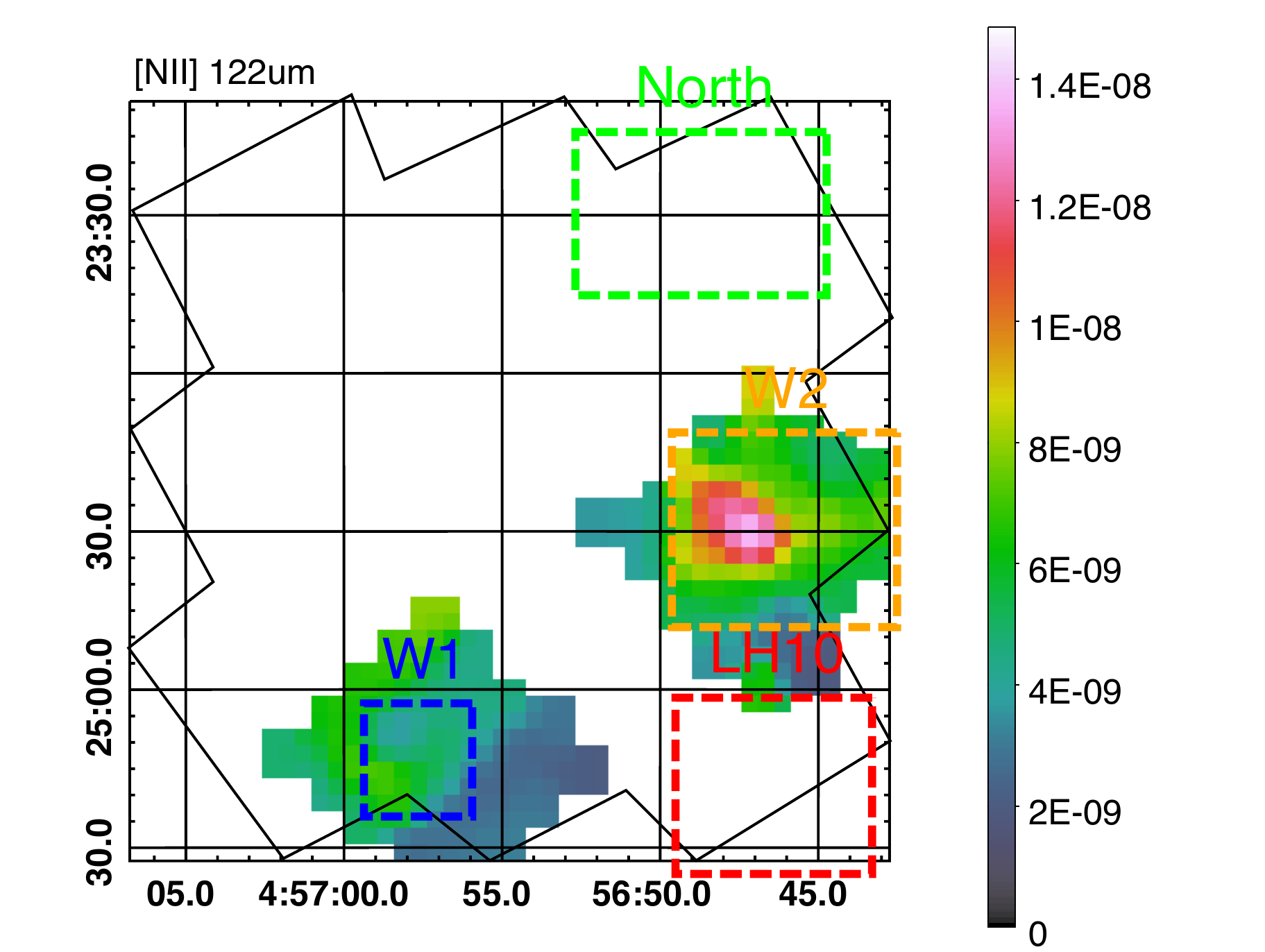}
\caption{\textit{Herschel}/PACS maps (in J2000 coordinates). The colorbar gives the flux in \wmsr. The dashed rectangles correspond to the zones described in Fig.\,\ref{fig:acs_all} and Sect.\,\ref{sec:morphology}. 
The spatial resolution is about $\approx9.5\arcsec$ for [N\3]$_{57}$, [O\1]$_{63}$, and [O\3]$_{88}$,  $\approx11.5\arcsec$ for [N\2]$_{122}$, [O\1]$_{145}$, and [C\2]$_{157}$. \label{fig:pacs_maps}}
\end{figure*}

The S/N is $>50$ in the [O\3]$_{88}$ line across the map, $>12$ in the [C\2] map, between $1$ and $100$ in the [O\1]$_{63}$ map, between $0$ and $17$ in the [O\1]$_{145}$ map, and between $0.5$ and $7$ in the [N\3] map. The [N\2]$_{122}$ line is barely detected in a few positions, while [N\2]$_{205}$ is not detected anywhere because of the known spectral leakage in the R1 band (see PACS Observers Manual). Final maps are shown in Fig.\,\ref{fig:pacs_maps}.

\subsection{Ancillary data}\label{sec:observations_ancillary}

\begin{figure*}
\includegraphics[angle=0,scale=0.48,clip=true]{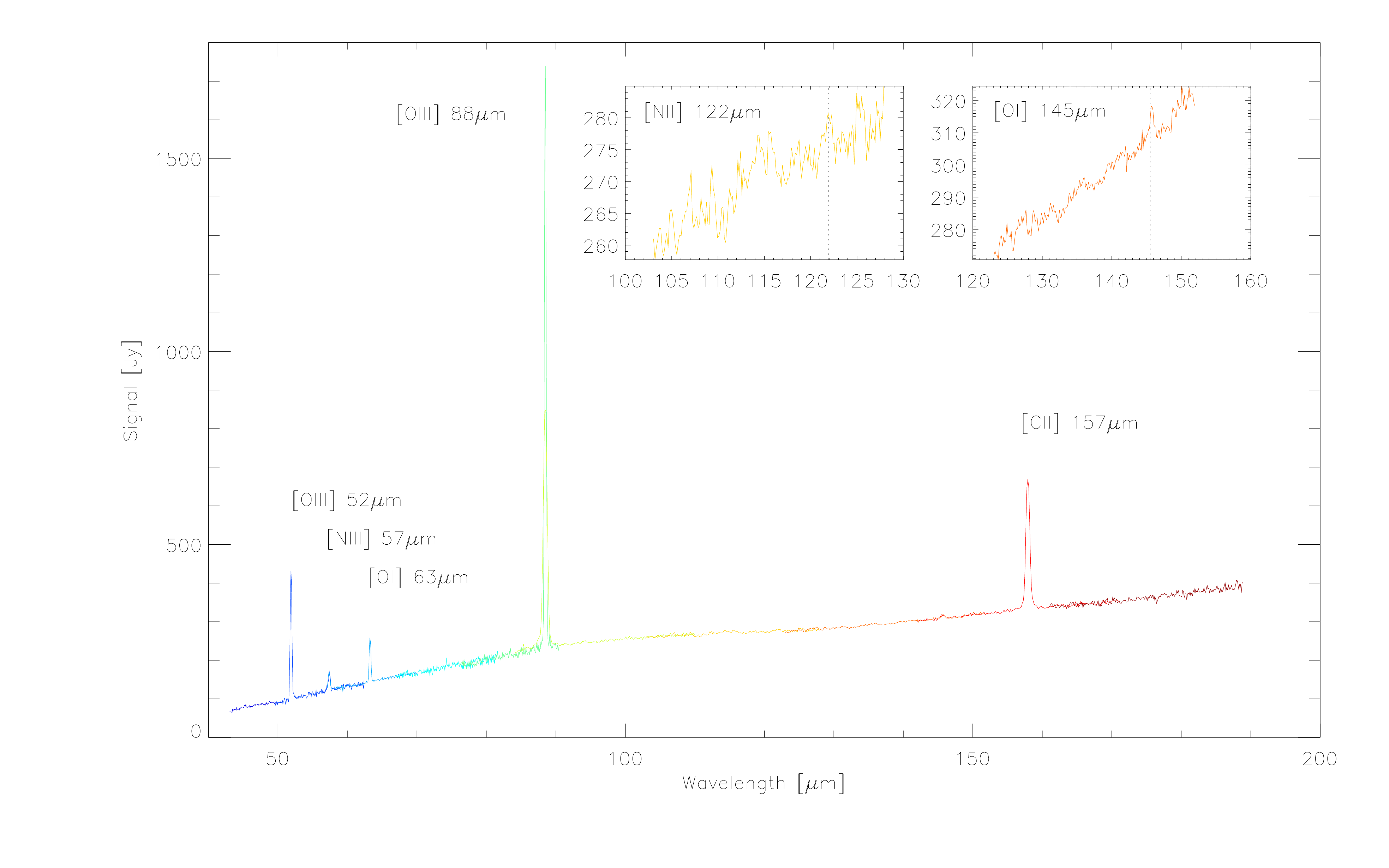}
\caption{ISO LWS spectrum of N\,11B. See Fig.\,\ref{fig:acs_all} for the location of the observation pointing. The colors correspond to the different spectral orders. The [O\3]$_{88}$ is observed in 2 spectral orders with different spectral resolutions. The displayed spectrum was extracted using a point-like source calibration. 
\label{fig:iso_spec}}
\end{figure*}

N\,11B was observed with the ISO space telescope (Kessler et al.\ 1996) in the LWS filter ($45-200$\mic) with a beam size between  $65\arcsec$ and $80\arcsec$. We retrieved the archival data, extracted the spectrum (Fig.\,\ref{fig:iso_spec}), and measured the line fluxes. The flux determinations are listed in Table\,\ref{tab:iso}. The values assuming a point-like source calibration agree within $10$\%\ with Vermeij et al.\ (2002)\footnote{As noted in Vermeij et al.\ (2002), the "N\,11A" LWS observation is in fact pointed toward N\,11B. }. We also extracted the line fluxes assuming an extended source calibration. The good alignment of the dust continuum across the spectral orders in Fig.\,\ref{fig:iso_spec} suggests that the dust distribution is point-like. 
We compared the ISO line fluxes with the integrated fluxes derived from the PACS maps in the same aperture (Fig.\,\ref{fig:acs_all}). The PACS fluxes agree remarkably well with the ISO results. The PACS fluxes are slightly underestimated because of the incomplete coverage of the ISO beam (Fig.\,\ref{fig:acs_all}). 

\begin{table}
\caption{Line fluxes derived from the ISO LWS spectrum.
\label{tab:iso}}
\centering
\begin{tabular}{llll}
\hline\hline
Line  & ISO\tablefootmark{a}  & ISO\tablefootmark{b}  & PACS\tablefootmark{c} \\
\hline
$[$O\3$]_{52}$  & $1.2\times10^{-13}$ &  $8.4\times10^{-14}$ & ...  \\
$[$N\3$]$   &  $1.5\times10^{-14}$ &  $1.1\times10^{-14}$ &  $9.9\pm5.4\times10^{-15}$ \\
$[$O\1$]_{63}$ &  $2.7\times10^{-14}$ &  $1.9\times10^{-14}$ &  $2.1\pm0.1\times10^{-14}$ \\
$[$O\3$]_{88}$  &  $1.7\times10^{-13}$ &  $8.5\times10^{-14}$ &  $9.8\pm1.7\times10^{-14}$ \\
$[$N\2$]_{122}$  &  $<9.1\times10^{-16}$ &  $<2.8\times10^{-16}$ &  $6.1\pm5.1\times10^{-16}$ \\
$[$O\1$]_{145}$  & $1.2\times10^{-15}$ &  $4.2\times10^{-16}$ &  $1.1\pm0.3\times10^{-15}$ \\
$[$C\2$]$ &  $2.9\times10^{-14}$ &  $1.0\times10^{-14}$ &  $2.7\pm0.7\times10^{-14}$ \\
\hline
\end{tabular}\\
\tablefoot{Line flux is given in \wm. }
\tablefoottext{a}{Point-like source calibration. Errors are $\sim25$\% for [O\1]$_{145}$ and $\sim10$\%\ for the other lines. The effective aperture radius is $\approx35\arcsec$ for [O\1]$_{145}$ and [C\2], $\approx39\arcsec$ for [N\2]$_{122}$, $\approx40\arcsec$ for [O\3]$_{88}$, $\approx42\arcsec$ for [O\3]$_{52}$ and [N\3], $\approx43.5\arcsec$ for [O\1]$_{63}$. }
\tablefoottext{b}{Extended source calibration. }
\tablefoottext{c}{Integrated PACS flux within the corresponding ISO aperture.}
\end{table}

\begin{figure}[h!]
\centering
\includegraphics[angle=0,scale=0.462,clip=true]{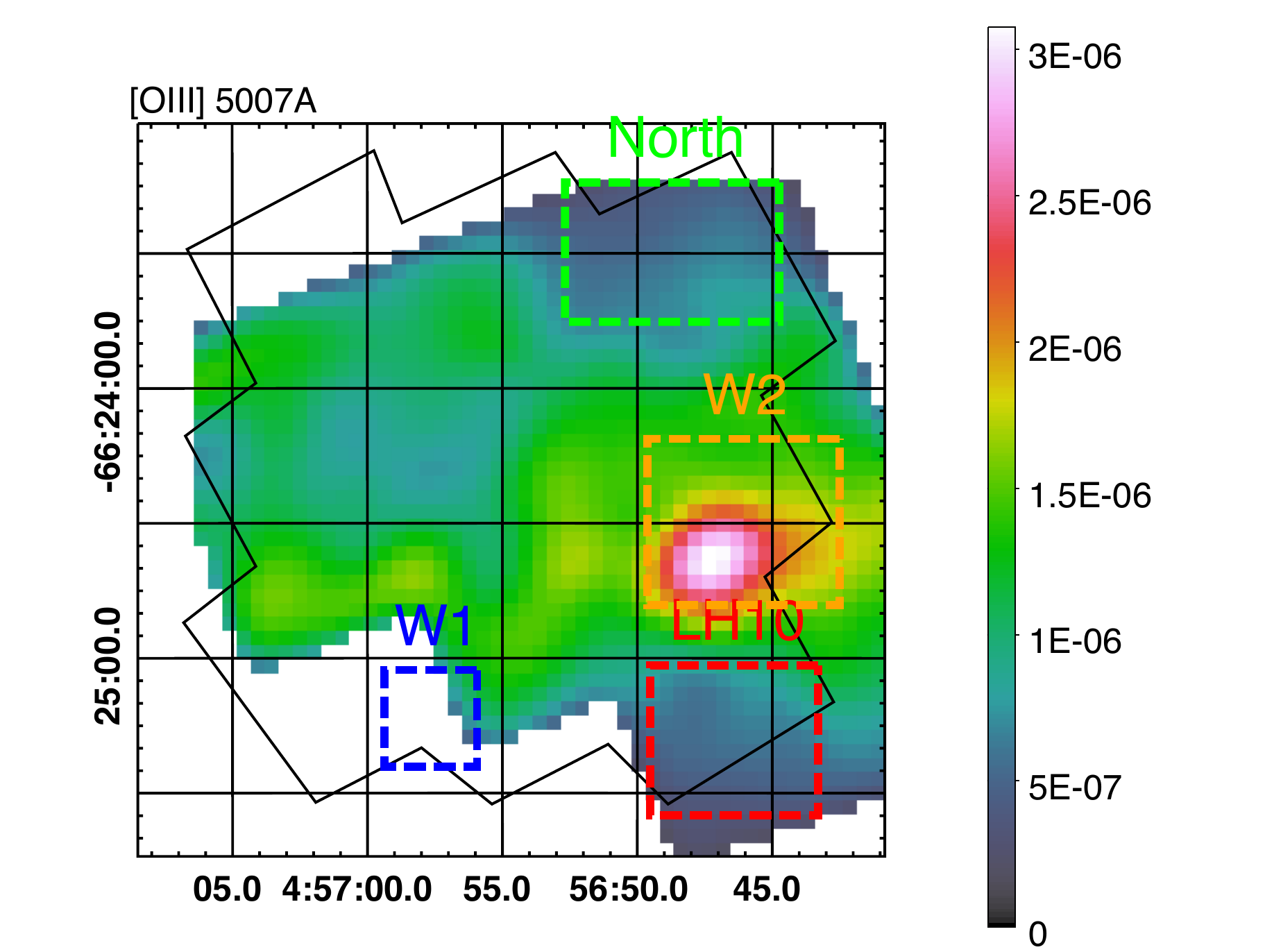}\\
\includegraphics[angle=0,scale=0.462,clip=true]{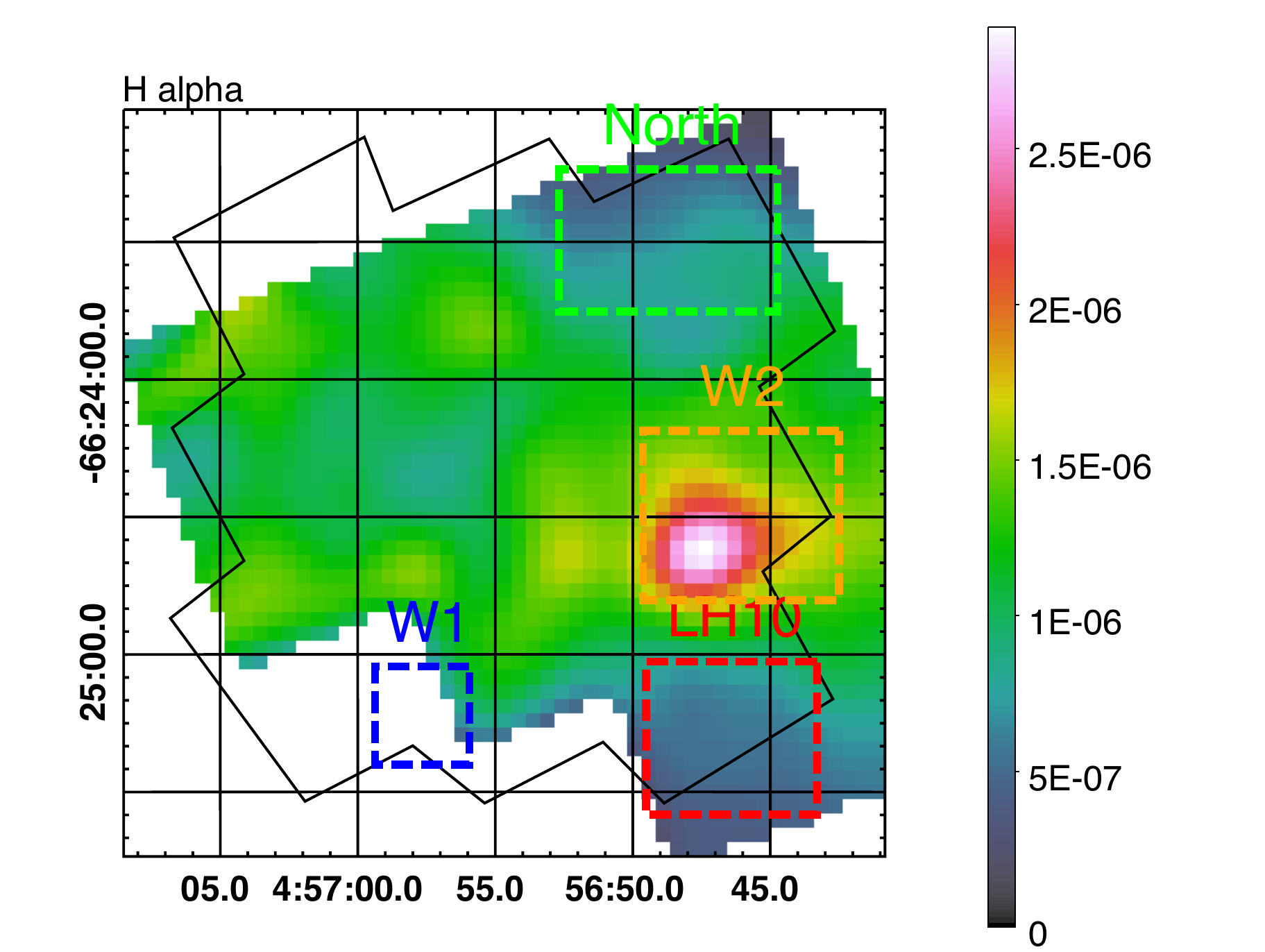}\\
\includegraphics[angle=0,scale=0.48,clip=true]{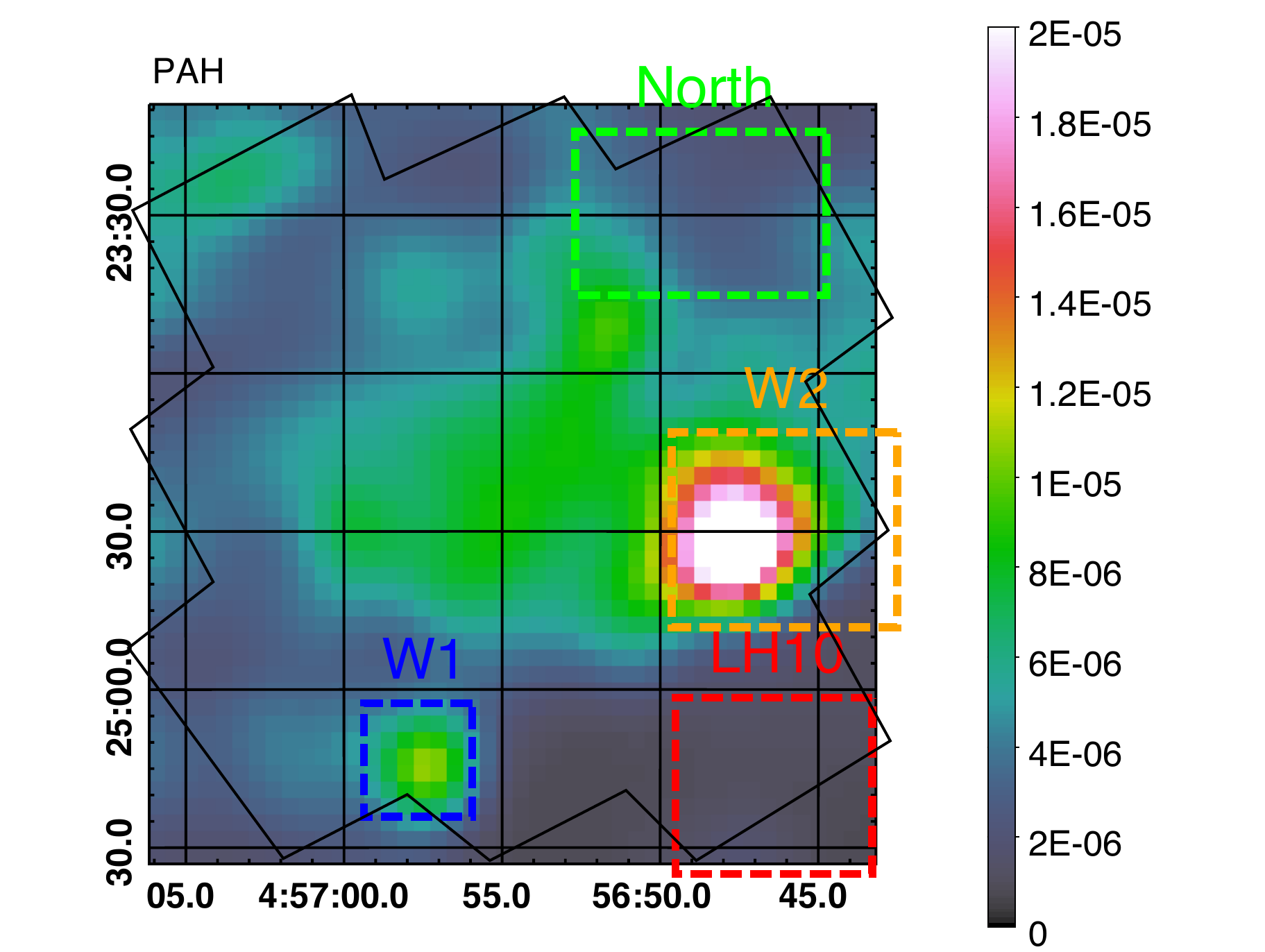}
\caption{[O\3]$_{\rm opt}$, H$\alpha$, and PAH maps, from top to bottom. The images were convolved at the resolution of the \textit{Herschel} [C\2] map. See Fig.\,\ref{fig:pacs_maps} for the figure description. \label{fig:ancillary_maps}}
\end{figure}

We also used the calibrated HST/WFPC2 images in the F502N and F658N narrow band filters, kindly provided to us by Y.\, Naz{\'e} (see also Naz{\'e} et al.\ 2001). The F502N band is dominated by the [O\3] 5007\AA\ (hereafter [O\3]$_{\rm opt}$) line emission. The continuum emission, which is dominated by dust-scattered light in N\,11B, represents only a small fraction ($\lesssim1$\%) of the emission within the narrow filter as compared to [O\3]$_{\rm opt}$ in the long-slit used by Tsamis et al.\ (2008). We assumed that the continuum emission remains negligible throughout the region. A similar assumption was used for the F658N band and the H$\alpha$ line. The WFPC2 images were convolved with a Gaussian PSF to reach a spatial resolution of $12\arcsec$, i.e., slightly larger than the PACS resolution ($\approx10\arcsec$). The WFPC2 images were then resampled to the PACS image pixel size (Sect.\,\ref{sec:fitandmap}). The final [O\3]$_{\rm opt}$ and H$\alpha$ maps are shown in Fig.\,\ref{fig:ancillary_maps}.

We derived the polycyclic aromatic hydrocarbon (PAH) emission by using the archival \textit{Spitzer}/IRAC images. We used band 4, centered at $8.0$\mic, which observes the PAH complex at 7.7\mic\ and 8.6\mic\ along with the warm dust continuum. To remove the warm dust contribution, we assumed a power-law emission on the Wien side of the spectral energy distribution (SED) and used the emission in bands\ 1 ($3.6$\mic) and\ 2 ($4.5$\mic) to extrapolate to the band 4 wavelength range for each pixel. The correction thereby applied is $\sim10$\%. 
We then calibrated the flux density in band\ 4 by using a template source spectrum, as explained in the \textit{Spitzer} synthetic photometry cookbook\footnote{\textit{http://irsa.ipac.caltech.edu/data/SPITZER/docs/dataanalysistools/}}. The template spectrum was built from an incomplete \textit{Spitzer}/IRS map of N\,11B, covering the LH\,10 and W2 regions (AORKEY 22469632).
From the IRAC transmission curve, we further estimated that about $\approx45$\% of the total PAH emission (including features from $5$\mic\ to $18$\mic) falls within band 4. We therefore corrected the PAH map by this factor. The final PAH map is presented in Fig.\,\ref{fig:ancillary_maps}.

\section{Distribution of the FIR lines}\label{sec:distrib}

The main observational results (noise level and integrated flux) are provided in Table\,\ref{tab:lines_results}. We discuss the detailed results for each line in the following.

\begin{table}
\caption{Summary of observations and integrated emission properties.
\label{tab:lines_results}}
\centering
\begin{tabular}{lllll}
\hline\hline
Line &  & RMS\tablefootmark{b} & SNR\tablefootmark{c} & $F_{\rm tot}$\tablefootmark{d} \\
       &    &  (\wm)                    &                                        & (\wm)      \\
\hline
$[$O\1$]_{63}$  & Map &  $2.1\times10^{-17}$ & $0/6/86$ & $3.9\times10^{-14}$   \\
$[$O\1$]_{145}$ & Map & $4.7\times10^{-18}$ & $0/2/35$ & $\gtrsim3.4\times10^{-15}$   \\
                            & Pointing &$1.5\times10^{-18}$ & $0/5/45$ & ... \\
$[$C\2$]$ & Map & $9.7\times10^{-18}$ & $10/43/95$ & $8.4\times10^{-14}$  \\
$[$O\3$]_{88}$  & Map & $2.0\times10^{-17}$ & $29/101/152$ & $3.8\times10^{-13}$   \\
$[$N\3$]$   &Map & $3.0\times10^{-17}$ & $0.7/4/14$ & $\gtrsim3.2\times10^{-14}$   \\
$[$N\2$]_{122}$ & Map & $8.6\times10^{-18}$ & $0/1.4/5$ & $\gtrsim3.2\times10^{-15}$  \\
                            & Pointing &  $1.4\times10^{-18}$ & $0.3/5/13$ & ...   \\
$[$N\2$]_{205}$ &Map & $5.5\times10^{-18}$ & $0/0.5/2$ & $\gtrsim1.0\times10^{-15}$  \\
 \hline
\end{tabular}\\
\tablefoottext{a}{Radial velocity.}
\tablefoottext{b}{Median root mean square (RMS) value per spaxel across the map (or across the footprint for the pointed observations).}
\tablefoottext{c}{Minimum, median, and maximum S/N ratio per spaxel.}
\tablefoottext{d}{Integrated flux in map.}
\end{table}

\subsection{[O\1]}\label{sec:ngtracers}

Both [O\1]$_{63}$ and [O\1]$_{145}$ have high critical densities (Table\,\ref{tab:lines}) and trace mostly the dense gas in PDRs. 
The two lines show nearly identical distributions (Fig.\,\ref{fig:pacs_maps}), most of the differences being due to the lower S/N of the [O\1]$_{145}$ map in the fainter and more diffuse regions. The [O\1]$_{145}$ emission is dominated by the region W2, while fainter emission is also seen toward the three compact regions on the eastern side, including the region W1. 

The [O\1]$_{63}$ line is detected throughout the entire map, its intensity varying by a factor of $\sim90$ (Table\,\ref{tab:lines_results}). The line ratio [O\1]$_{145}$/[O\1]$_{63}$ varies from $\approx0.05$ to $\approx0.11$ across the map. 
The integrated [O\1]$_{63}$ emission in the map is $3.9\times10^{-14}$\,W\,m$^{-2}$, about $\approx39$\%\ of which comes from W2, and $\approx44$\%\ come from both W1 and W2. This fraction is remarkably large as compared to the area covered by W1+W2 within the PACS map ($\sim13$\%).

\subsection{[C\2]}\label{sec:cii}

The [C\2] line is thought to be the dominant cooling mechanism in low-density PDRs and in the diffuse ISM for temperatures $\lesssim8\times10^3$\,K (Dalgarno \&\ McCray 1972; Tielens \&\ Hollenbach 1985a, 1985b; Wolfire et al.\ 1995). Depending on the ionization fraction of the gas, excitation of the ${^2}P_{3/2}$ level  is due to inelastic collisions with neutral hydrogen atoms and molecules or with electrons (Dalgarno \&\ McCray 1972; Stacey 1985; Kulkarni \&\ Heiles 1987). Higher electronic states ($^4P$, $^2D$, and $^2S$) lie more than $6.2\times10^4$\,K above the ground state, temperatures that are not attained in most astrophysical environments (see Crawford et al.\ 1983).

[C\2] varies by a factor of $20$ across the map (Fig.\,\ref{fig:pacs_maps}). Although it shows a flatter distribution than [O\1]$_{63}$, [C\2] is the brightest where [O\1]$_{63}$ is the brightest, i.e., toward W2. The emission around W2 is, however, more extended than the [O\1]$_{63}$ emission, following the main optical arc I1 well (Fig.\,\ref{fig:acs_all}). The [O\1]$_{63}$/[C\2] ratio shows strong variations, from $1.5-2$ toward W1 and W2 down to $\approx0.3$ everywhere else. The total [C\2] emission in the map is $8.4\times10^{-14}$\,W\,m$^{-2}$, about $\approx19$\%\ of  which comes from W2. 
This fraction is only slightly greater than the area fraction covered by W1+W2 within the PACS map ($\sim13$\%), suggesting that most of the [C\2] emission is located in relatively extended and diffuse regions. 

The [C\2] line was observed in N\,11B with the NASA Kuiper Airborne Observatory (KAO) by Israel \&\ Maloney (2011),  who find a peak emission of $2.2\times10^{-7}$\wmsr\ with a beam of $\sim68$\arcsec, and by Boreiko \&\ Betz (1991) who found $1.0\times10^{-7}$\wmsr\ in a $43$\arcsec\ beam. With BICE, Mochizuki et al.\ (1994) measured $6\times10^{-8}$\wmsr\ in a $12.4$\arcmin\ beam. The peak emission we detect ($8\times10^{-7}$\wmsr\ with a beam of $\sim11.5$\arcsec) is significantly larger by a factor of $\sim3.5$ than in Israel \&\ Maloney (2011). We attribute this difference to the relatively smaller beam of PACS ($\approx12\arcsec$; Sect.\,\ref{sec:obs}).

\subsection{High-excitation ionized gas tracers}\label{sec:he_tracers}

Both [O\3]$_{88}$ and [N\3] trace the gas photoionized by massive stars. In the N\,11B region, we identify at least two O3 stars, and probably as many as a few dozen later-type O stars able to create O$^{++}$ and N$^{++}$ ions (Sect.\ \ref{sec:ionizing_sources}).

The critical density of [O\3]$_{88}$ is $n_{\rm cr}\approx510$\cc\ (as compared to $\approx3000$\cc\ for [N\3]), which makes it the best tracer available for low-density, high-excitation, ionized gas. The electron density in N\,11B is expected to be around $25$\cc\ on a global scale, and reaches up to a few hundred per cm$^3$ in low-ionization bright optical arcs (Sect.\,\ref{sec:density}). We therefore expect most of the gas to be able to de-excite radiatively through both [O\3]$_{88}$ and [N\3].  

Both [O\3]$_{88}$ and [N\3] show a fairly similar spatial distribution with a relatively narrow range of values (both within a factor of $7$; Fig.\,\ref{fig:pacs_maps}). Most of the emission is concentrated between W2 and the stellar cluster LH\,10 to the west, and around the ``footprint" stellar cluster to the east. The total [O\3]$_{88}$ emission in the map is $3.8\times10^{-13}$\,W\,m$^{-2}$, $\approx14$\%\ of which comes from W2, which is only barely larger than the actual area covered by W2 within the PACS map. [O\3]$_{88}$ is therefore dominated by extended emission, making it the brightest FIR line over the entire map area. [O\3]$_{88}$ is brighter than [C\2] by a factor of $1.2$ to $20$ throughout the map, and its integrated flux is four times brighter than [C\2]. The predominance of [O\3]$_{88}$ as compared to [C\2] is also found in the pointed ISO observation, with [O\3]$_{88}$ being about seven times brighter than [C\2] (Sect.\,\ref{sec:observations_ancillary}).

The peak of [O\3]$_{88}$ lies directly south of W2, where it is slightly offset with respect to the location of the main ionizing front ``I1" (Fig.\,\ref{fig:acs_all}). The [N\3] peak lies between the [O\3]$_{88}$ peak and the PDRs. Considering the ionization potentials of N$^+$ and O$^+$ ($35.1$\,eV and $29.6$\,eV respectively), the difference in the peak location suggests that the ionization structure on the western half of N\,11B is dominated by the LH\,10 cluster in the south. In the eastern half, the emission is concentrated toward W1, the ``footprint" cluster, and the optically bright filaments (Fig.\,\ref{fig:acs_all}).

\begin{figure}
\includegraphics[angle=0,scale=0.9,clip=true]{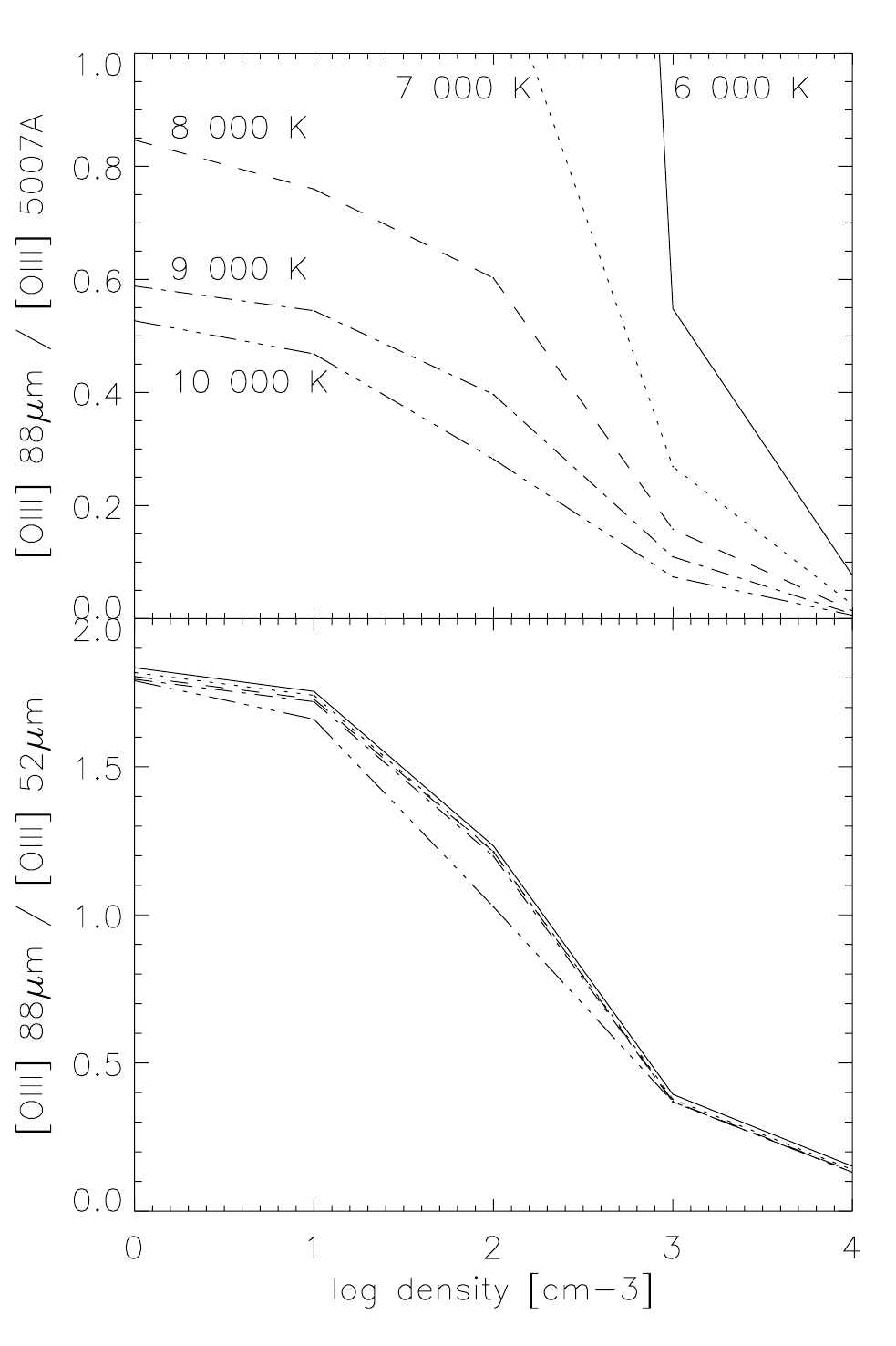}
\caption{Density diagnostic plot using the [O\3]$_{\rm 88/opt}$ ratio (\textit{top}) and the [O\3]$_{88/52}$ ratio (\textit{bottom}). The [O\3]$_{\rm 88/opt}$ ratio depends mostly on the electron temperature for $T\lesssim8000$\,K. Results were computed with Cloudy using a half-solar metallicity (Sect.\,\ref{sec:igmodels}).  \label{fig:oiii_densitydiag}}
\end{figure}

In the following (Sect.\,\ref{sec:oiii}) we compare [O\3]$_{88}$ to the optical [O\3]$_{\rm opt}$ line at $5007$\AA\ (Fig.\,\ref{fig:ancillary_maps}). The [O\3]$_{\rm opt}$ line has a much higher critical density of $n_{\rm cr}\approx8.6 \times10^4$\cc\ ($\approx510$\cc\ for [O\3]$_{88}$), and thus traces a wider range of gas densities, provided there is no significant extinction by dust. The [O\3]$_{\rm opt}$ and [O\3]$_{88}$ lines also have different excitation temperatures. Their theoretical ratio is shown in Fig.\,\ref{fig:oiii_densitydiag} as a function of density and temperature (see also Osterbrock \&\ Ferland 2006 and Palay et al.\ 2012). Tsamis et al.\ (2001) found $T_e\sim9400$\,K by observing the [O\3] optical lines $\lambda4363$, $\lambda4959$, and $\lambda5007$ in a long slit. For such high temperatures, the [O\3]$_{\rm 88/opt}$ ratio mostly depends on density. 

The [O\3]$_{52}$ line was not observed with PACS, but it was observed and detected with ISO (Sect.\,\ref{sec:observations_ancillary}). The observed [O\3]$_{88/52}$ ratio lies between $1$ and $1.4$. The exact value depends on the asumed flux calibration (extended or point-like). Within the ISO beam, we estimate that only $\sim20$\%\ of the [O\3]$_{88}$ flux comes from W2, while the rest comes from extended emission. We thus consider that the [O\3]$_{88/52}$ ratio should be close to the extended source calibration value, i.e., $\approx1$. Figure\,\ref{fig:oiii_densitydiag} shows that such a ratio implies an electron density of around $100$\cc. This value is somewhat higher than the density derived from the emission measure in the global ISM of N\,11B (Sect.\,\ref{sec:density}). We note, however, that $100$\cc\ is an upper limit since the [O\3]$_{88}$ emission is not completely uniform.  

\subsection{Low-excitation ionized gas tracers}

The [N\2]$_{122}$ and [N\2]$_{205}$ lines trace the low-excitation ionized gas (the N$^+$ ion exists for energies $>14.5$\,eV). The critical density of [N\2]$_{122}$ ($n_{\rm cr}\approx400$\cc) is similar to that of [O\3]$_{88}$ (Table\,\ref{tab:lines}). [N\2]$_{122}$ was only barely detected toward a few positions in the mapping observation because of a too short integration time. The upper limit on the total emission in the map is $<5.5\times10^{-15}$\,W\,m$^{-2}$. The follow-up pointed observations successfully detected [N\2]$_{122}$ toward W2 and W1, with peak emissions of $1.4\times10^{-8}$\wmsr\ and  $8.5\times10^{-9}$\wmsr, respectively. 

Unfortunately, [N\2]$_{205}$ was not detected in the mapping observation. The upper limit on the total emission in the map is $<3.6\times10^{-15}$\,W\,m$^{-2}$. Since the [N\2]$_{122}$ is systematically brighter than [N\2]$_{205}$ for densities over $10$\cc\ (Oberst et al.\ 2006), the upper limit on [N\2]$_{205}$ is not a useful constraint on the gas density and on the models discussed in this paper.

\section{Physical conditions of the ionized gas}\label{sec:physcond}

\subsection{Contribution from shocks}\label{sec:shocks}

The presence of shocks from supernovae remnants or from stellar winds could modify the electronic level population and therefore line ratios. A diffuse soft X-ray emission component was detected in N\,11B (Mac Low et al.\ 1998; Naz{\'e} et al.\ 2004), and wind-blown bubbles around individual stars or groups of stars were detected kinematically by Naz{\'e} et al.\ (2001). These bubbles are all X-ray emitters despite their low expansion velocities of $10-15$\kms. On the other hand, as noted in Naz{\'e} et al.\ (2001), the young age of the LH\,10 association, as opposed to the central association of N\,11 (LH\,9), suggests that the most massive stars did not have time to evolve into supernovae. Strong interstellar shocks are therefore not expected in N\,11B. 

To verify these assumptions, we compared the observed line ratio [O\3]$_{\rm 88/opt}$ to the predictions from the shock model described in Allen et al.\ (2008). For LMC abundances and a density of $1$\cc, the [O\3]$_{\rm 88/opt}$ ratio in the shocked gas is always lower than $\approx0.23$ throughout the range of velocity and magnetic field values in the model. A somewhat lower ratio would be obtained for densities higher than $1$\cc. Accounting for the extinction by dust (which affects the optical line emission at least toward W2 and LH\,10; Sect.\,\ref{sec:oiii}), our observations show that [O\3]$_{\rm 88/opt}\gtrsim0.25$ across the nebula (Fig.\,\ref{fig:lineratios}c). As a result one can see that (1) shocks will tend to lower the observed [O\3]$_{\rm 88/opt}$ ratio, and (2) although shocks might exist, they do not dominate the ionization structure of the N\,11B nebula.

\subsection{Photoionization modeling}\label{sec:igmodels}

\begin{figure*}
\includegraphics[angle=0,scale=0.65,clip=true]{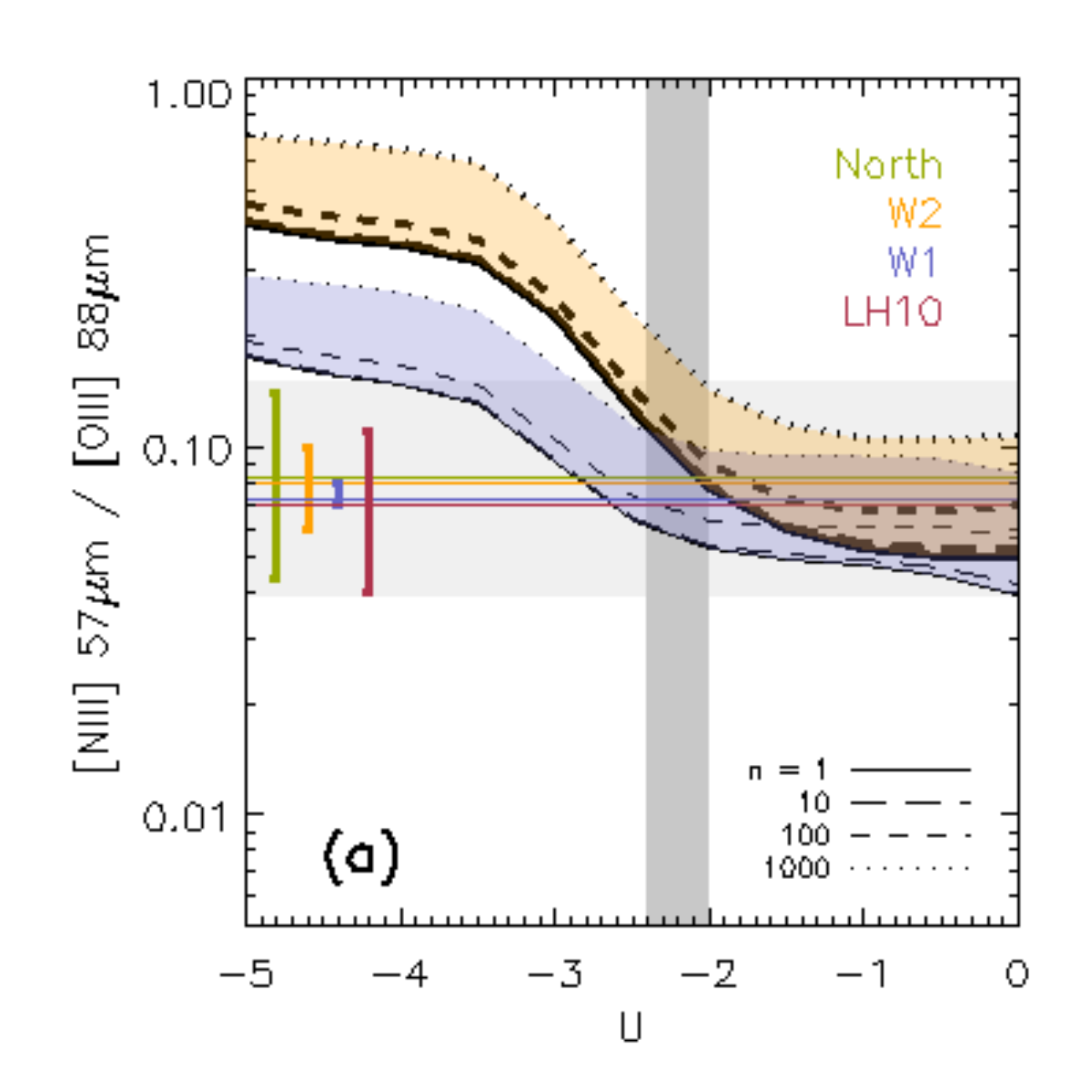}
\includegraphics[angle=0,scale=0.65,clip=true]{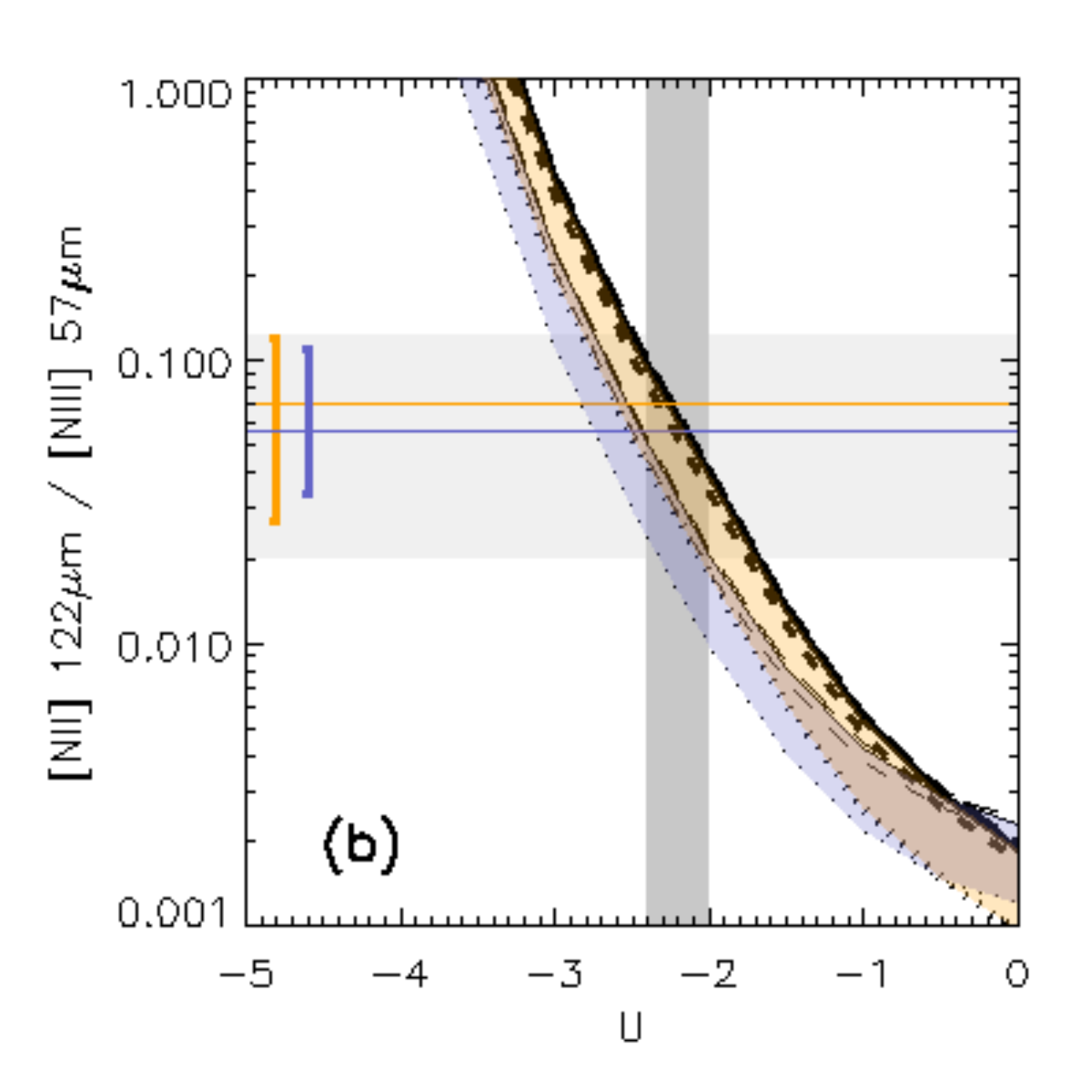}\\
\includegraphics[angle=0,scale=0.65,clip=true]{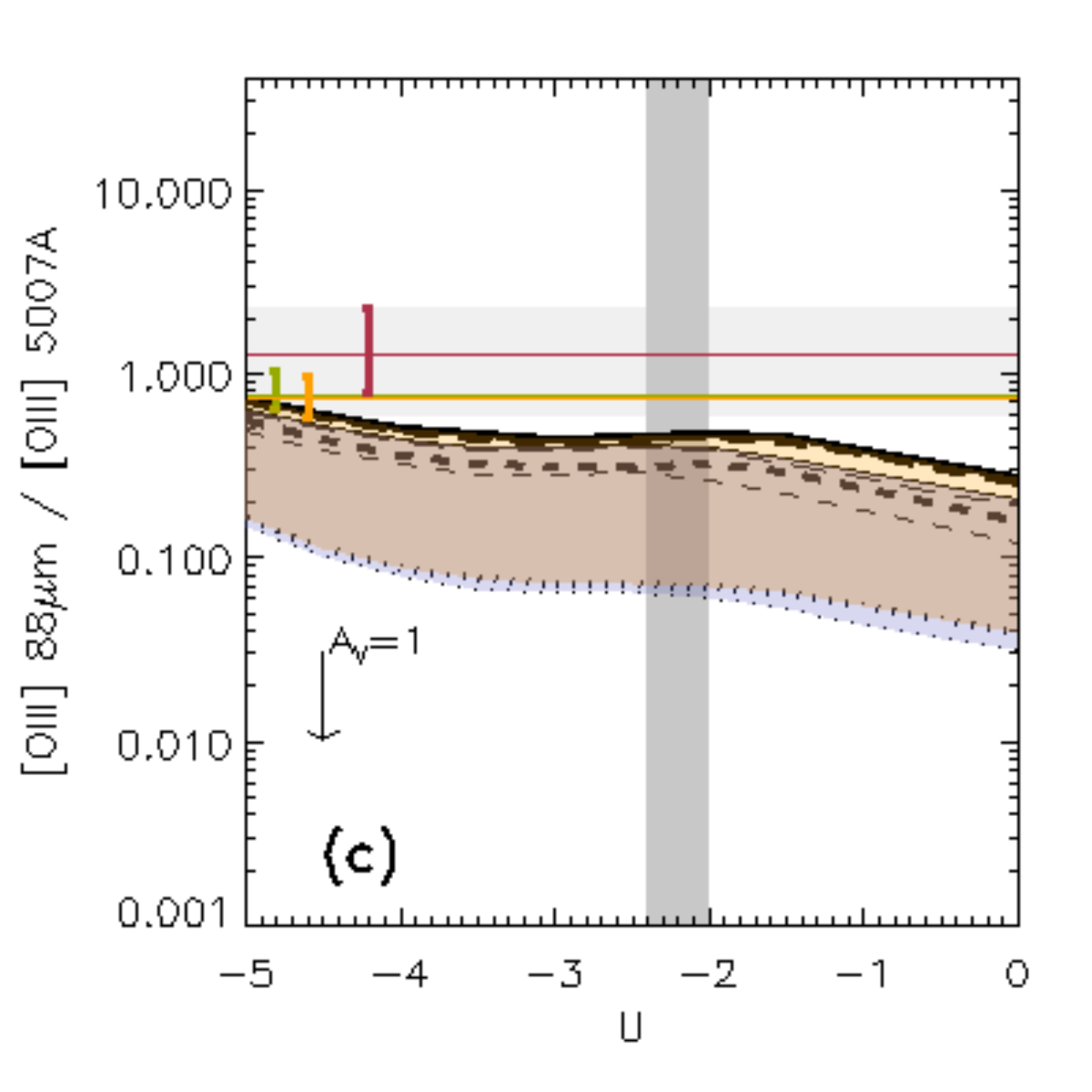}
\includegraphics[angle=0,scale=0.65,clip=true]{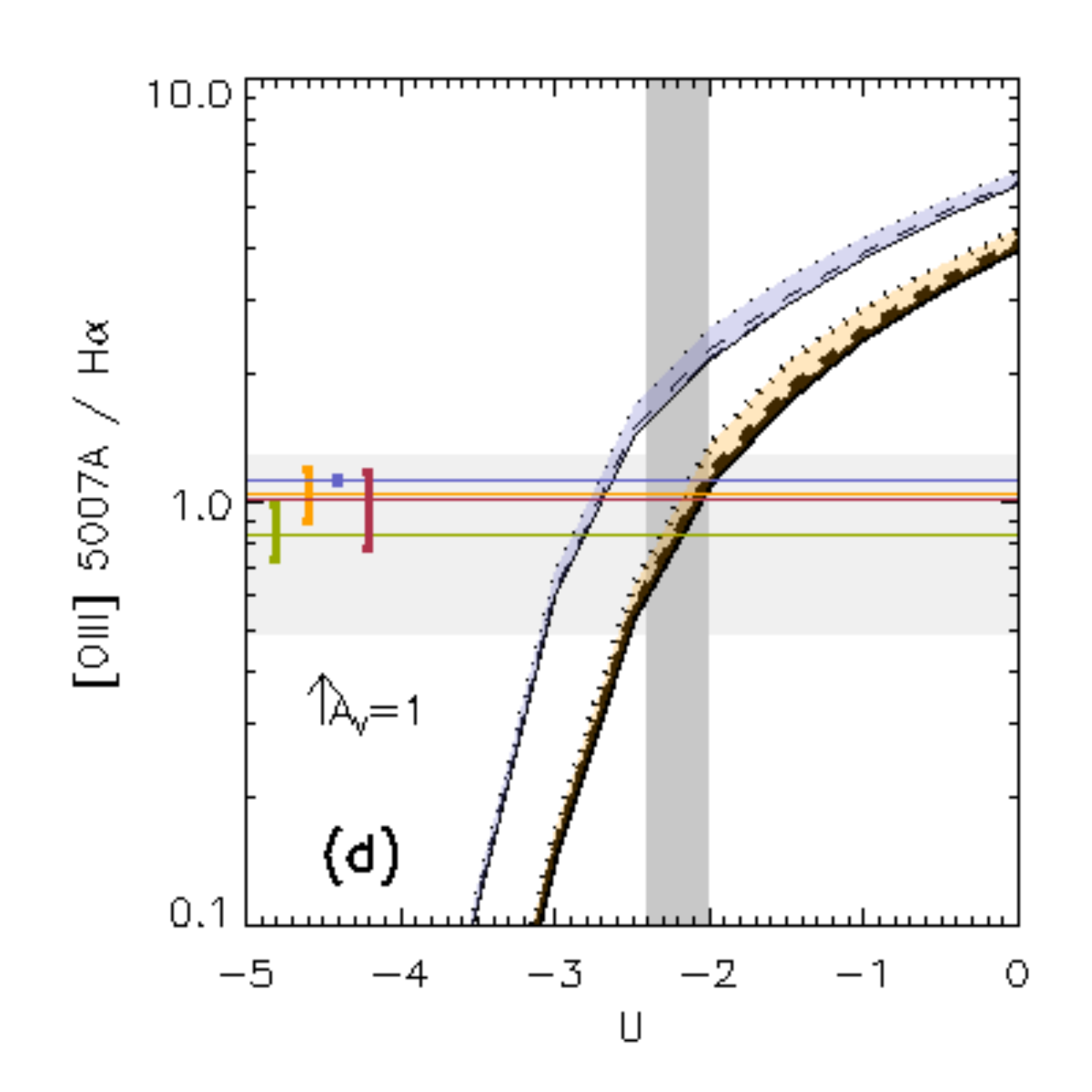}
\caption{The line flux ratios plotted against the ionization parameter $U$ for radiation temperature of $T_{\rm eff}=40000$\,K (orange models) and $45000$\,K (blue models). The horizontal lines show the observed values toward specific regions illustrated in previous figures, while the associated vertical bars show the range of values within each region. The light gray horizontal rectangle shows the range of values observed across the entire N\,11B nebula. The dark gray vertical rectangle indicates the range of $U$ values agreeing the best with the observed line ratios (see text). The vertical arrow shows the effect of dust extinction on the observations for $A_V=1$. The W1 region was not covered by the [O\3]$_{\rm opt}$ HST/WFPC2 observation (Fig.\,\ref{fig:ancillary_maps}), and [N\2]$_{122}$ was only detected by the \textit{Herschel}/PACS follow-up pointed observations (Fig.\,\ref{fig:pacs_maps}). 
\label{fig:lineratios}}
\end{figure*}

In the following, we model the physical conditions in the N\,11B nebula with the photoionization code Cloudy (c10.00; Ferland et al.\ 1998). A single stellar source is used to model the radiation field. We used the stellar atmosphere grid WM-Basic (Pauldrach et al.\ 2001 and references therein) because it includes non-LTE effects. The stellar grid was computed with half-solar metallicity and main-sequence parameters. The intensity of the radiation field varies through the ionization parameter $U$, which is defined as
\begin{equation} 
U = \frac{ \Phi({\rm H}) } { n({\rm H})\ c },
\end{equation}
where $\Phi({\rm H})$ is the flux of ionizing photons per unit surface and $n({\rm H})$ is the hydrogen density. The other varying parameters are the stellar radiation effective temperature $T_{\rm eff}$ and the gas density $n$ (assumed to be constant). The main parameters for the models are listed in Table\,\ref{tab:models}. Models are stopped at the ionization front, i.e., when the electron fraction reaches $0.5$. Figure\,\ref{fig:lineratios} shows the line ratios predicted by the models, together with the observations toward specific regions in the map (the compact regions W1 and W2, the stellar cluster LH\,10, and the relatively diffuse northern region; Sect.\,\ref{sec:morphology}).

\begin{table}
\caption{Main parameters for the models.
\label{tab:models}}
\begin{tabular}{ll}
\hline\hline
Parameter & Value  \\
\hline
Gaseous abundances\tablefootmark{a}  & O:\,$8.40$, N:\,$6.94$, C:\,$7.91$  \\
Ionization parameter $U$  & $[-5, 0]$ \\
Stellar temperature  & $[30000, 50000]$\,K \\
Stellar atmosphere metallicity & $0.5$\,Z$_\odot$ \\
Density (constant)  & $[1, 10^3]$\cc\ \\
Filling factor & $1$ \\
 \hline
 \end{tabular}\\
\tablefoot{Chemical abundances, expressed as $12+\log({\rm X/H})$, are discussed in Sect.\,\ref{sec:abundances}.}
\end{table}

The observed [N\3]/[O\3]$_{88}$ ratio is compatible with ionization parameters $\gtrsim-3$ and with stellar temperatures between $\sim40000$\,K and $\sim45000$\,K (Fig.\,\ref{fig:lineratios}a). Such temperatures correspond to O7-O5 main sequence stars, many of which are detected in N\,11B (Sect.\,\ref{sec:spatialOIII}). Only a couple of more massive stars than O5 were detected that must drive $T_{\rm eff}$ slightly higher. Inversely, radiation temperatures colder than $40000$\,K fail to reproduce the observed [N\3]/[O\3]$_{88}$ ratio because corresponding models produce too little [O\3]$_{88}$ emission. 

Considering that the gaseous nebula in N\,11B should have a density higher than $>10$\cc\ (Sect.\,\ref{sec:density}), the average [N\3]/[O\3]$_{88}$ ratio in each region (Fig.\,\ref{fig:lineratios}a) is compatible with an ionization parameter $\sim-2$ for a radiation temperature of $T_{\rm eff}=40000$\,K and $\sim-3$ for $T_{\rm eff}=45000$\,K. Still, [N\3]/[O\3]$_{88}$ shows significant variations across N\,11B, by a factor of $\sim5$. The lowest value is found in LH\,10 and agrees with both low-density and high ionization-parameter. Inversely, the highest value is found in the northern region and it agrees with a low ionization-parameter ($\sim-3$), which is expected from the larger distance to the known ionizing clusters. 

For illustration, an ionization parameter of $-2$ would correspond to a $25$\cc\ cloud located $6$\,pc from a O5\,V star ($U=-1$ implies a distance of $2$\,pc while $U=-3$ implies $18$\,pc) or to a $200$\cc\ cloud located $2$\,pc away. 
A lower limit of $5$\,pc can be set on the distance between the W2 ionization fronts and most stars in LH\,10 by measuring the projected distance in Fig.\,\ref{fig:acs_all}. Considering the large number of ionizing stars in LH\,10 (Sect.\,\ref{sec:ionizing_sources}), it is surprising to find such a low ionization parameter. However, we note that the flux of all the ionized gas tracers (H$\alpha$, [O\3]$_{\rm 88, opt}$, [N\3], and [N\2]$_{122}$) is in fact only twice greater toward W2 than toward the surrounding regions, so half of the ionization must come from an extended gas component, which has likely a lower $U$ value because of geometrical dilution. We thus expect a very low-density component that drives $U$ lower (see also Sect.\,\ref{sec:spatialOIII}). Choosing a density of $n_e=25$\cc\ and a distance of $30$\,pc (size of the PACS map coverage), $\log\ U=-2$ corresponds to about $26$ O5\,V sources, which is a realistic number considering the O star census in N\,11B (Sect.\,\ref{sec:spatialOIII}).

[N\2]$_{122}$/[N\3] is another useful tracer of the physical conditions in the ionized gas, as it usually provides an estimate of $T_{\rm eff}$ (e.g., Rubin et al.\ 1994). This ratio is also geometry-dependent because [N\2]$_{122}$ emits close to the ionization front, so that the nebula should be radiation-bounded for the derived physical conditions to be meaningful. We now test this hypothesis. [N\2]$_{122}$ was only detected toward the brightest regions W1 and W2 (Fig.\,\ref{fig:pacs_maps}). Figure\,\ref{fig:lineratios}b shows that [N\2]$_{122}$/[N\3] toward W1 and W2 is compatible with an ionization parameter in the range $[-2.5, -1.8]$ ($n_e\sim100$\cc\ and $T_{\rm eff}=40000$\,K). We obtain the same range from the [N\3]/[O\3]$_{88}$ diagnostic plot, implying that [N\2]$_{122}$/[N\3] provides a good estimate of the physical conditions toward W1 and W2; i.e., the nebula must be radiation-bounded toward these two positions. This agrees with the optical images that show multiple ionization fronts (Sect.\,\ref{sec:morphology}). If the nebula was not radiation-bounded, we should have observed a lower [N\2]$_{122}$/[N\3] ratio, which in turn would have translated into a larger $U$ parameter. Unfortunately, we cannot conclude anything about the other regions, since [N\2]$_{122}$ was detected only toward W1 and W2.

The diagnostic plots in Fig.\,\ref{fig:lineratios} give useful indications about the excitation conditions within the compact source W1. The ratios [N\2]$_{122}$/[N\3] and [N\3]/[O\3]$_{88}$ toward W1 are similar to the values within W2. Assuming that the same radiation field  temperature holds for both regions ($\sim45000$\,K), this implies roughly the same ionization parameter. Given the small size of W1 ($1$\,pc; Fig.\,\ref{fig:acs_all}), we could have expected a much larger $U$. That $U$ is the same toward W1 and W2 is a strong indication that sources within W1 do not dominate the ionization of the region. Instead, it is likely that the ionization fronts are due to sources outside W1. These sources are very likely the massive stars in the ``footprint" cluster and in LH\,10. This was hypothesized by Naz{\'e} et al.\ (2001) based on the orientation of the ionization fronts as seen from the optical images. Sources within W1 must be much colder than $40000$\,K, resulting in a larger $U$ parameter but a much weaker and, in fact, negligible [O\3]$_{88}$ flux.

\subsection{Evidence of obscured ionized gas}\label{sec:oiii}

We now investigate the dust extinction by comparing [O\3]$_{88}$ and [O\3]$_{\rm opt}$. The FIR line is not affected by dust extinction and potentially provides a better handle on the spatial distribution of the ionized gas. Furthermore, understanding the extinction across N\,11B allows us to use the [O\3]$_{\rm opt}$/H$\alpha$ ratio as a diagnostic tool for the physical conditions. 

The relative variations of [O\3]$_{\rm opt}$ and [O\3]$_{88}$, which intrinsically depend on the electron density and temperature (Sect.\,\ref{sec:distrib}), is then modified by dust extinction, which significantly affects the emission from the optical line emission. The variation in [O\3]$_{88}$ and [O\3]$_{\rm opt}$ between LH\,10 and W2 is remarkably different, the FIR line emission being only $2.5$ times lower toward W2 than toward LH\,10, while the optical line is ten times weaker  toward W2 than toward LH\,10 (Figs.\,\ref{fig:pacs_maps} and \,\ref{fig:ancillary_maps}). This could be due a priori to a variety of parameters (different line critical densities, different electron temperature, extinction by dust). 

Figure\,\ref{fig:lineratios}c shows that [O\3]$_{\rm 88/opt}$ is systematically underestimated by all models. The most relevant models for comparison to observations have densities around ($\sim10-100$\cc; long- and short-dashed lines) and $U\sim-2$ (Sect.\,\ref{sec:igmodels}). As shown in Fig.\,\ref{fig:oiii_densitydiag}, [O\3]$_{\rm 88/opt}$ varies with the electron density and temperature. Tsamis et al.\ (2003) find a large discrepancy between optical recombination vs.\ collisionally-excited lines of O$^+$, but they note that this difference cannot be explained either by temperature fluctuations in a chemically and density homogenous nebula or by high-density clumps. 

Another caveat concerns the metal abundance. Although the observed [O\3]$_{\rm 88/opt}$ ratio could be reconciled with models by using a higher metallicity (since heavy elements provide more cooling, which in turn reduces the optical line emission), we consider this hypothesis as being unrealistic given the well constrained abundances in N\,11B (Tsamis et al.\ 2003). For illustration, a solar abundance would be required for the models to reach the observed [O\3]$_{\rm 88/opt}$ toward W1 and LH\,10. 

Given the different wavelength ranges of [O\3]$_{88}$ and [O\3]$_{\rm opt}$, extinction by dust plays the most important role. The effect of dust extinction with $A_V=1$ is shown in Fig.\,\ref{fig:lineratios}c. The average color excess towards stars across the N\,11B nebula is $E($B$-$V$)\approx0.2$, while some stars have a significantly greater reddening (Parker et al.\ 1992; Lee 1990). This can be compared to $E($B$-$V$)\approx0.05$ toward the central cluster LH\,9 of N\,11, which is essentially foreground galactic extinction (Parker et al.\ 1992). Tsamis et al.\ (2003) find that the Balmer ratios in the optical spectra (long slit of $5.6\arcmin\times1.5\arcsec$) agree fairly well with the theoretical values in the case B with a Galactic foreground extinction. From the same Balmer ratios, we estimate that the extinction cannot be much more than $E({\rm B}-{\rm V})\lesssim0.15$ in the area probed by the long slit. From this value of $E($B$-$V$) = 0.15$, we estimate that as much as $\approx38$\%\ of the [O\3]$_{\rm opt}$ emission is extinguished by dust by using the extinction curve from Cardelli, Clayton \&\ Mathis (1989) with a total-to-selective extinction $R_{\rm V} = 3.1$ (Howarth 1983). This leads to a lower [O\3]$_{\rm 88/opt}$ by a factor $\approx1.6$, which is high enough to reconcile most of the observations with the models. 

Toward a more detailed analysis, we find that [O\3]$_{\rm 88/opt}$ ratio toward the stellar cluster LH\,10 is a factor of $\approx2$ above any models. In particular, the observed ratios lie above the models with the low densities expected in that region (Sect.\,\ref{sec:density}). Assuming that the difference between the models and the observations is driven by interstellar dust extinction, such a factor translates into a visual extinction of $A_{\rm V}\approx1$ or $E($B$-$V$)\approx0.21$. This determination is in good agreement with the average color excess $E($B$-$V$)=0.2$ toward the stars. 

The situation toward W2 is more complex. If the line emission is dominated by low-density ionized gas, then almost no dust extinction is required since the observed [O\3]$_{\rm 88/opt}$ is close to the $10$\cc\ track. On the other hand, if the emission is dominated by denser gas (Sect.\,\ref{sec:morphology}), then a factor $1.5-2$ could be required, translating into $E($B$-$V$)\sim0.12-0.2$. Finally, the [O\3]$_{\rm 88/opt}$ ratio toward the northern region is also close to the lowest density tracks, requiring only little dust extinction ($A_V\sim0.5$). 

We thus get the picture of a low-density nebula emission that is hardly attenuated by dust extinction, except toward the stellar cluster LH\,10 and possibly W2. Is the required extinction compatible with the amount of dust available? We now determine the maximum dust extinction across the region. The maximum extinction is given by the total amount of dust available. We built the extinction map from the modeling of the spatial distribution of the dust mass derived by Hony et al.\ (in preparation). The grain emission is constrained by the \textit{Spitzer} and \textit{Herschel} data of N\,11. The grain composition is the ``Amorphous carbon model" of Galliano et al. (2011), since it is the model that gives gas-to-dust mass ratios that are consistent with the elemental abundances in the LMC. In addition, it provides a conservative estimate, since it gives a lower dust mass, hence a lower $A_V$. The dust mass derived from SED modeling is turned into an $A_V$ by using
\begin{equation}
  A_V = \kappa_{\mbox{ext}}(V)\times M_{\mbox{dust}} / A_{\mbox{pix}} \times 1.086,
  \label{eq:Av}
\end{equation}
where $A_{\mbox{pix}}=124\;\rm pc^2$ is the surface of the pixel used for the modeling, and $\kappa_{\mbox{ext}}(V)=3175\;\rm m^2/kg$ is the $V$ band opacity of the grain mixture (Fig.\,A.1 of Galliano et al.\ 2011). Equation\,(\ref{eq:Av}) corresponds to the maximum extinction since it corresponds to the $A_V$ of the entire LMC toward the line of sight. The $A_V^{\rm max}$ calculated from the dust modeling ranges from $\sim1$ to $\sim16$ in the area observed by the PACS FIR lines, with a median value of $3$. $A_V^{\rm max}$ is $\sim1.4$ toward LH\,10, $\sim5-16$ toward W2, and $\sim3-4$ toward the northern region. We call here that the actual extinction $A_V$ corresponds to the dust fraction between and the relevant emission source (here the emission-lines from the ionized gas). Without a precise knowledge of the morphology of the region, this fraction lies anywhere from $0$ to $1$. We nevertheless conclude that there is enough dust to explain an extinction of $A_V\sim1$ toward LH\,10 and W2. 

Correcting for dust extinction with $A_V\sim1$ toward LH\,10 and W2, we find that [O\3]$_{\rm opt}$/H$\alpha$ toward these two regions is in reality significantly greater than toward the northern region (Fig.\,\ref{fig:lineratios}d). This is simply explained by the larger distance between the stellar cluster and the northern region, which reduces the ionization parameter.  

\section{Spatial extent of the [O\3] emission}\label{sec:spatialOIII}

We observe that the [O\3]$_{88}$ emission is spread throughout the N\,11 nebula. 
Is the [O\3]$_{88}$ emission compatible with the number of known massive stars and their spatial distribution? The ISM density structure plays a dominant role in controlling the mean free path of FUV photons. The presence of [O\3]$_{88}$ emission significantly far away from any massive star is a strong indication that photons are able to travel through a low-density material and/or that the ISM is significantly porous. 
We now attempt to model the [O\3]$_{88}$ emission around each massive star in order to constrain an upper limit on the density above which FUV photons are not able to travel to large distances. 

We consider the O stars identified by Parker et al.\ (1992) (Table\,\ref{tab:oiii_radius}) since cooler stars are not able to ionize significant amounts of oxygen into O$^{++}$ (Sect.\,\ref{sec:ionizing_sources}). It must be noted that unidentified embedded massive stars can contribute to the [O\3]$_{88}$ emission. Furthermore, for several stars, the identification could be uncertain because of multiple stars combined in the same spectrum. While the spectral type derived from the composite spectrum can be correct, the number of sources (and thus the number of ionizing photons) might be different. The number of O stars given in Table\,\ref{tab:oiii_radius} should be considered as a lower limit. 

\begin{table}
\caption{O stars with known spectral types.
\label{tab:oiii_radius}}
\begin{tabular}{lllll}
\hline\hline
Name\tablefootmark{a} & Type\tablefootmark{a} & $r$\tablefootmark{b}  &  $f$\tablefootmark{c} & $F$([O\3]$_{88}$) \\
 &  &  (pc) &  &($\times10^{-14}$\ W\,m$^{-2}$)  \\
\hline
\multicolumn{2}{l}{Eastern half\tablefootmark{d}} & $22.6\ (20.0)$ & $0.5$& $24.4$ \\
3173 (N11-080) & O7\,V &$7.2\ (6.0)$   &    $1.0$ & $1.2$ \\
3224\tablefootmark{e} (N11-050) & O4\,V &$13.4\ (11.2)$   &     $0.7$ & $4.7$ \\
3204 (N11-048) & O6.5\,V &$11.0\ (9.0)$   &    $1.$ & $3.5$ \\
3168 (N11-032) & O7\,II & $12.2\ (11.3)$   &     $1.$ & $6.2$ \\
3223 (N11-013) & O8\,V & $5.8\ (4.1)$   &     $1.$ & $0.2$ \\
3209\tablefootmark{e} & O3\,III & $18.8\ (16.2)$   &   $0.7$ & $16.6$ \\
\hline
\multicolumn{2}{l}{Western half\tablefootmark{d}} & $27.8\ (24.5)$ &  $0.4$ & $29.3$ \\
3061 (N11-031) & ON2\,III & $18.8\ (16.3)$   &    $0.3$& $6.5$  \\
3058 (N11-060) & O3\,V & $17.4\ (15.0)$   &    $0.3$ & $6.6$  \\
3053 (N11-018) & O6\,II & $17.7\ (15.0)$   &    $0.3$ & $6.2$  \\
3042 (N11-087) & O9.5\,V &$3.2\ (2.2)$   &     $0.$  & $0.$ \\
3100 (N11-038) & O5\,III & $10.2\ (8.2)$   &   $0.5$  & $1.4$ \\
3070 & O6\,V & $11.0\ (9.0)$   &     $0.3$ & $1.3$ \\
3126 & O6.5\,V & $10.2\ (8.2)$   &    $0.7$  & $2.0$ \\
3120\tablefootmark{e} & O5.5\,V & $12.0\ (10.0)$   &   $0.7$ & $3.8$ \\
3073 & O6.5\,V & $10.2\ (8.2)$   &   $0.4$ & $1.2$  \\
3089 & O8\,V & $6.8\ (4.9)$   &    $0.5$ & $0.2$  \\
3102 & O7\,V &$9.1\ (7.0)$   &      $0.6$  & $1.0$ \\
3103 & O9.5\,IV &$3.4\ (2.3)$   &   $0.9$  & $0.07$ \\
3115 & O9\,V & $4.7\ (3.3)$   &   $0.9$  & $0.2$ \\
3123 & O8.5\,V & $5.8\ (4.1)$   &   $0.9$  & $0.2$ \\
 \hline
\end{tabular}\\
\tablefoottext{a}{Spectral type and name are from Parker et al.\ (1992). The label in parentheses refers to Evans et al.\ (2006). When available, the spectral type was updated using Evans et al.\ (2006).}
\tablefoottext{b}{Radius  of the [O\3]$_{88}$ emitting sphere containing $99$\%\ of the total [O\3]$_{88}$, assuming an homogenous density of $16$\cc. The value between parentheses give the radius of the sphere containing $75$\%\ of the total [O\3].}
\tablefoottext{c}{Fraction of the [O\3]$_{88}$ emission falling into the PACS map (assuming a density of $16$\cc).}
\tablefoottext{d}{The model for each half of N\,11B considers all stars located in the same location, it is not a combination of the models from individual stars.}
\tablefoottext{e}{Possible composite spectrum in Parker et al.\ (1992).}
\end{table}

\begin{figure*}
\centering
\includegraphics[angle=0,scale=0.8,clip=true]{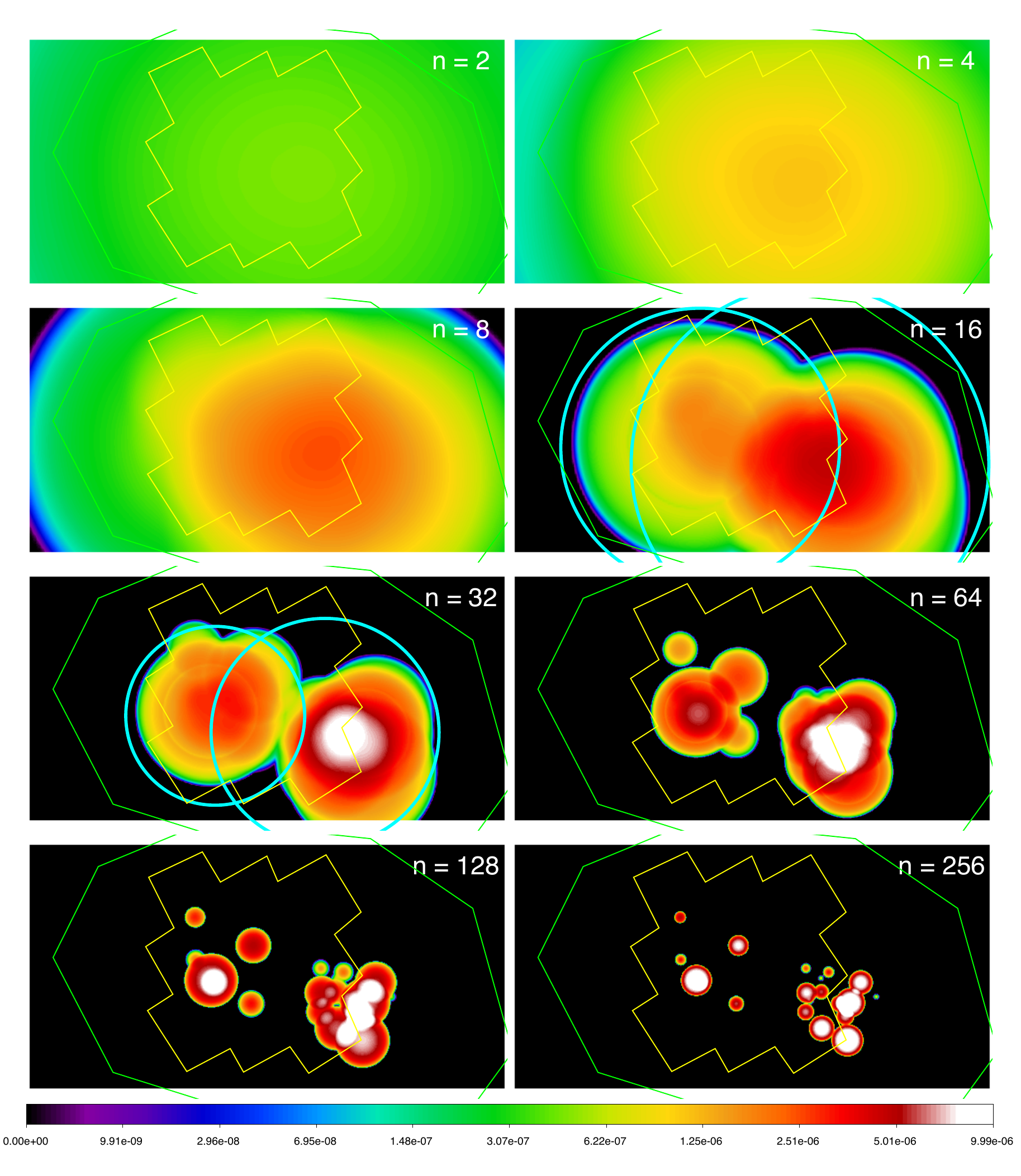}
\caption{[O\3]$_{88}$ distribution modeled around each star with an homogeneous density ranging from $2$\cc\ to $256$\cc. The yellow polygon indicates the PACS coverage and the green polygon indicates the approximate extent of the optical body. The cyan circles in the $n=16$ and $32$\cc\ panels represent the extent of the [O\3]-emitting zone when considering a cluster instead of the individual stars (see text). 
\label{fig:acs_spheres2}}
\end{figure*}

We assume in the following that stars excite their own surrounding nebula (i.e., photoionization is radiation-bounded) and that the density is homogenous. The stellar properties (number of ionizing photons, temperature) are taken from Sternberg et al.\ (2003). Figure\,\ref{fig:acs_spheres2} shows the [O\3]$_{88}$ emitting spheres around each O star in N\,11B. 
The emissivity from the Cloudy models around each star was computed assuming a uniform spherical geometry and projected on the sky at the distance of N\,11B. The final maps (Fig.\,\ref{fig:acs_spheres2}) were created by coadding the individual models. 

It can be seen that models with a density of $\gtrsim100$\cc\ predict a small ionized volume as compared to the physical scale of N\,11B. As the density decreases, FUV photons are able to travel farther away and ionize the ISM several tens of parsecs from the star. A density around $8-16$\cc\ seems to reproduce the physical extent of N\,11B well. The [O\3]$_{88}$ extent for each O star is given in Table\,\ref{tab:oiii_radius} for a density of  $16$\cc. The largest diameter at a density of $16$\cc\  reaches $\approx40$\,pc, for PGMW\,3061, 3058, 3053, and 3209, which is close to the physical size of N\,11B ($\approx80\times60$\,pc from the H$\alpha$ image). We conclude that it is possible for the O stars to excite [O\3]$_{88}$ throughout the PACS map, provided some low-density channels exist that are as low as $\lesssim16$\cc. This value is significantly less than the densities from the optical doublet diagnostics ($\sim100$\cc, Sect.\,\ref{sec:density}), and it suggests that the optical line measurements are biased toward the brightest and possibly densest regions. 

Although the absorption of UV photons is mostly due to the gas rather than to the dust, dust does play an important role in determining the mean free path of UV photons. It is therefore interesting to consider a cluster of stars ionizing the ISM around it rather than a collection of separate individual stars. Models with stars gathered in clusters remove the assumption that each star excites their own surrounding nebula. We therefore now consider two main regions powered by their own stellar cluster. The western half of N\,11B is dominated by the LH\,10 southern cluster and the eastern half by the ``footprint" cluster (Fig.\,\ref{fig:acs_spheres2}). 
The nebular density best agreeing with the physical size of N\,11B is around $\lesssim16$\cc. The radius corresponding to each cluster is reported in Table\,\ref{tab:oiii_radius} and shown in Fig.\,\ref{fig:acs_spheres2}. 
The volume ionized by the clusters is slightly smaller than the sum of the individual volumes around each star for a given region (by $17$\%\ and $9$\%\ lower for the western and eastern halves, respectively). Although the exact location of the cluster centroid is quite uncertain since stars are in reality not co-spatial, the overall [O\3] extent of the two clusters encompasses the N\,11B nebula relatively well. 

We conclude that massive stars in N\,11B are able to produce [O\3]$_{88}$ emission at large distances in the global low-density medium of N\,11B. Although higher density material does exist, the low-density regions provide channels for the FUV photons to permeate into the ISM. For instance, the northern region shows significant [O\3]$_{88}$ emission, while there are no ionizing sources, implying that FUV photons were able to cross most of the N\,11B nebula. The global density of $\lesssim16$\cc\ that we derive is an upper limit, since projection effects prevent us from asserting the depth of the nebula in detail so that distances are in fact lower limits to the actual distances. Slightly larger distances would then require slightly lower densities in order to reach further. Another caveat should be added that we assumed a gas filling-factor of $1$. If the gas were to fill only a fraction of the nebula, UV photons would travel even farther for a given density.

\section{Origin of the [C\2] emission}\label{sec:cii_origin}

\begin{figure}
\includegraphics[angle=0,scale=0.56,clip=true]{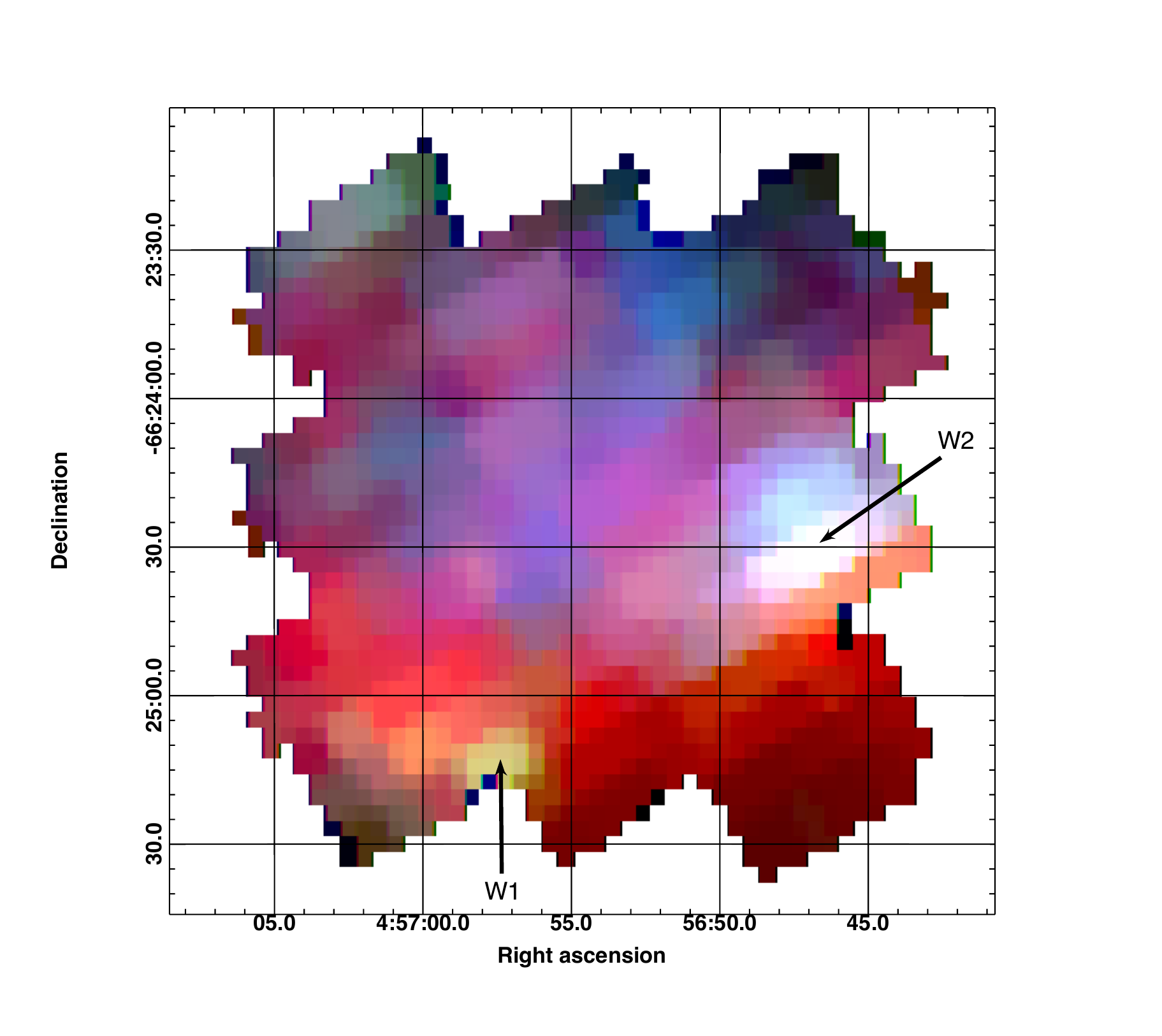}
\caption{Three-color image of the following PACS lines, [C\2] (blue), [O\1]$_{63}$ (green), and [O\3]$_{88}$ (red). 
\label{fig:cii_oi_oiii}}
\end{figure}

Because of its low excitation temperature, [C\2] can potentially emit from several ISM phases (see introduction). The [C\2] emission peaks toward W2, but it is dominated by extended emission (Sect.\,\ref{sec:cii}). The [C\2] distribution is characterized by a sharp decrease in flux toward the stellar cluster LH\,10, contrasting with the flat [O\3]$_{88}$ emission (Fig.\,\ref{fig:pacs_maps}). 
In Fig.\,\ref{fig:cii_oi_oiii}, it can be seen that the spatial distributions of [O\3]$_{88}$ and [C\2] are anti-correlated to first order, except toward W2 where all the FIR tracers are concentrated. The [O\3]$_{88}$ emission thus appears to surround lower ionization regions, which might be associated to photodissociation regions. In the following we now investigate the possible contribution from the ionized gas and from PDRs to the [C\2] emission. We use [N\2]$_{122}$ and [O\3]$_{88}$ to trace the low-density ionized gas, and [O\1]$_{63}$ and PAH emission to trace PDRs.

\subsection{Ionized gas component}\label{sec:cii_ig}

\begin{figure}
\includegraphics[angle=0,scale=0.65,clip=true]{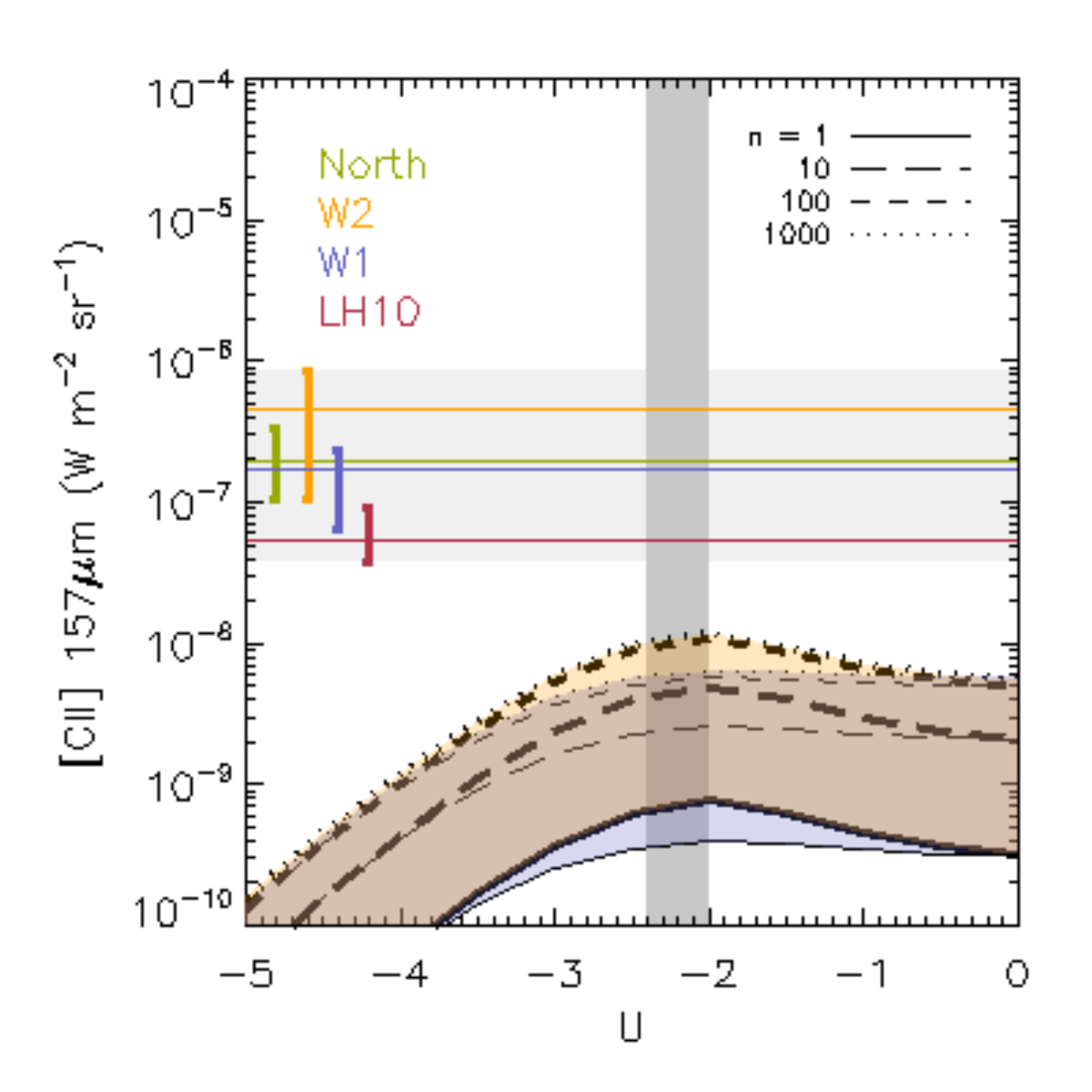}\\
\includegraphics[angle=0,scale=0.65,clip=true]{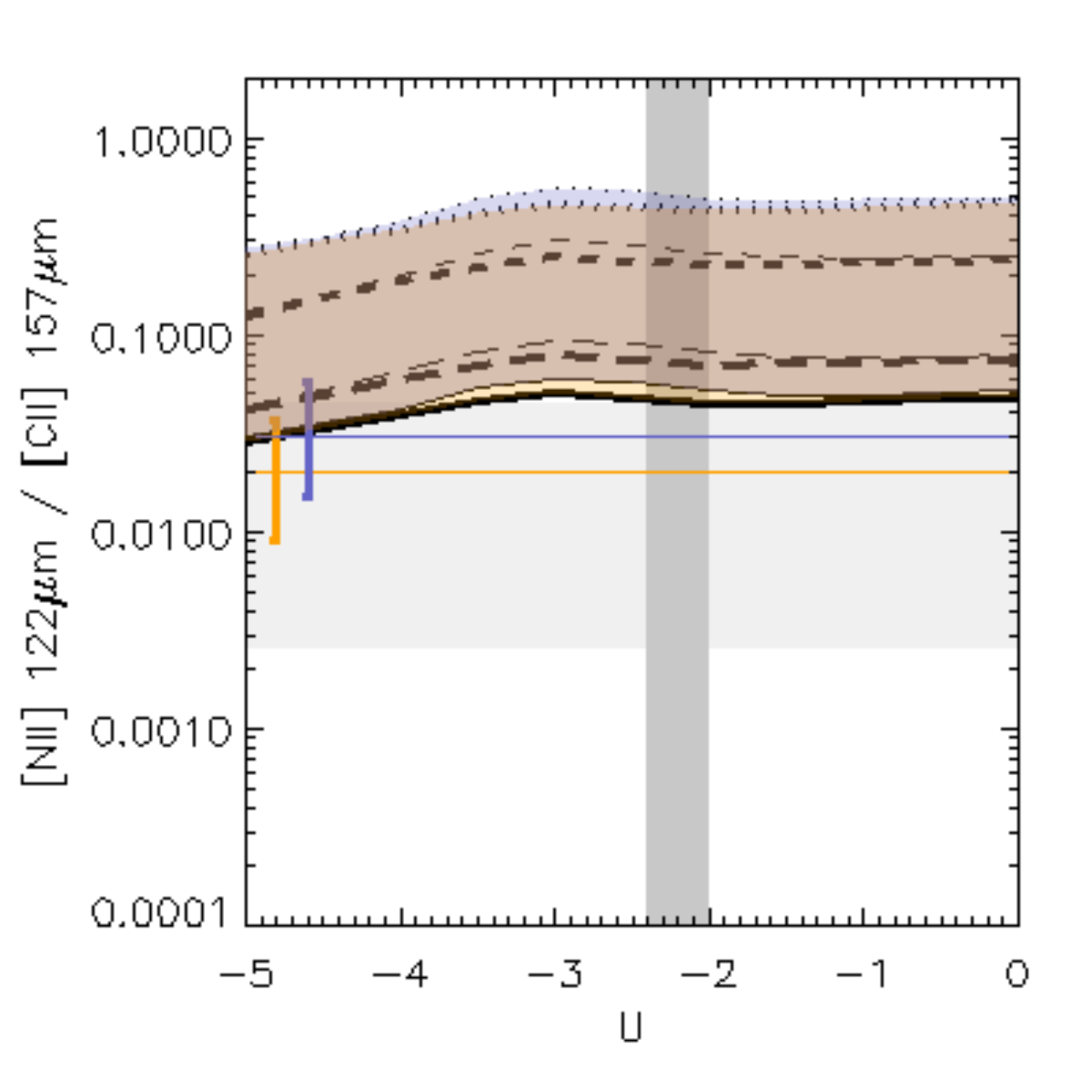}
\caption{The integrated emission of [C\2] (\textit{top}) and the [N\2]$_{122}$/[C\2] ratio (\textit{bottom}) are plotted against the ionization parameter $U$. Models show the expected values in the \emph{ionized gas} of the H\2\ region.
See Fig.\,\ref{fig:lineratios} for the plot description. 
\label{fig:cii}}
\end{figure}

Most of the ionized gas in N\,11B has a density around or lower than the critical density of [C\2] ($80$\cc\ for collisions with $e^-$; Table\,\ref{tab:lines}), so we expect most of the C$^+$ in the ionized phase to de-excite radiatively in the ionized gas. Figure\,\ref{fig:cii} shows that the [C\2] emission is systematically underestimated by the models. 
We wish to call here that our models correspond to the photoionized gas in the H\2\ region and that calculations were stopped at the ionization front (see Sect.\,\ref{sec:igmodels}). 
An important warning is that [C\2] in the ionized gas is expected to emit close to the ionization front, so that our results strongly depend on the geometry and on the radiation-bounded nature of the nebula. While the compact regions W1 and W2 appear to be radiation-bounded (Sect.\,\ref{sec:igmodels}), the situation is uncertain for the more diffuse and extended regions. Therefore, the model predictions for [C\2] emission should be considered as upper limits except toward W1 and W2.  
With these caveats in mind, we assume a density between $10$\cc\ and $100$\cc, and from Fig.\,\ref{fig:cii} we estimate that the fraction of [C\2] originating in the ionized gas is $\sim15$\%\ toward LH\,10 and $\sim4$\%\ toward the northern region, while it is $\lesssim4$\%\ toward W1 and $\lesssim2$\%\ toward W2.

The [N\2]$_{122}$ line is a good tracer of the diffuse ionized gas from which [C\2] could also arise (e.g.; Oberst et al.\ 2006). [N\2]$_{122}$ and [C\2] involve similar energies (ions existing for energies $>14.5$\,eV and $>11.3$\,eV respectively), but they have significantly different critical densities in the ionized gas ($\approx400$\cc\ and $\approx80$\cc, respectively, for collisions with $e^-$). Even for densities as low as $\sim50$\cc, therefore, the de-excitation of [C\2] in the ionized gas is affected by collisions, and the [N\2]$_{122}$/[C\2] ratio thus increases with density until the critical density of [N\2]$_{122}$ is reached. [N\2]$_{122}$ was detected only toward the compact regions W1 and W2 with pointed observations (Sect.\,\ref{sec:distrib}). The PACS footprint size unfortunately barely samples the extended emission around these two regions (Fig.\,\ref{fig:pacs_maps}). Figure\,\ref{fig:cii} shows that [N\2]$_{122}$/[C\2] toward W1 and W2 is significantly lower than the models with $n_e\sim100$\,\cc, confirming that most of [C\2] arises in PDRs toward these regions. The ISO observation (Sect.\,\ref{sec:observations_ancillary}) shows that [N\2]$_{122}$/[C\2]$\lesssim0.03$ in the $80"$ beam, which includes W2. It thus seems that our results remains unchanged on a larger scale; i.e., [N\2]$_{122}$/[C\2] is lower than expected from the ionized gas, and most of [C\2] toward W1 and W2 originates in PDRs. 

\subsection{PDR component}\label{sec:cii_pdr}

Our results suggest that most of the [C\2] emission in N\,11B does not originate in the ionized gas (Sect.\,\ref{sec:cii_ig}). However, the uncertainty on the ionization structure and the low S/N ratio of the ionized gas tracer [N\2]$_{122}$ observation prevents a conclusive result for the most diffuse regions (i.e., other than the regions W1 and W2). We now investigate the relation between gas heating tracers (e.g., TIR and PAH emission) and gas cooling tracers (e.g., [C\2] and [O\1]$_{63}$) to understand the origin of [C\2] and also to constrain the dominant gas heating source in PDRs.

\subsubsection{Gas heating}\label{sec:heatingtracers}

The gas heating in PDRs and in neutral atomic clouds is dominated by the photoelectric effect on dust grains (e.g., Hollenbach \&\ Tielens 1999; Weingartner \&\ Draine 2001), in which a photoelectron is ejected from the grain following the absorption of a FUV photon. The power absorbed by dust that goes into ISM heating, referred to as the photoelectric efficiency $\epsilon_{\rm PE}$, is $\sim1$\%, and is the largest for small grains and neutrally charged grains (e.g., Tielens \& Hollenbach 1985a). The gas heating is expected to originate about equally in PAHs (and PAH clusters) and in very small grains with radii $\lesssim100$\AA\ (Bakes \&\ Tielens 1994).

\begin{figure}
\centering
\includegraphics[angle=0,scale=0.48,clip=true]{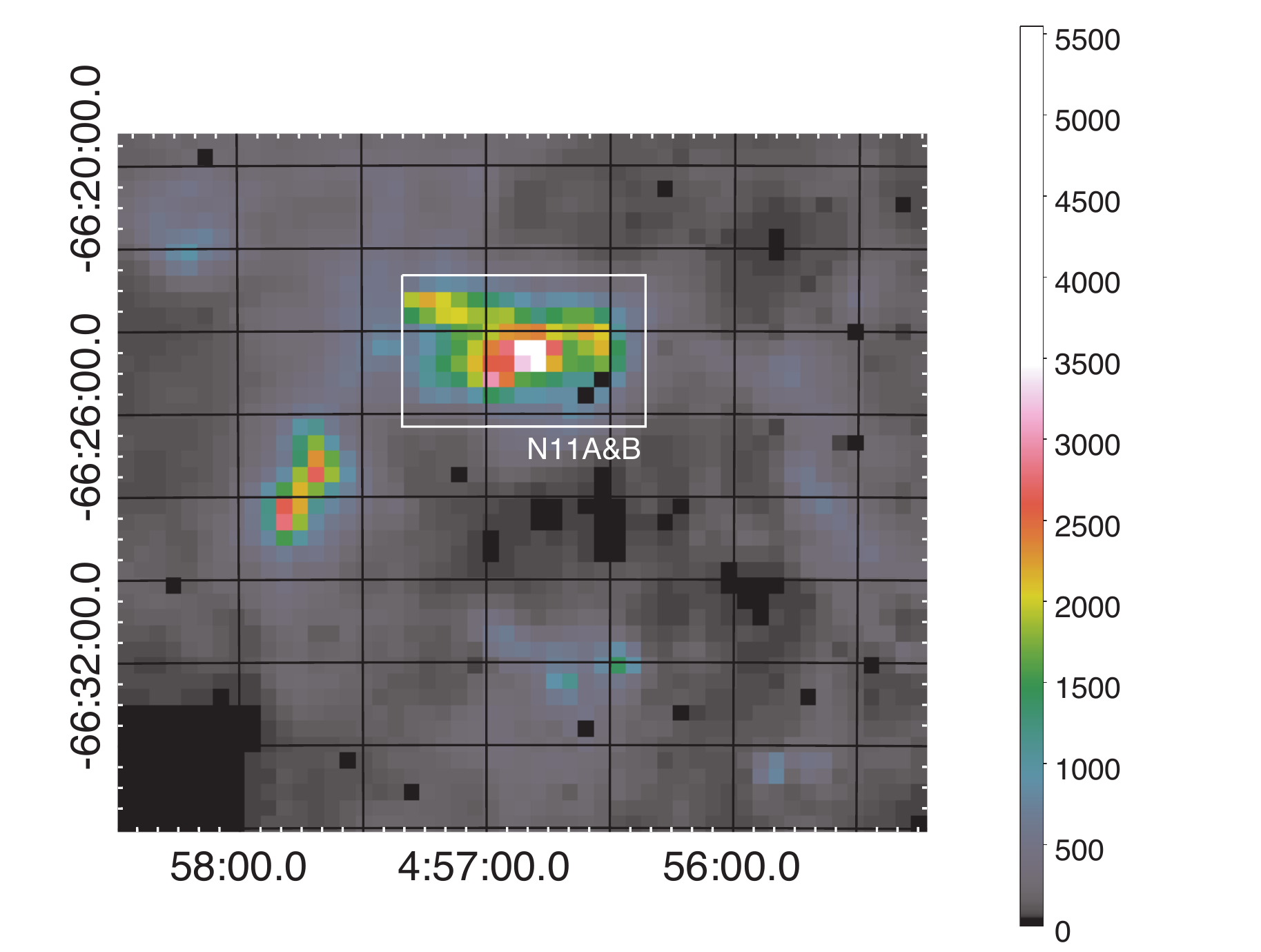}
\includegraphics[angle=0,scale=0.58,clip=true]{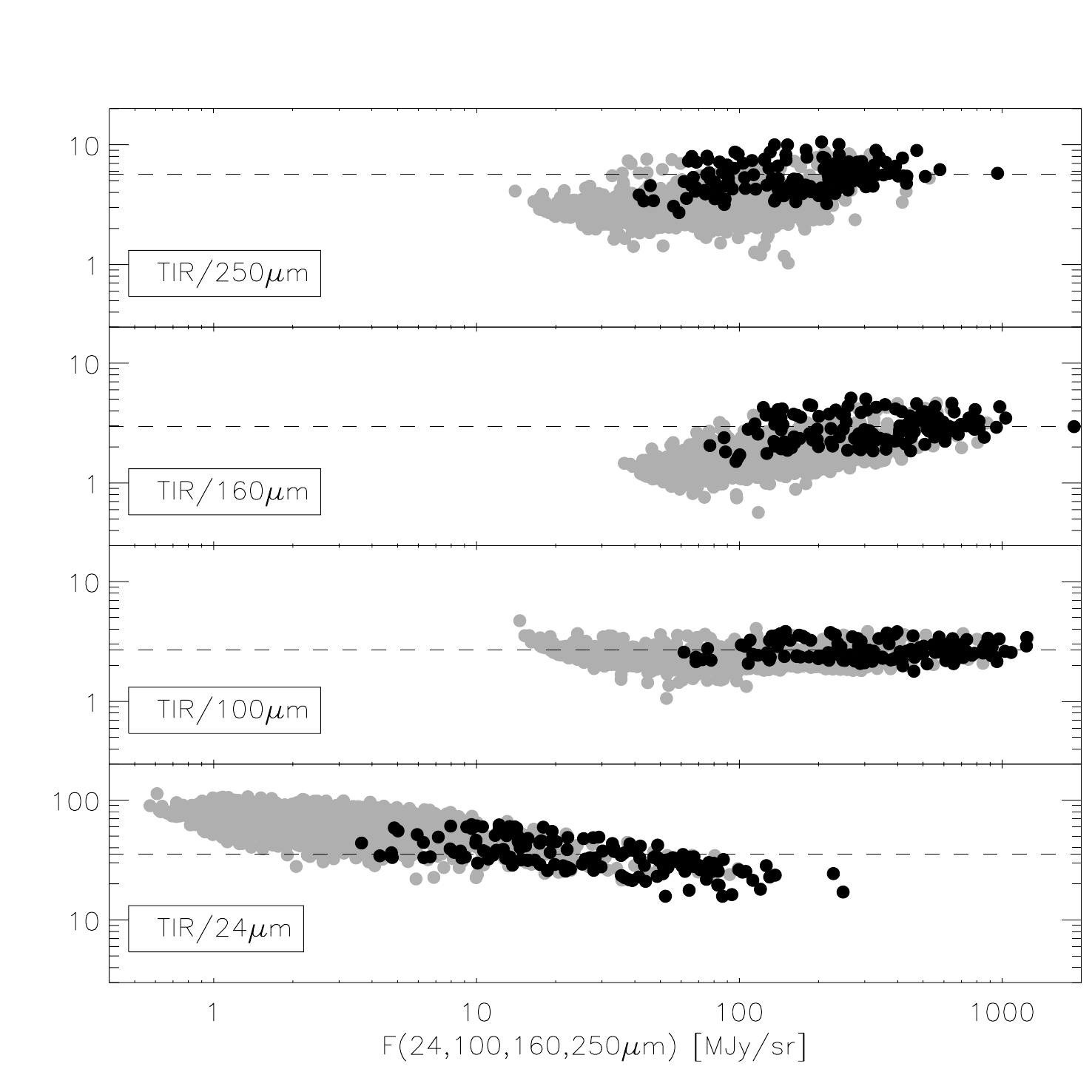}
\includegraphics[angle=0,scale=0.5,clip=true]{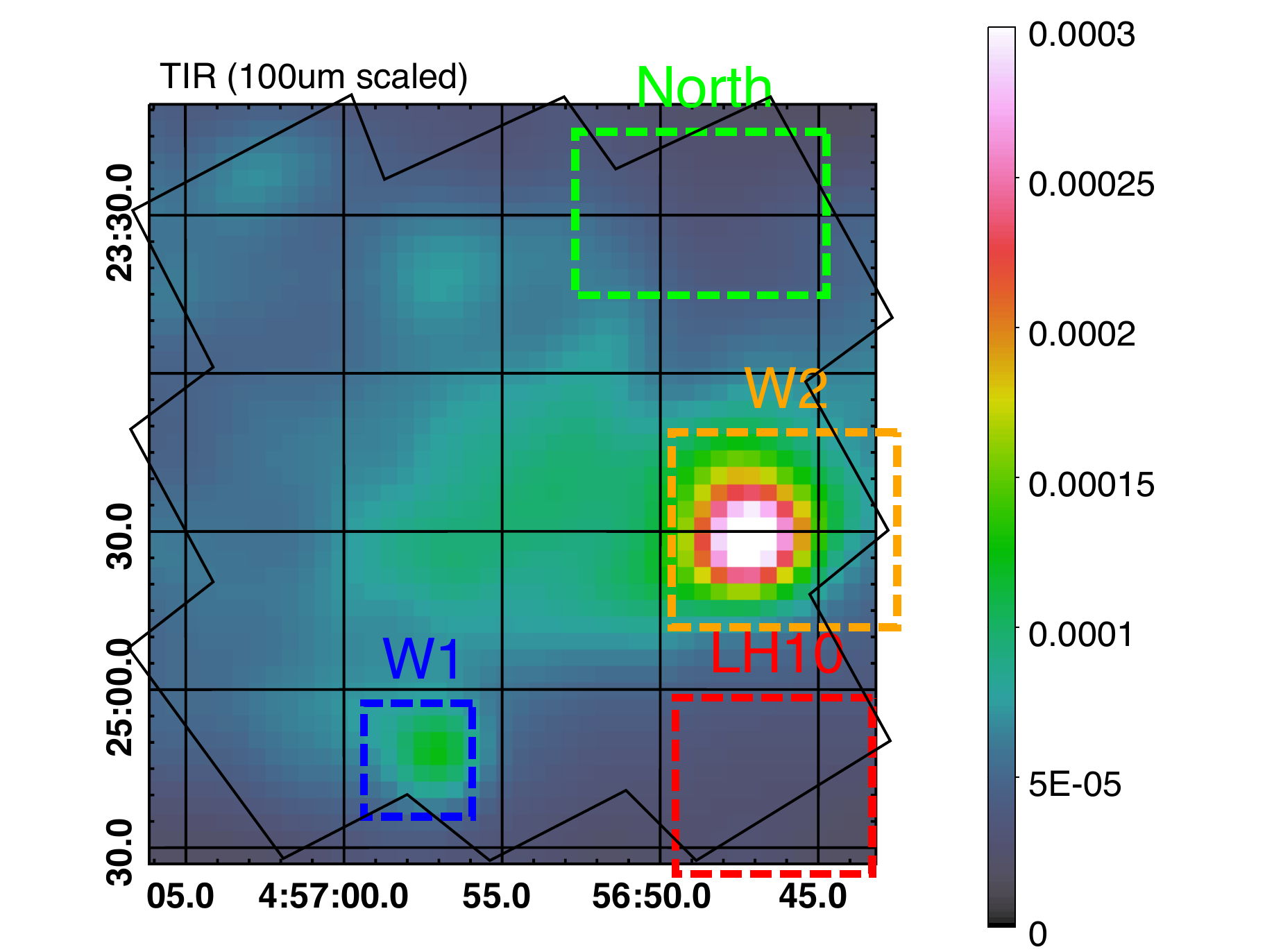}
\caption{\textit{Top} $-$ TIR map of N\,11 with a resolution of $23\arcsec$ (see Sect.\,\ref{sec:oiii}). \textit{Middle} $-$ Correlations between TIR and individual dust continuum bands. Black points correspond to positions in N\,11A and N\,11B and gray points to the rest of the nebula (see top panel). The dashed line shows the average value. \textit{Bottom}$-$ TIR map using the $100$\mic\ dust continuum as a tracer. See Fig.\,\ref{fig:pacs_maps} for the figure description. \label{fig:tir_map}}
\end{figure}

Although TIR emission mostly traces classical large grains in thermal equilibrium with the interstellar radiation field, the TIR emission is a good proxy for the gas heating rate in PDRs assuming a given photoelectric efficiency. The [C\2]/TIR ratio has often been used to approximate $\epsilon_{\rm PE}$, under the assumptions that TIR traces the gas heating rate and that [C\2] dominates the gas cooling (e.g., Rubin et al.\ 2009 and Israel et al.\ 2012 in the LMC). The TIR map we used to estimate the dust extinction in Sect.\,\ref{sec:oiii} is too coarse to compare to the FIR line maps (spatial resolution of $23\arcsec$ as compared to $\approx11.5\arcsec$; Fig.\,\ref{fig:pacs_maps}). We therefore examined the correlation between TIR and single dust continuum bands in order to derive a higher resolution proxy for TIR. The following bands were considered: MIPS $24$\mic, PACS $100$\mic\ and $160$\mic, and SPIRE $250$\mic, with resolutions of $6\arcsec$, $7\arcsec$, $12\arcsec$, and $17\arcsec$ respectively. The continuum images were convolved to $23\arcsec$ resolution and resampled to match the TIR image. The correlation between TIR and each continuum band is shown in Fig.\,\ref{fig:tir_map}. We find that TIR correlates best with the $100$\mic\ continuum on the scale of the entire  N\,11 nebula and most importantly within the star-forming regions N\,11A and N\,11B (Fig.\,\ref{fig:tir_map}). The scatter is also the smallest for TIR/$100$\mic. The linear regression fit shows that the flux density ratio is ${\rm TIR/100{\mu}m} = 2.70\pm0.27$. In the following, we use the $100$\mic\ dust continuum image to trace TIR to a spatial resolution compatible with the [C\2] and [O\1]$_{63}$ mapping observations shown in Fig.\,\ref{fig:tir_map}.

PAH molecules and PAH clusters are expected to be important contributors to gas heating. Rubin et al.\ (2009) find a good correlation between [C\2] and the Spitzer/IRAC $8.0$\mic\ band throughout the LMC and proposes that PAHs dominate the photoelectric heating rate in PDRs. In the following, we revisit these findings by using the PAH map presented in Sect.\,\ref{sec:observations_ancillary} and by examining both [C\2] and [O\1]$_{63}$ as potential coolants.

\subsubsection{Relation between dust heating and gas cooling}\label{sec:heatingcooling}

\begin{figure*}
\centering
\includegraphics[angle=0,scale=0.45,clip=true]{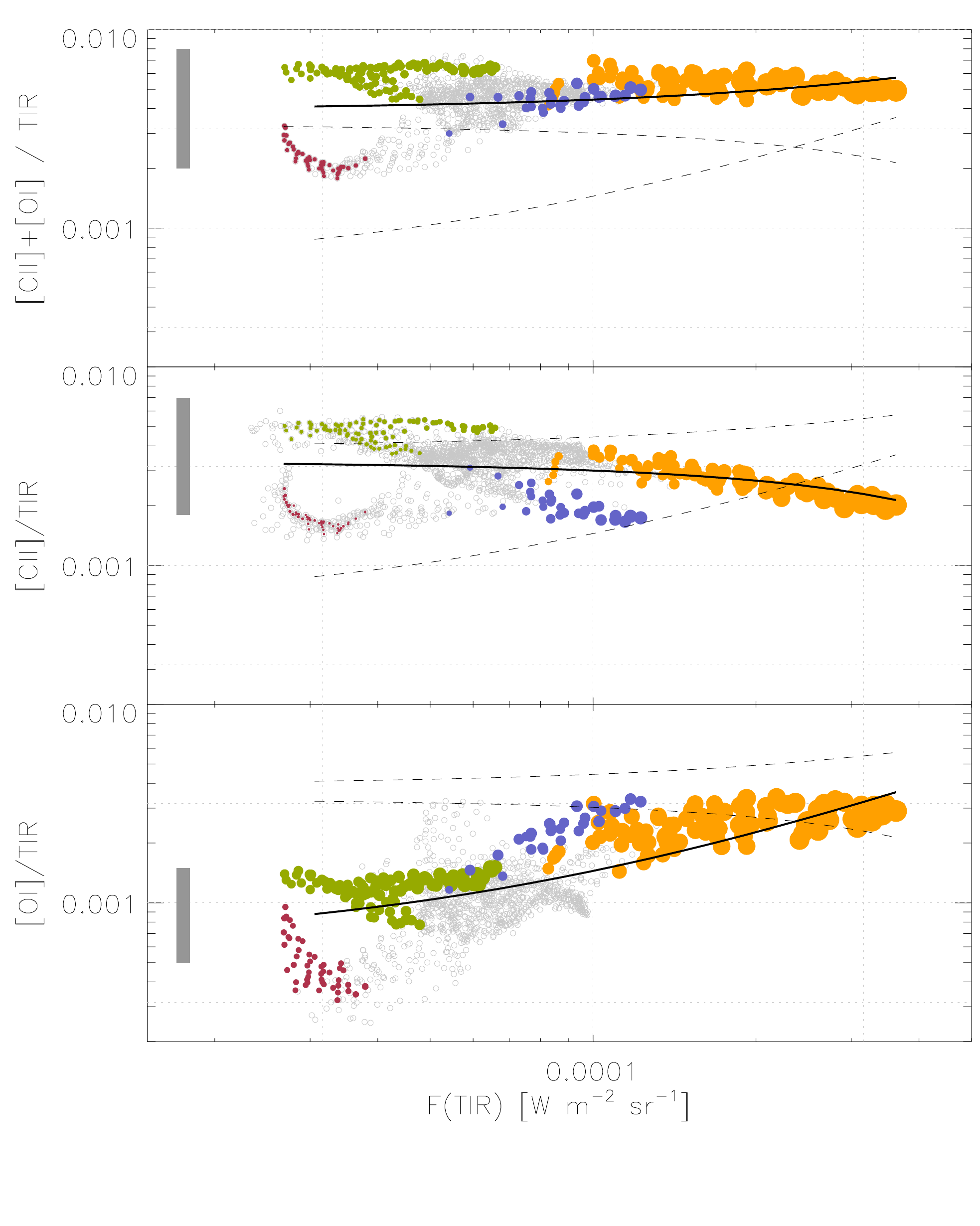}
\includegraphics[angle=0,scale=0.45,clip=true]{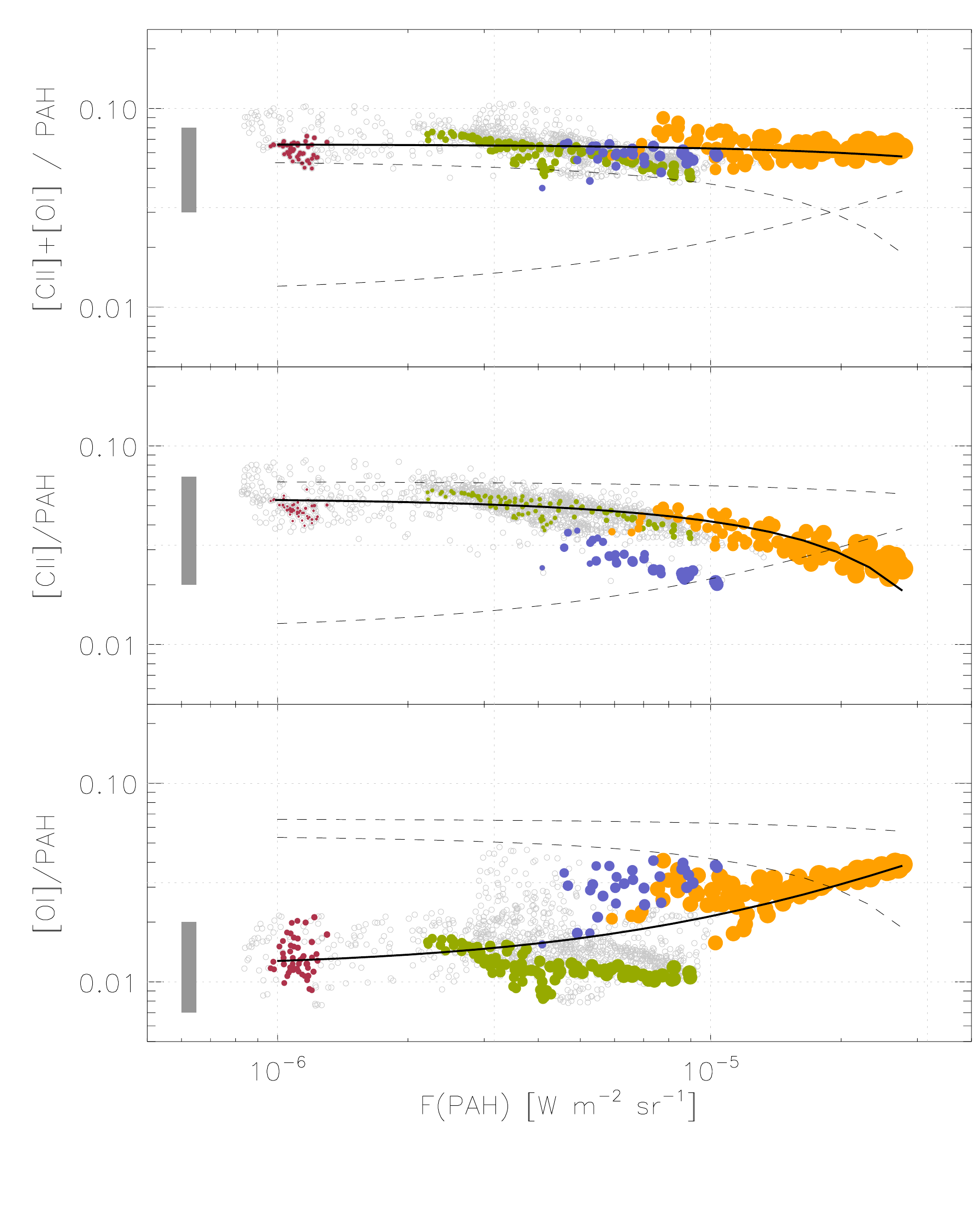}
\caption{
Pixel-to-pixel correlation between the PACS cooling lines ([C\2], [O\1]$_{63}$, and their sum) and 2 quantities used to trace the gas heating, i.e., the TIR emission (\textit{left panel}) and the PAH emission (\textit{right panel}).
The solid curve represents the linear regression for a given plot and the dashed curves the linear regression of the other ratios for comparison. The colors correspond to specific regions illustrated in Fig.\,\ref{fig:tir_map}. The symbol size is proportional to [C\2]+[O\1]$_{63}$ (\textit{top}), [C\2] (\textit{middle}), and [O\1]$_{63}$ (\textit{bottom}). The gray rectangles on the left side show the values observed by Croxall et al.\ (2012) in 2 super-solar metallicity nearby galaxies (see text). 
\label{fig:ciipah}}
\end{figure*}

Figure\,\ref{fig:ciipah} shows the correlation between gas cooling tracers ([C\2], [O\1]$_{63}$, and their sum) and dust heating tracers (TIR, PAH). We use here the 100\mic\ dust continuum as a proxy for TIR (Sect.\,\ref{sec:heatingtracers}).
The ratio of cooling lines to dust heating tracers provides a proxy for the fraction of the power absorbed by dust that is transferred into gas heating. The [O\1]$_{63}$/(TIR,PAH) ratio shows a large scatter (factor of $\approx15$ for [O\1]$_{63}$/TIR, $\approx5$ for [O\1]$_{63}$/PAH), with the lowest values toward the stellar cluster LH\,10 and the northern region (Fig.\,\ref{fig:ciipah}), while the highest values are found in W1 and W2. The [C\2]/(TIR,PAH) ratio is significantly tighter than [O\1]$_{63}$/(TIR,PAH), with a factor of $\approx6$ for [C\2]/TIR and $\approx4$ for [C\2]/PAH. 
Finally, the use of the sum [C\2]+[O\1]$_{63}$ to trace the gas cooling provides the tightest relation with both TIR and PAH emission, with a factor of $\approx4$ for ([C\2]+[O\1]$_{63}$)/TIR and $\approx2$ for ([C\2]+[O\1]$_{63}$)/PAH. We note that the relatively large scatter of ([C\2]+[O\1]$_{63}$)/TIR is caused by the LH\,10 region and its surroundings (see also Sect.\,\ref{sec:lh10}). When ignoring LH\,10, the scatter of ([C\2]+[O\1]$_{63}$)/TIR is similar to that of ([C\2]+[O\1]$_{63}$)/PAH. 

The [C\2]/(TIR,PAH) and [O\1]$_{63}$/(TIR,PAH) ratios are anti-correlated in most regions. The ratios are also anti-correlated with respect to the TIR and PAH emission, with [C\2]/(TIR,PAH) decreasing with TIR and PAH emission, while the inverse is observed for [O\1]$_{63}$/(TIR,PAH). The anti-correlation between [C\2]/(TIR,PAH) and [O\1]$_{63}$/(TIR,PAH) is due to different PDR conditions, with the cooling being dominated either by [C\2] in the most diffuse regions or by [O\1]$_{63}$ in the densest regions and/or the regions with intense UV field (e.g., Kaufman et al.\ 2006). This result is compatible with the fact that [O\1]$_{63}$/(TIR,PAH) is relatively brighter toward W1 and W2, which are relatively dense regions located near massive stars (Sects.\,\ref{sec:morphology}, \,\ref{sec:density}). The cooling via [O\1]$_{63}$ represents $\sim25$\%\ of the total cooling rate across N\,11B except toward W1 and W2 where it reaches $\sim50$\%. On large scales, we therefore expect [C\2] to be a fairly good proxy for the total cooling, which confirms the estimates made by Rubin et al.\ (2009). 
We keep in mind that the [O\1]$_{63}$ line could be affected by optical depth effects, more so than [O\1]$_{145}$ and [C\2] (e.g., Abel et al.\ 2007). However, only low $A_V$ values are found in N\,11B ($A_V\lesssim1$; Sect.\,\ref{sec:oiii}), suggesting that optical depths effects are negligible at the spatial resolution probed by PACS ($\approx9.5-13"$). 
[C\2] and [O\1]$_{63}$ thus compensate for each other and, assuming these two lines are the dominant coolants at low $A_V$ (e.g., Tielens et al.\ 1985a; Hollenbach \& Tielens 1999), the sum [C\2]+[O\1]$_{63}$ provides a proxy for the total gas cooling.

\subsubsection{PDR conditions}\label{sec:pdrcontitions}

The flat correlation between [C\2]+[O\1]$_{63}$ and PAH (or, although to a lesser extent, TIR), together with the fact that W1 and W2 are dominated by PDRs (Sect.\,\ref{sec:cii_ig}), suggests that [C\2] and [O\1]$_{63}$ originate in PDRs throughout N\,11B. The scatter of ([C\2]+[O\1]$_{63}$)/PAH is a factor of $\approx2$, implying that the fraction of [C\2] originating in PDRs is at least $\sim50$\%\ throughout N\,11B. 
The spatial anticorrelation between [C\2] and the warm photoionized gas probed by [O\3]$_{88}$ (Fig.\,\ref{fig:cii_oi_oiii}) suggests that PDRs are mainly located within the ionized gas nebula (i.e., they are not located at the edge of the N\,11B region).

The small scatter of ([C\2]+[O\1]$_{63}$)/PAH further implies that the photoelectric efficiency in PDRs is uniform within a factor of $\approx2$. This implies that the grain charging parameter, related to the UV field intensity and to the electron density (e.g., Tielens \&\ Hollenbach 1985a), is also uniform across the region. 
Rubin et al.\ (2009) find a bias toward lower [C\2]/TIR ratios in star-forming regions of the LMC, which they ascribes to relatively low photoelectric efficiency. The [C\2]/TIR in the entire N\,11 nebula found by the authors, $\approx0.56$\%, agrees with the highest values we measure in N\,11B. It is therefore likely that the large beam used by Rubin et al.\ (2009), $14.9$\arcmin, was filled by the most diffuse regions where the cooling is dominated by [C\2]. Our results also imply that [C\2]/TIR alone (or [C\2]/PAH) does not trace the photoelectric efficiency well on small enough scales.

We do not find any significant evidence of a lower photoelectric efficiency in the regions with the greatest expected UV field intensity, such as W2 or LH\,10, as compared to more quiescent regions, such as the northern region. We note, however, that the ionization parameter in most regions, including W2, indicates a dilution of the UV field on large physical scales (Sect.\,\ref{sec:igmodels}). It is therefore possible that variations in the UV field intensity across N\,11B are not large enough to modify significantly the grain charging parameter. 
From Fig.\,\ref{fig:ciipah}, the average ratios are ([C\2]+[O\1]$_{63}$)/PAH$=7$\%\ and ([C\2]+[O\1]$_{63}$)/TIR$=0.55$\%, which can be considered as proxies for the photoelectric efficiency, if either PAHs or dust probed by TIR dominate the gas heating. We investigate the dominant heating source in Sects.\,\ref{sec:lh10} and\ \ref{sec:heating}.

\subsubsection{LH\,10}\label{sec:lh10}

The case of the stellar cluster LH\,10 is critical to understanding the dust component responsible for the gas heating in PDRs. LH\,10 is dominated by warm photoionized gas, and the gas heating is due to photoionization in addition to the photoelectric effect. Most important, relatively warmer dust contributes to the TIR emission that is not associated with the heating of the gas in PDRs probed by [C\2] and [O\1]$_{63}$.
In Sect.\,\ref{sec:heatingcooling}, we found that LH\,10 drives the scatter on the ([C\2]+[O\1]$_{63}$)/TIR ratio up. In contrast, the ([C\2]+[O\1]$_{63}$)/PAH ratio shows a small scatter throughout N\,11B (factor of $\approx2$). This shows that the PAH emission describes the gas heating well even toward the stellar cluster where other dust components exist (in particular stochastically heated small grains and warm grains in thermal equilibrium with the interstellar radiation field). 

The dust model presented in Sect.\,\ref{sec:oiii} indicates that the PAH abundance varies by a factor of $\sim10$ across N\,11B, with a minimal value toward LH\,10 where most PAHs are likely destroyed by the radiation field. Still, ([C\2]+[O\1]$_{63}$)/PAH remains constant within a factor of $\approx2$ as compared to other regions in N\,11B. 
PDRs therefore exist toward LH\,10 and PAH emission is a good proxy for the gas-heating rate in these PDRs even toward such an extreme environment. The PAH/TIR ratio is the lowest toward LH\,10, and only a fraction of TIR must originate in PDRs, about $\sim40$\%. We investigate the dominant heating source in Sect.\,\ref{sec:heating}.

\subsubsection{Dominant gas heating source}\label{sec:heating}

The gas heating should be dominated by very small grains and PAHs (Sect.\,\ref{sec:heatingtracers}). 
The $24$\mic\ dust continuum, which is often used to trace warm dust grains and stochastically heated small grains, correlates poorly with [C\2]+[O\1]$_{63}$, with a scatter $\sim5$ times larger than the correlation with PAH emission (not shown in Fig.\,\ref{fig:ciipah}). 
On the other hand, we find a remarkable correlation between [C\2]+[O\1]$_{63}$ and PAH emission, suggesting that PAHs could dominate the gas heating over the dust components probed by TIR emission (Sect.\,\ref{sec:heatingcooling}). However, a similarly small scatter exists with TIR emission if we ignore the LH\,10 region (Sect.\,\ref{sec:lh10}). Croxall et al.\ (2012) also find a smaller scatter with ([C\2]+[O\1]$_{63}$)/PAH than with ([C\2]+[O\1]$_{63}$)/TIR and concludes that the integrated dust emission does not describe well the photoelectric heating where [C\2] and [O\1]$_{63}$ originate. Instead, photoelectrons ejected from PAHs would dominate the gas heating (see also Helou et al.\ 2001; Rubin et al.\ 2009). 
We consider in the following PAHs and dust grains probed by TIR emission as potential gas-heating sources.

The total gas-cooling rate $\Lambda_{\rm g}$, traced by [C\2]+[O\1]$_{63}$, is equal to the gas heating rate resulting from the photoelectric effect on dust grains probed by TIR emission ($\Gamma_{\rm g}^{\rm TIR}$) and on PAHs ($\Gamma_{\rm g}^{\rm PAH}$):
\begin{equation}
\Lambda_{\rm g} = \Gamma_{\rm g}^{\rm TIR} + \Gamma_{\rm g}^{\rm PAH} = \epsilon_{\rm PE,TIR}\ \Gamma_{\rm d}^{\rm TIR} + \epsilon_{\rm PE,PAH}\ \Gamma_{\rm d}^{\rm PAH},
\end{equation}
where $\Gamma_{\rm d}^{\rm TIR}$ and $\Gamma_{\rm d}^{\rm PAH}$ are the dust heating rates for dust grains and PAHs, and $\epsilon_{\rm PE,TIR}$ and $\epsilon_{\rm PE,PAH}$ are the corresponding photoelectric efficiencies. Assuming that both dust components contribute to the gas heating, we conclude that upper limits on 
$\epsilon_{\rm PE,TIR}$ and $\epsilon_{\rm PE,PAH}$ are given by the ([C\2]+[O\1]$_{63}$)/(TIR,PAH) ratio, i.e., $\approx0.55$\%\ and $\approx7$\%\ respectively.

\begin{figure}
\centering
\includegraphics[angle=0,scale=0.45,clip=true]{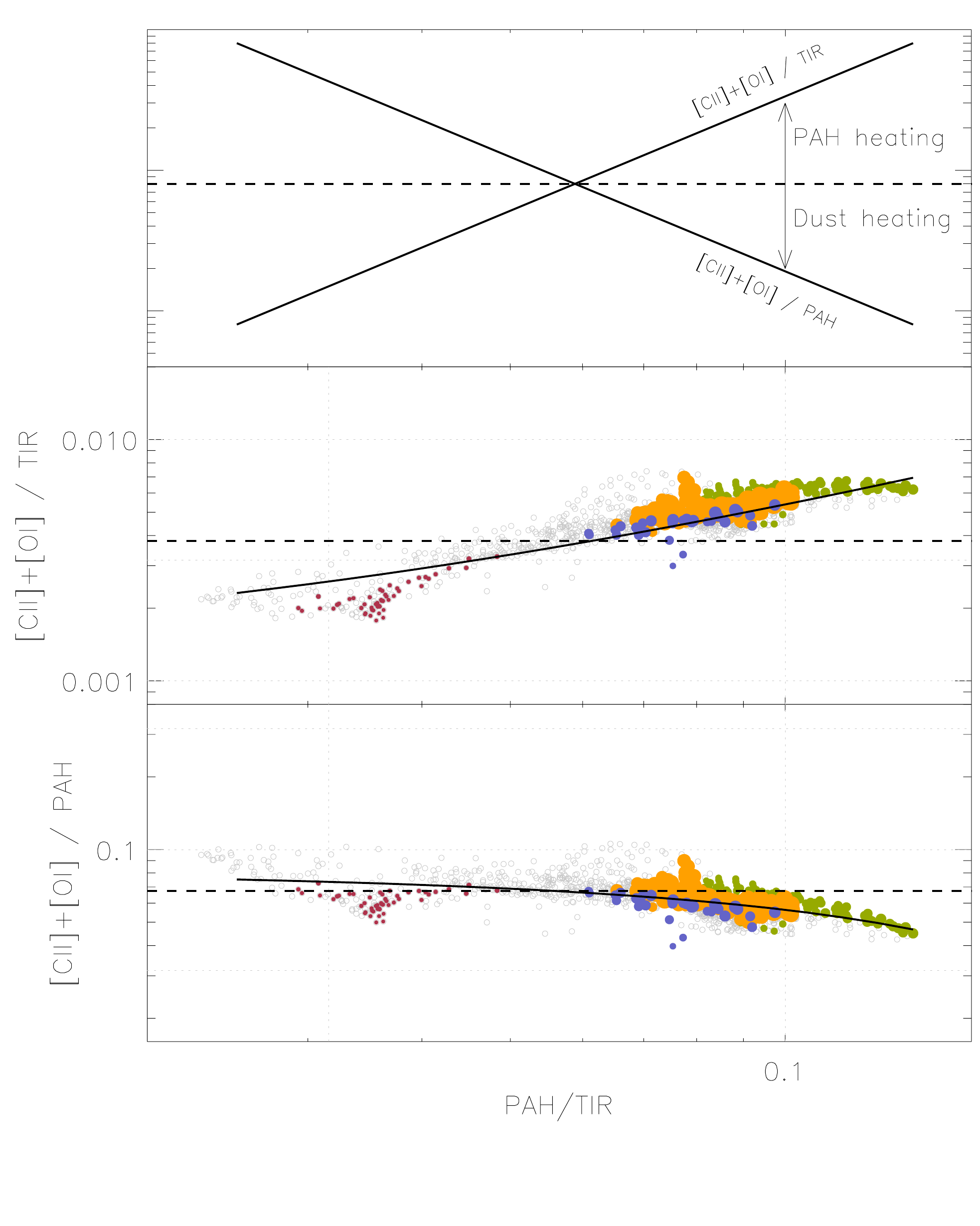}
\caption{
\textit{Top} $-$ Expected trend between ([C\2]+[O\1]$_{63}$)/(TIR,PAH) against PAH/TIR, assuming that either PAHs or the dust grains traced by TIR dominate the gas heating. 
\textit{Middle and bottom} $-$ Pixel-to-pixel correlation between ([C\2]+[O\1]$_{63}$)/(TIR,PAH) and PAH/TIR. The solid line represents the linear regression and the dotted line represents the median value. See Fig.\,\ref{fig:ciipah} for the plot description. 
\label{fig:cii_heating}}
\end{figure}

The PAH/TIR ratio varies by a factor of $\approx5$ across N\,11B (Fig.\,\ref{fig:cii_heating}), enabling us to study the influence of PAH and TIR emission as gas-heating tracers. 
Figure\,\ref{fig:cii_heating} shows that ([C\2]+[O\1]$_{63}$)/PAH and ([C\2]+[O\1]$_{63}$)/TIR are a smooth function of the PAH/TIR ratio. The standard deviation around the ([C\2]+[O\1]$_{63}$)/PAH linear regression is remarkably small, with only $6$\%, while it is $15$\%\ for ([C\2]+[O\1]$_{63}$)/TIR. 
The top panel in Fig.\,\ref{fig:cii_heating} shows the trend of the ([C\2]+[O\1]$_{63}$)/TIR and ([C\2]+[O\1]$_{63}$)/PAH ratios versus PAH/TIR. When PAHs dominate the gas heating, the ([C\2]+[O\1]$_{63}$)/TIR ratio increases with PAH/TIR, while the inverse is true for ([C\2]+[O\1]$_{63}$)/PAH. 
We normalized each ratio to its average value and calculated that the linear regression slope is $\approx-3.6$ for ([C\2]+[O\1]$_{63}$)/PAH, while it is $\approx10.3$ for ([C\2]+[O\1]$_{63}$)/TIR. From this, we conclude that PAHs dominate the gas heating over the dust component probed by TIR emission. This implies that $\epsilon_{\rm PE,PAH}$ is close to $\approx7$\%, and therefore a factor of $\gtrsim12$ more than $\epsilon_{\rm PE,TIR}$. It is possible that $\epsilon_{\rm PE,TIR}$ is larger if we would only account for the gas heating from the smallest grains (Sect.\,\ref{sec:heatingtracers}).

Our results suggest that classical dust grains, which are dominating the TIR emission, do not dominate the gas heating. The fraction of the TIR emission corresponding to smaller grains certainly scales better with [C\2]+[O\1]$_{63}$ than the total TIR. Unfortunately, in the lack of constraints on the abundance of small grains, we cannot conclude anything about the relative importance of PAHs and very small grains as gas-heating sources.

\section{Conclusions}

We present the \textit{Herschel}/PACS spectroscopic observation of the H\2\ region N\,11B in the LMC. The spatial distribution of the FIR lines shows remarkable differences:
\begin{itemize}
\item The emission of [O\3]$_{88}$, [N\3], and [C\2] is quite flat across the map, suggesting it is dominated by extended emission. This contrasts with the [O\1]$_{63}$ emission, which is concentrated in a few compact regions. [O\1]$_{63}$ also shows an extended component, about ten times fainter than the peak emission.
\item The spatial distribution of [O\3]$_{88}$ appears to be anti-correlated with that of [C\2].
\end{itemize}

In the present paper, we investigate the extended emission of the FIR tracers, while in a second paper, we will investigate the compact regions and the PDR properties. The main results of the present study follow\\
\begin{enumerate}
\item \ The [O\3]$_{88}$ line, with its low critical density, is an excellent probe of the extended ionized gas. It is the brightest line across the region, in particular a factor of four brighter than [C\2] (in some regions $20$ times brighter). 
\item \ By comparing [O\3]$_{88}$ to the optical line [O\3] 5007\AA, we find an extinction of $A_V\sim1$ toward W2 and the stellar cluster LH\,10, implying that a significant fraction of the ionized gas is hidden in the optical. The dust mass measured in N\,11B is compatible with such extinctions. The extinction $A_V$ is between $\sim0.5$ and $\sim1$ toward the other regions. 
\item \ The spatial extent of [O\3]$_{88}$ emission could be reproduced by considering all the confirmed O stars, provided the density of the ionized gas is generally lower than $\lesssim16$\cc.   
\item \ We used the [N\3]/[O\3]$_{88}$ and [N\2]$_{122}$/[N\3] ratios to estimate the physical conditions in the ionized gas. The average ionization parameter is compatible with relatively low-density ($\lesssim25$\cc) gas located on average $\lesssim30$\,pc away from the ionizing sources.
\item \ By comparing [C\2] to the low-excitation, low-density ionized gas tracer [N\2]$_{122}$, we find that [C\2] mainly arises from PDRs across N\,11B. 
\item \ We examined the relation between the gas cooling in PDRs via [C\2] and [O\1]$_{63}$ and the gas heating traced by TIR emission and PAH emission. We find a tight correlation between [C\2]+[O\1]$_{63}$ and the PAH emission. Assuming [C\2] and [O\1]$_{63}$ are the dominant coolants at low $A_V$, the sum of [C\2] and [O\1]$_{63}$ provides a proxy for the total cooling, with [C\2] being the dominant coolant in most regions characterized by a diffuse ISM, and [O\1]$_{63}$ contributing significantly (up to $\approx50$\%) in the brightest and densest regions. 
\item \ The scatter of the ([C\2]+[O\1]$_{63}$)/TIR ratio is greater than that of ([C\2]+[O\1]$_{63}$)/PAH, which is mostly due to the stellar cluster LH\,10 where another dust component, unrelated to PDRs, drives TIR to higher values. 
\item \ We find that PAHs dominate the gas heating in PDRs, with a photoelectric efficiency of $\sim7$\%.
\end{enumerate}

\begin{acknowledgements}
PACS has been developed by a consortium of institutes led by MPE (Germany) and including UVIE (Austria); KU Leuven, CSL, IMEC (Belgium); CEA, LAM (France); MPIA (Germany); INAF-IFSI/OAA/OAP/OAT, LENS, SISSA (Italy); IAC (Spain). This development has been supported by the funding agencies BMVIT (Austria), ESA-PRODEX (Belgium), CEA/CNES (France), DLR (Germany), ASI/INAF (Italy), and CICYT/MCYT (Spain).
HCSS, HSpot, and HIPE are joint developments by the \textit{Herschel} Science Ground Segment Consortium,
consisting of ESA, the NASA \textit{Herschel} Science Center, and the HIFI, PACS and SPIRE consortia. We thank Y.\ Naz{\'e} for providing us with the calibrated HST images. This research made use of Montage, funded by the National Aeronautics and Space Administration's Earth Science Technology Office, Computation Technologies Project, under Cooperative Agreement Number NCC5-626 between NASA and the California Institute of Technology. Montage is maintained by the NASA/IPAC Infrared Science Archive.
\end{acknowledgements}

\appendix

\section{The PACSman suite for \textit{Herschel}/PACS spectroscopy data analysis}\label{sec:pacsman}

\texttt{PACSman}\footnote{Available at \textit{http://www.myravian.fr} or by email contact (\textit{myravian@gmail.com}).} is an IDL package designed to provide an alternative for several reduction and analysis steps performed in HIPE on PACS spectroscopic data.
The following operations are currently included: transient correction, line fitting, map projection, and map analysis. All of the observation modes are supported: wavelength switching (now decommissioned), unchopped scan, and chop/nod.

\subsection{Transient correction}

Transients are cosmic ray hits to the detector that modify the pixel response over a given period of time (see the works of Coulais \&\ Abergel 1999; 2000 for the ISOPHOT and ISOCAM detectors on ISO). Two types of transients are observed in the PACS data, ``dippers" that result in a lower pixel response and ``faders" that result in a larger pixel response. The response after a dipper hit normalizes after a short time (about the timescale of a line scan), while the response after a fader hit takes considerably longer to settle (near the observation itself or even more). A fader is often seen after the calibration block, which observes a bright source. The transient correction is only necessary for unchopped scan observations since the fast chopping in chop/nod observations allows reliable removal of the transients (dippers and faders). 

\subsubsection{Faders}

Faders are responsible for large jumps in the flux, which can render a spectral pixel useless unless its signal is aligned with the others. HIPE includes a task called \textit{specLongTermTransient} that minimizes the scatter of points by using a series of exponential functions to model the ``faders" transients (Fadda \&\ Jacobson 2011). However, it is possible that several transients overlap in time, which makes it difficult to find the number of exponential functions required and the parameters of each function. This is the main reason we developed a special routine in \texttt{PACSman} that performs the transient correction on the signal timeline for each of the 16 spectral pixels and each of the 25 spaxels. 

\begin{figure}
\includegraphics[angle=0,scale=0.36,clip=true]{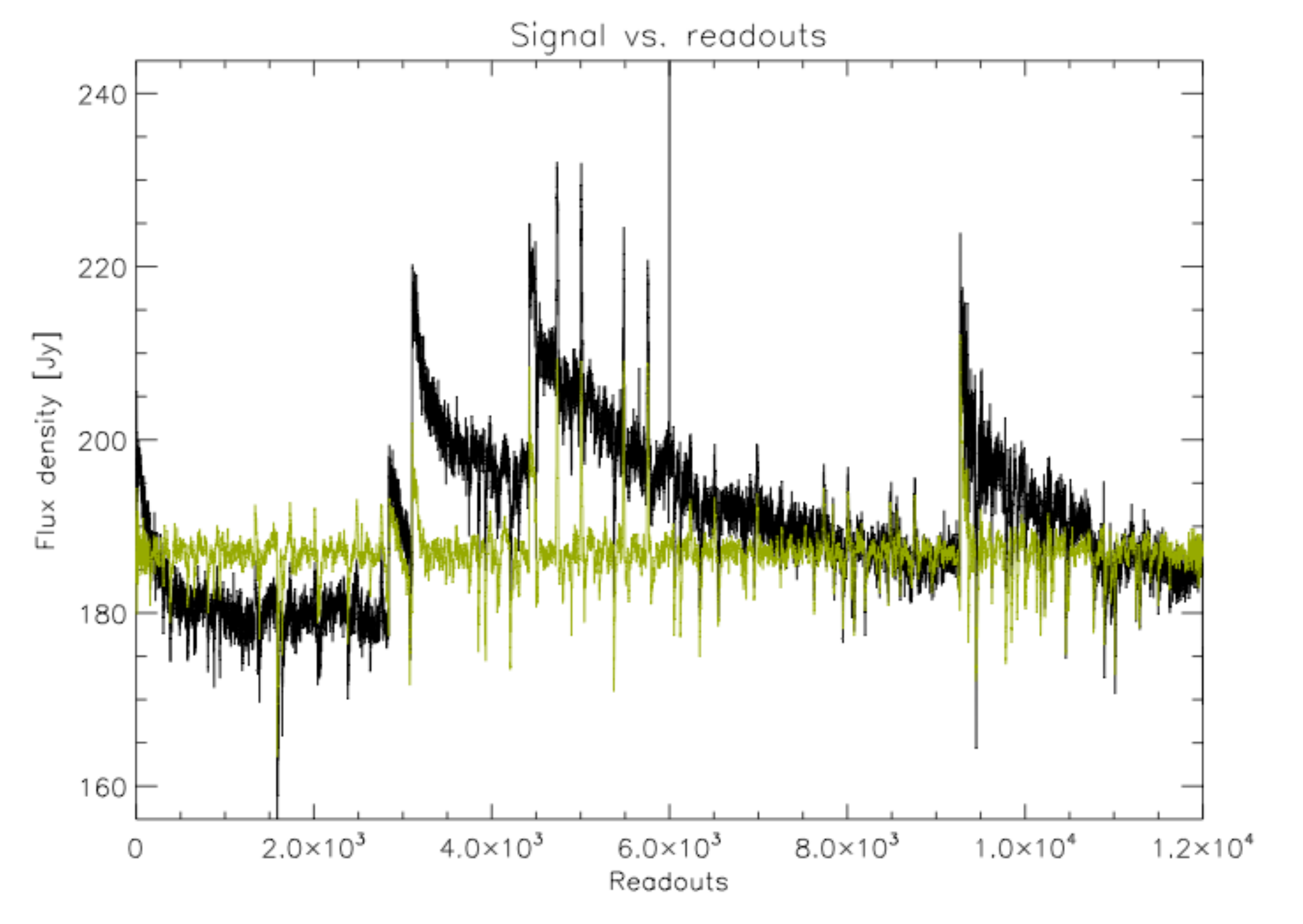}
\caption{Example of fader transients occurring in unchopped scan observations. The signal is plotted against the readout number for a given spaxel and spectral pixel. The irregular black curve is the unflagged data, while the flat green curve shows the corrected signal. Several transients are observed in this timeline, including the systematic transient after the calibration block on the extreme left.
\label{fig:fader_example}}
\end{figure}

\texttt{PACSman} includes a correction tool that uses a multiresolution algorithm to straighten the signal (Starck et al.\ 1999). The smallest resolution scales are subtracted to remove the noise and the largest scale is subtracted to remove any fader-type signal (Fig.\,\ref{fig:fader_example}). The main advantage of multiresolution over an exponential curve fitting is its ability to remove the signal from several faders that are overlapping in time. Furthermore, it provides an unbiased approach since the faders do not need to be identified beforehand. 

\subsubsection{Dippers}

\begin{figure}
\includegraphics[angle=0,scale=0.34,clip=true]{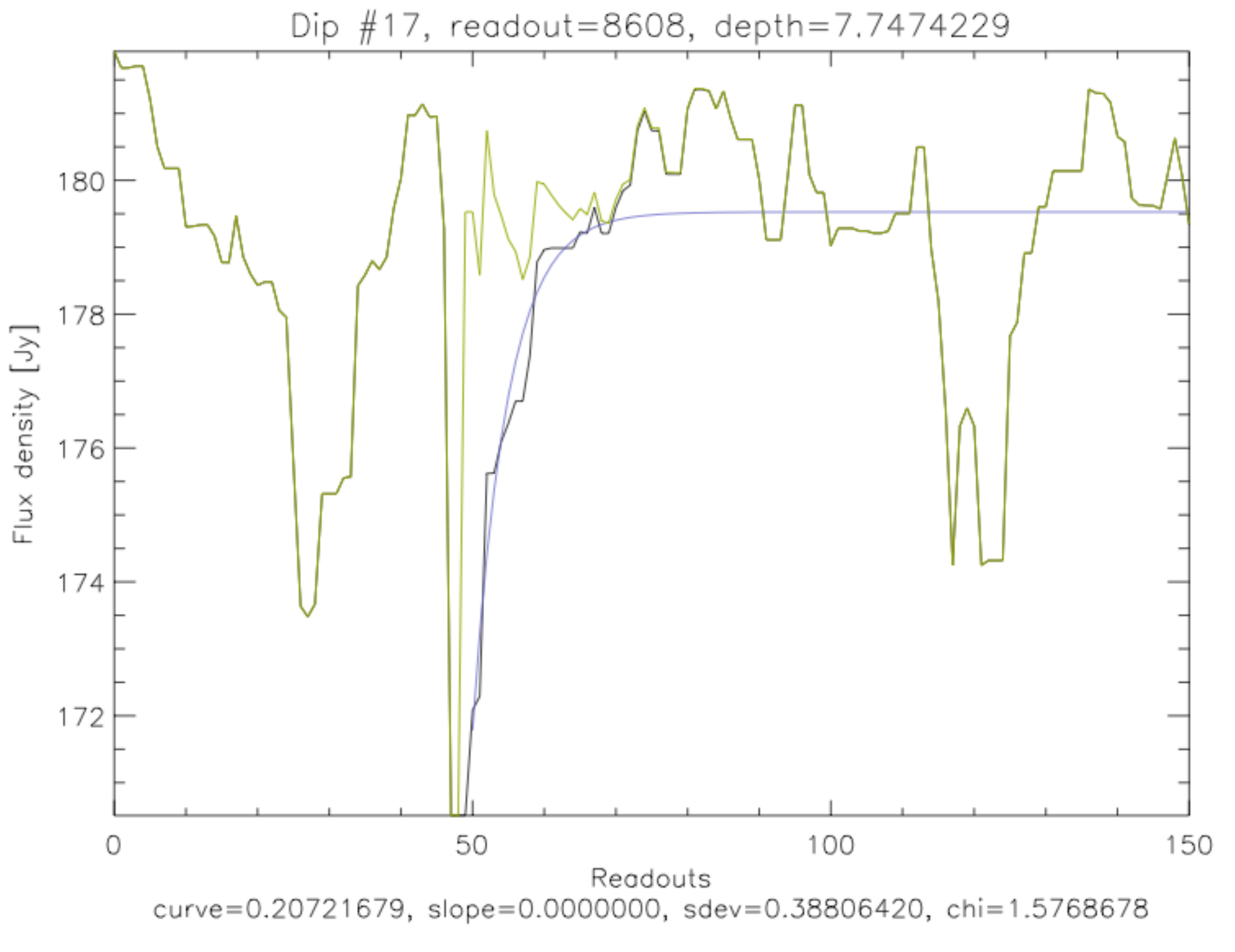}
\caption{Example of a dipper fit. The profile is fitted by an exponential curve whose parameters are compared to the expected behavior of a dipper (in particular the timescale). 
\label{fig:dipper_example}}
\end{figure}

The presence of dippers dominates the S/N in the timeline (thus in the spectra as well). Dippers are searched for after each flagged pixel. An exponential curve is fitted with free parameters. If the timescale parameter is consistent with a dipper signal, the fit is used to correct the signal (Fig.\,\ref{fig:dipper_example}).

\subsection{Line fitting}

The input data for line fitting and map projection (Sect.\,\ref{sec:pacsman_map}) is the Level 1 cube reduced with HIPE. At this level of reduction, the data is calibrated in flux and in wavelength. The bad pixels are masked according to the HIPE reduction criteria (outliers, glitches, saturated, noisy, etc.). 
The data cube contains one spectral window for each of the 25 spaxels and 16 spectral pixels (and for a given raster position when mapping). The HIPE reduction corrects for the flat-fielding of the spectral pixels so that the combination of the spectral windows from the 16 spectral pixels provides one spectrum per spaxel. The data cloud for a given spaxel typically contains a few million points (for line spectroscopy). 

The line fitting in \texttt{PACSman} is performed on the full data cloud; i.e., the data is not rebinned beforehand. This is to keep all the possible information on the data and on the masks during the line fitting process. The data cloud does not provide propagated errors for the individual measurements. We estimate the error by calculating the dispersion of measurements in a given wavelength bin (defined as $1/5$ of the line FWHM). By default, a $5\sigma$ clipping is performed prior the line fitting. 

Line fitting is then performed with the MPFIT package\footnote{http://www.physics.wisc.edu/~craigm/idl/fitting.html}. 
The continuum is adjusted first using spectral windows on each side of the line. The result is then used as an initial guess for the final fit that combines the continuum polynomial and the Gaussian curve. Default constraints exist that can be changed on the line FWHM, line central wavelength, and continuum range around the line. It is essential that constraints exist to provide reliable upper limits in faint lines. For a chop/nod observation, the two nods can either be combined in a single spectrum or considered individually. The fitting is performed as follows: (1) the continuum is fitted first around the line, and (2) the line+continuum profile is fitted using the continuum fit as a first guess. Example of fits are shown in Fig.\,\ref{fig:fits}.

\subsection{Map projection}\label{sec:pacsman_map}

For a mapping observation, the raster footprints ($5\times5$ spaxel image of the line flux) have to be combined to create the final map. Maps are projected on a subpixel grid with a pixel size three times smaller (linear scale) than the spaxel size. The pixel size is about $3\arcsec$. This allows recovering the best spatial resolution possible (see PACS ICC calibration document PICC-KL-TN-038). 

\begin{figure*}
\centering
\includegraphics[angle=0,scale=0.5,clip=true]{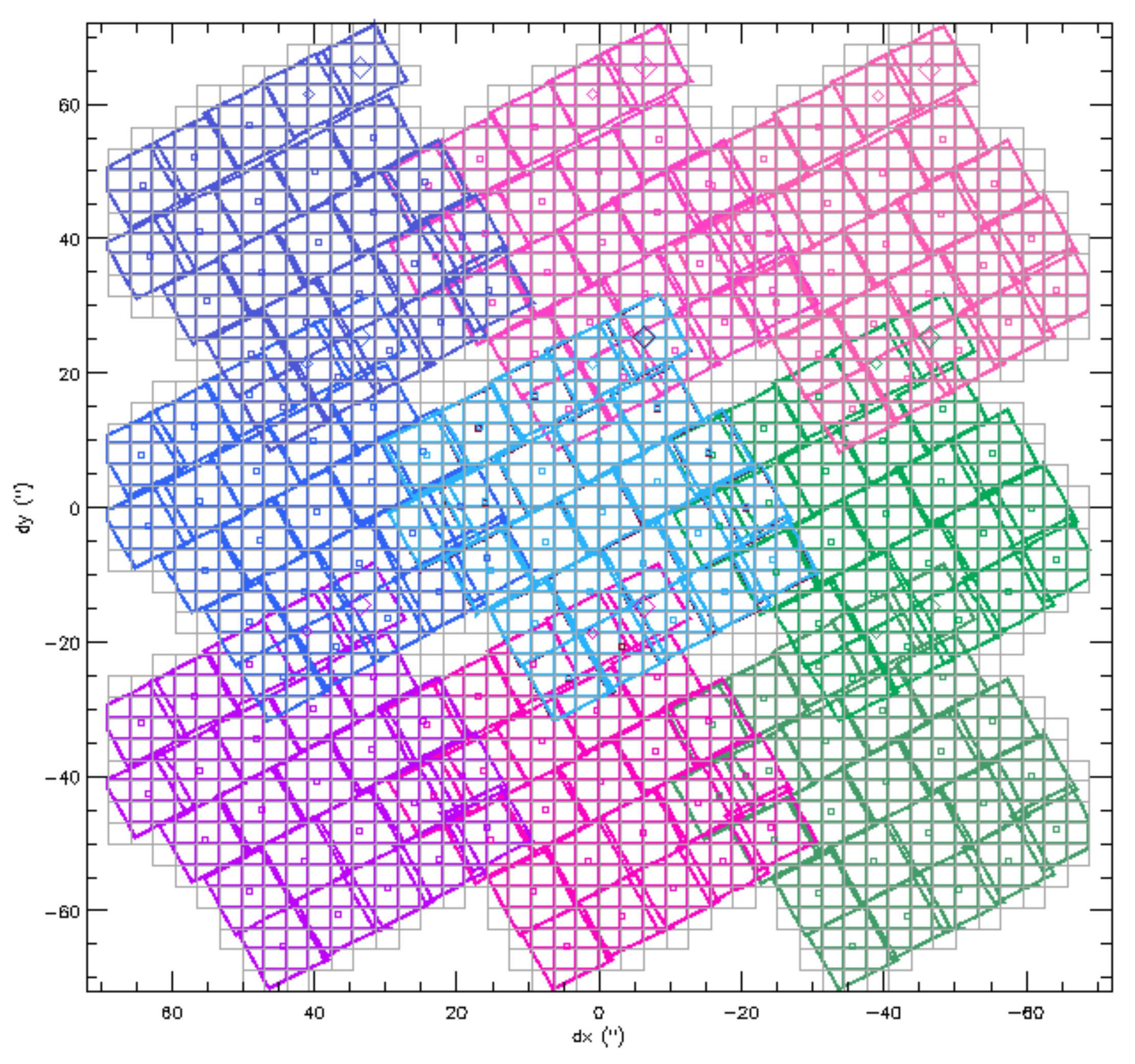}
\caption{Example of a map projection. The observation ($3\times3$ raster footprints) is shown with color symbols. The squares show the spaxel center ($5\times5$ spaxels for one raster position), the large diamond shows the spaxels (1,1), and the dots show the center of the subspaxel ($1/3$ of the size of a spaxel). The information from different spaxels and rasters can be combined to populate the projected grid. Subpixels where there is no coverage are shown as gray squares.
\label{fig:projected_grid}}
\end{figure*}

The grid orientation is chosen to be orthogonal with the raster maps so that the splitting of the spaxels requires as little rotation as possible. 
Since the spaxel positions and sizes are not regular, the orthogonality is only approximate. 
For a given pixel in the projected grid, the contributing fraction of each spaxel in the map is calculated assuming a uniform surface brightness within a spaxel (Fig.\,\ref{fig:projected_grid}). Several determinations are thus obtained for each pixel in the projected grid. The final image simply averages these determinations to obtain one value per pixel. 

\subsection{Map analysis}

\begin{figure*}
\includegraphics[angle=0,scale=0.4,clip=true]{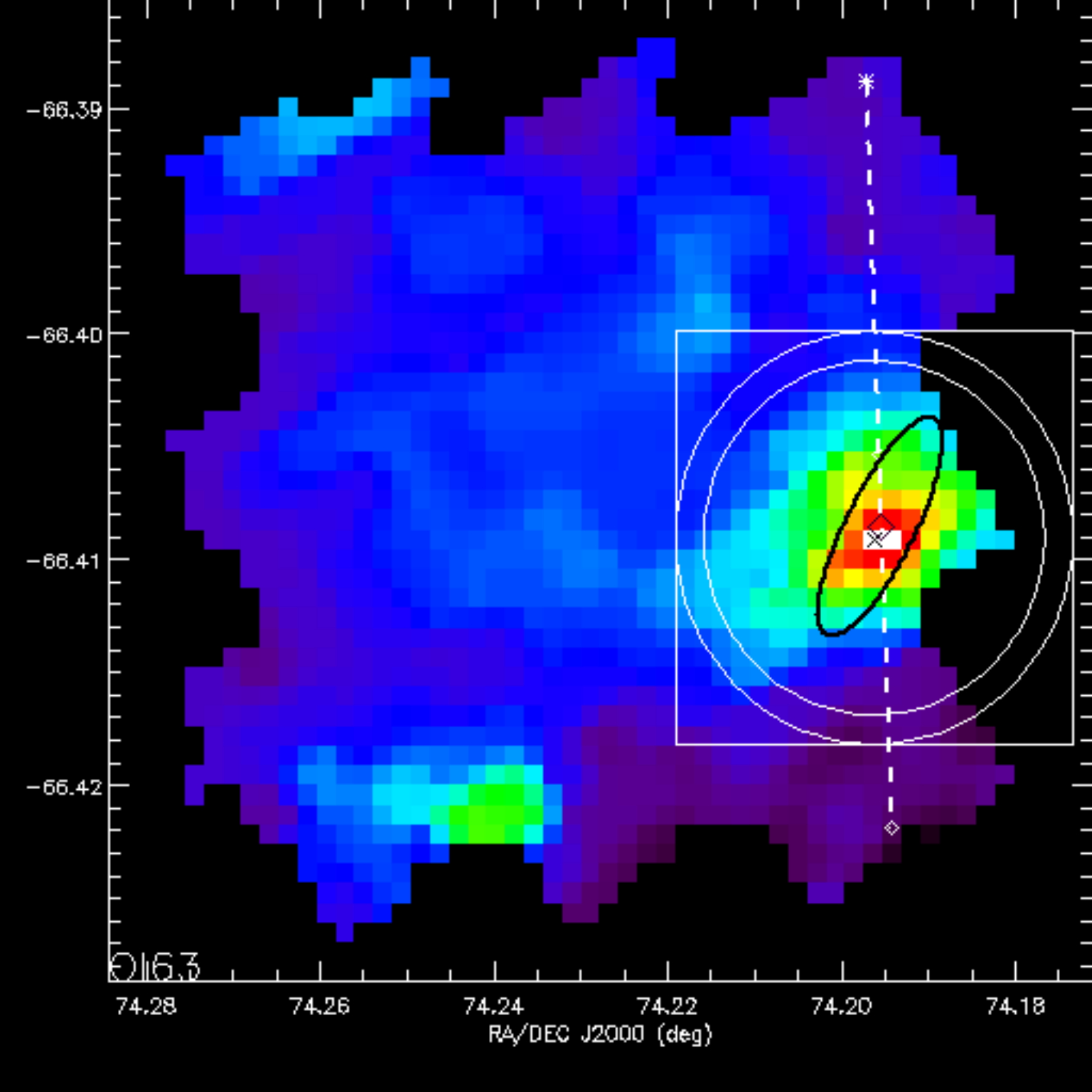}
\includegraphics[angle=0,scale=0.41,clip=true]{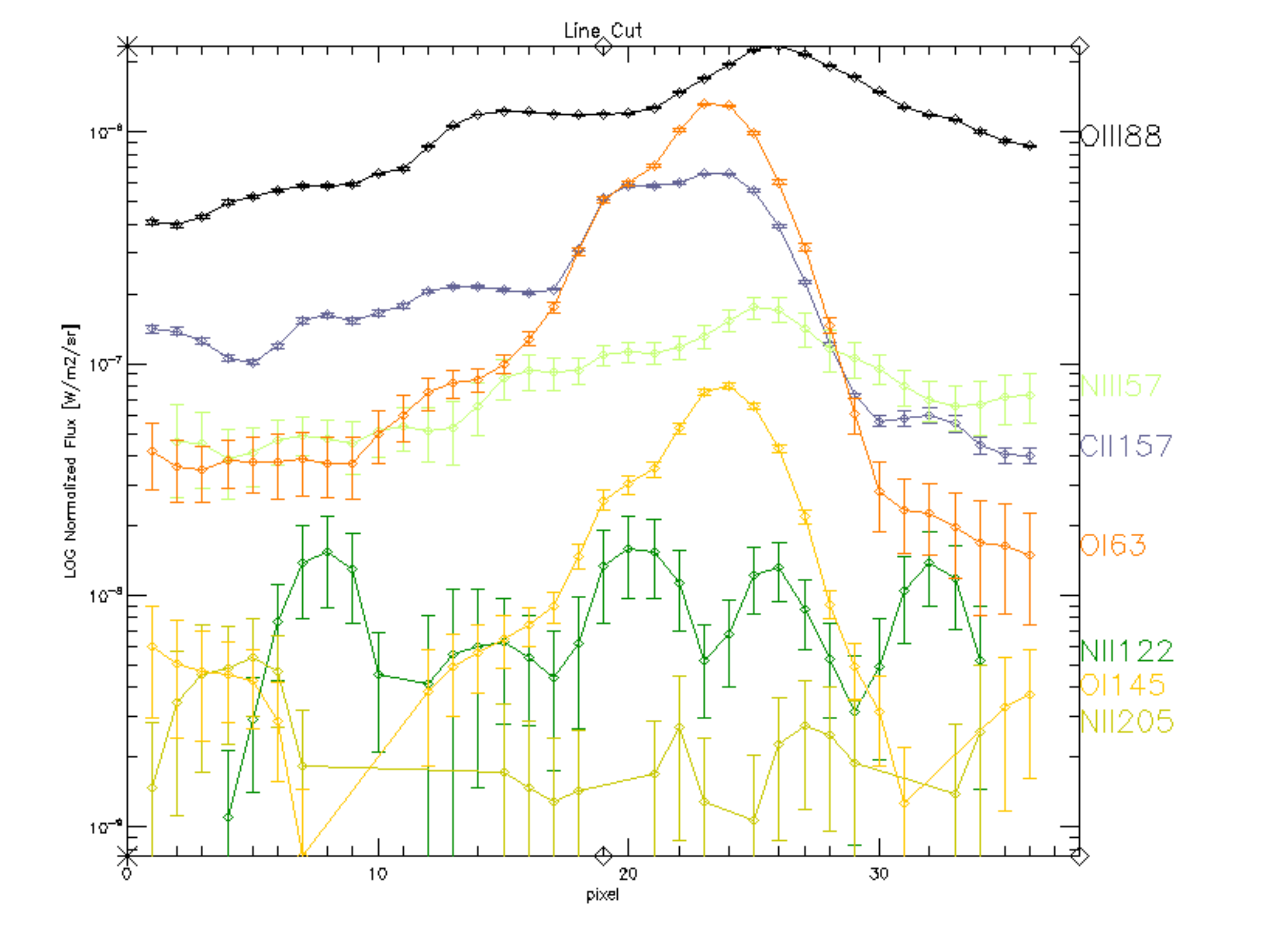}
\caption{Example of a line cut across the map. The circles on the left image show the aperture used for extraction while the line shows the profile cut. 
\label{fig:linecut}}
\end{figure*}

A tool is provided to analyze the spectral maps. Several tasks can be performed, such as aperture extractions with annulus subtraction (see Figs.\,\ref{fig:linecut}). A 2D Gaussian surface can also be fitted to the data to estimate the source spatial FWHM. Finally, spatial profiles can also be plotted along a given cut.

\end{document}